\RequirePackage{fix-cm}
\documentclass[5p,sort&compress]{article}
\usepackage{arxiv}

\usepackage{graphicx}
\usepackage{siunitx}
\usepackage{tikz}
\usetikzlibrary{arrows,arrows.meta,positioning,calc,math}
\tikzset{
 arrow/.style={-latex, shorten >=1ex, shorten <=1ex}
 }
\usepackage{pgfplots}
\pgfplotsset{compat=1.16}
\usepackage{geometry}

\usepackage{xcolor}
\definecolor{azure}{rgb}{0.0, 0.5, 1.0}
\definecolor{awesome}{rgb}{1.0, 0.13, 0.32}
\definecolor{asparagus}{rgb}{0.53, 0.66, 0.42}
\definecolor{cadetgrey}{rgb}{0.57, 0.64, 0.69}

\usepackage{xfrac}

\usepackage{booktabs}

\usepackage[hidelinks]{hyperref}
\usepackage{bm}
\usepackage{mathptmx}      % use Times fonts if available on your TeX system
\usepackage{amssymb}
\usepackage{amsmath}
\usepackage{nicefrac}

\usepackage{subfig}

\usepackage{algorithmicx}
\usepackage{algpseudocode}
\usepackage{algorithm}
\usepackage{moresize}

\usepackage{placeins}
\newcommand{\xb}{\mathbf{x}}% TODO revisit
\newcommand{\fb}{\mathbf{f}}
\newcommand{\ub}{\mathbf{u}}
\newcommand{\R}{\mathbb{R}}

\newcommand{\vb}{\mathbf{v}}
\newcommand{\bb}{\mathbf{b}}
\newcommand{\ab}{\mathbf{a}}
\newcommand{\Nb}{\mathbf{N}}
\newcommand{\Kb}{\mathbf{K}}
\newcommand{\Bdx}{\,\mathrm{d}\xb}
\newcommand{\Bds}{\,\mathrm{d}s}
\newcommand{\Bspan}[1]{\mathrm{span}\big\langle{#1}\big\rangle}

\newcommand{\traction}{\bar{\mathbf{t}}}

\newcommand{\mySubScriptSize}{\ssmall}
\newcommand{\hPum}{h_\text{\mySubScriptSize{PUM}}}
\newcommand{\hPd}{h_\text{\mySubScriptSize{PD}}}
\newcommand{\deltaCoarse}{\delta_\text{\mySubScriptSize{coarse}}}
\newcommand{\deltaFine}{\delta_\text{\mySubScriptSize{fine}}}
\newcommand{\hCoarse}{\hPd^\text{\mySubScriptSize{coarse}}}
\newcommand{\hFine}{\hPd^\text{\mySubScriptSize{fine}}}

\begin{document}

%\begin{frontmatter}

\author{Matthias Birner \\
Institute for Numerical Simulation, University of Bonn, Bonn, Germany \\
Fraunhofer SCAI, Sankt Augustin, Germany\\
\AND
        Patrick Diehl \\
        Center for Computation \& Technology, Louisiana State University, Baton Rouge, LA
        \AND
        Robert Lipton \\
        Department of Mathematics, Louisiana State University, Baton Rouge, LA\\
        Center for Computation \& Technology, Louisiana State University, Baton Rouge, LA
        \AND
        Marc Alexander Schweitzer\\
        Institute for Numerical Simulation, University of Bonn, Bonn, Germany \\
Fraunhofer SCAI, Sankt Augustin, Germany
}

\title{A Fracture Multiscale Model for Peridynamic enrichment within the Partition of Unity Method%\thanks{Grants or other notes
}

%\author[ins,scai]{Matthias Birner\corref{mycorrespondingauthor}}
%\ead{matthias.birner@scai.fraunhofer.de}

%\author[cct]{Patrick Diehl}
%\ead{pdiehl@cct.lsu.edu}
%\ead[url]{http://orcid.org/0000-0003-3922-8419}

%\author[lsu]{Robert Lipton}
%\ead{lipton@lsu.edu}

%\author[ins,scai]{Marc Alexander Schweitzer}
%\ead{schweitzer@ins.uni-bonn.de}

%\address[ins]{Institute for Numerical Simulation, University of Bonn, Bonn, Germany}

%\address[scai]{Fraunhofer SCAI, Sankt Augustin, Germany}

%\address[cct]{Center for Computation \& Technology, Louisiana State University, Baton Rouge, LA}

%\address[lsu]{Department of Mathematics, Louisiana State University, Baton Rouge, LA}

%\begin{keyword}
%Peridynamic \sep Partition of Unity \sep %Multiscale \sep Global-local Method
%\end{keyword}

\maketitle

\begin{abstract}
Partition of unity methods (PUM) are of domain decomposition type and provide the opportunity for multiscale and multiphysics numerical modeling. Different physical models can exist within a PUM scheme for handling problems with zones of linear elasticity and zones where fractures occur. Here, the peridynamic (PD) model is used in regions of fracture and smooth PUM is used in the surrounding linear elastic media. The method is a so-called global-local enrichment strategy, see \cite{duarte2007global, birner2017global}. The elastic fields of the undamaged media provide appropriate boundary data for the localized PD simulations. The first steps for a combined PD/PUM simulator are presented. In part I of this series, we show that the local PD approximation can be utilized to enrich the global PUM approximation to capture the true material response with high accuracy efficiently. Test problems are provided demonstrating the validity and potential of this numerical approach.
\end{abstract}

%\end{frontmatter}




%\begin{abstract}
%\keywords{Peridynamic \and Partition of Unity\and Coupling}
%\subclass{MSC code1 \and MSC code2 \and more}
%\end{abstract}

\cleardoublepage
\newpage
\twocolumn

%%%%%%%%%%%%%%%%%%%%%%%%%%%%%%%%%%
\section{Introduction}
%%%%%%%%%%%%%%%%%%%%%%%%%%%%%%%%%%
Peridynamics (PD) is a non-local generalization of classical continuum mechanics~\cite{silling2005meshfree} and naturally allows for the direct modelling of fracture phenomena and damage within its original formulation. It can give good agreement with experimental data~\cite{diehl2019review} in complex deformation scenarios. \textcolor{black}{Another notable approach are phase-field models~\cite{bourdin2000numerical} which similar benefits and challenges as PD models. For a comparison of these two models, we refer to~\cite[Section 5]{diehl_lipton_wick_tyagi_2021}}. However, the computational expense incurred by exclusive use of peridynamics for large problems  where damage and fracture only appear across  small regions
and most of the body behaves linearly elastic, can be needlessly high. 
%\emph{i.e.}\ in a way that is well captured by the classical PDE model
Therefore, adaptive approaches have been proposed for such problems, \emph{e.g.}\ force-based blending~\cite{seleson2013force,seleson2015concurrent,silling2020Couplingstresses}, the Arlequin approach~\cite{han2012coupling}, the variable horizon method~\cite{silling2015variable, NIKPAYAM2019308}, or discrete coupling~\cite{liu2012coupling,shojaei2016coupled,zaccariotto2018coupling,madenci2018coupling,kilic2010coupling,GALVANETTO201641,zaccariotto2017enhanced}. These methods directly couple PD discretizations and finite element (FE) models of linear elasticity to overcome the computational expense. These so-called concurrent coupling approaches introduce a non-physical interface (or overlapping handshake zone) where the two computational models, the PD and linear elastic discretizations, must agree on the computed deformation behavior in a delicate way, and it is not obvious how computational artifacts at this interface can be avoided in general. For example, in~\cite{ONGARO2021113515} an out-of-balance force in the coupling region due to a lack of balance between the local and non-local traction is reported. For more details, we refer to the review paper~\cite{d2019review}.

An alternative to the concurrent coupling approaches are so-called hierarchical coupling techniques where scale separation is assumed, and a damage model together with an elastic model are utilized successively,  exchanging information in a global and accumulated fashion, see e.g. \cite{FishBook} and references therein.
For instance, in multiscale finite element methods like FE$^2$ methods two FE calculations are carried out in a nested manner, one at the macroscale and the other at the microscale (in the vicinity of every quadrature point of the macroscale model). Obviously, such approaches yield a tremendous computational cost and are typically not trivial to implement, which prohibits their widespread use.
A number of multiscale PD concepts have been proposed~\cite{littlewood2016peridynamic,xu2016multiscale,bobaru2009convergence,rahman2014multiscale,xu2016multiscale,Rahman_2014,doi:10.1063/1.4971634}, however, many open questions remain. With respect to partition of unity methods, attempts for multiscale coupling with molecular dynamics and smoothed particle hydrodynamics~\cite{talebi2014computational} were made.

We will pursue yet another approach to overcome the computational expense of PD discretization, based on the partition of unity (PU) approach \cite{schweitzer2003phd}. 
The overall process we envision is to construct a combined PD/PU simulator which automatically determines the regions where PD descritizations and linear elastic FE discretization and should be employed so that the resulting local PD approximation can be utilized to construct a respective multiscale enrichment function for the global partition of unity method (PUM) approximation to incorporate the true material response including fracture growth with high accuracy efficiently.
In this paper, we present the first steps towards the realization of this process.
Namely, we study the compatibility of the different material models and the methods required for combining them. To accomplish this, we seek to outline the transfer of information between them.
To this end, we study several reference cases (with and without fracture growth) to validate the fundamental assumptions of our approach.
As this paper provides a proof of concept, the reference cases considered are not too complex in geometry and loading.

The paper is structured as follows:
Section 2 introduces the preliminaries of the PUM and PD theory.
Section 3 sketches the approach to use peridynamics to construct appropriate multiscale enrichments on a local sub-domain for the global PUM.
In Section 4 numerical results are presented, which validate isolated steps of the overall multiscale enrichment scheme.
Specifically, we validate that PD and PUM simulate a similar enough physical problem despite different model formulations, in order to be able to combine them.
The issues here primarily are the different treatment of boundary conditions and the underlying material models.
Moreover, we show examples of information transfer in both directions: using the PD computed crack path in a PUM simulation and using a PUM displacement as boundary conditions on a PD problem.
The results clearly show the validity and potential of the presented approach.
Finally, Section 5 concludes the paper with some remarks on future research steps.

%%%%%%%%%%%%%%%%%%%%%%%%%%%%%%%%%%
\section{Preliminaries}
\label{sec:preliminaries}
%%%%%%%%%%%%%%%%%%%%%%%%%%%%%%%%%%

%%%%%%%%%%%%%%%%%%%%%%%%%%%%%%%%%%
\subsection{Classical continuum mechanics}
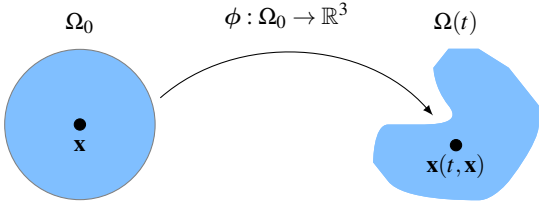
\begin{figure}[htbp]
\centering
\begin{tikzpicture}
\draw[fill=azure, opacity=0.5](1,1) circle (1);
 \node at (1,2.35) {{\small $\Omega_0$}};
 \draw[fill=black](1,1) circle (0.075);
 \node at (1,0.725) {{\small $\xb$}};
 \node at (6,1) {\includegraphics[scale=1.25]{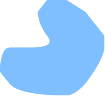}};
 \node at (6,2.35) {{\small $\Omega(t)$}};
 \draw[fill=black](6,0.725) circle (0.075);
 \node at (6,0.45) {{\small $\xb(t,\xb)$}};
 \node at (3.75,2.45) {{\small $\bm{\phi}:\Omega_0\rightarrow \R^3$}};
 \draw [arrow, bend angle=45, bend left] (1.95,1.25) to (5.8,1);
\end{tikzpicture}
\caption[The continuum in the reference configuration $\Omega_0$ and after the deformation $\bm{\phi} : \Omega_0 \rightarrow \R^3$ in the current configuration $\Omega(t)$ at time $t$.]{The continuum in the reference configuration $\Omega_0$ and after the deformation $\bm{\phi} : \Omega_0 \rightarrow \R^3$ with $\det(\text{grad}\;\bm{\phi}) > 0$ in the current configuration $\Omega(t)$ at time $t$.}
\label{fig::chapter2:01}
\end{figure}

%%%%%%%%%%%%%%%%%%%%%%%%%%%%%%%%%%
Before we introduce the two models, we will look briefly into the basics of classical continuum mechanics (CCM). Figure~\ref{fig::chapter2:01} shows a continuum in the reference configuration $\Omega_0 \subset \R^3$ where a material point is identified with its position $\xb \in \R^3$. The reference configuration $\Omega_0$ refers to the shape of the continuum at rest with no internal forces. The deformation $\bm{\phi} : [0,T] \times \R^3 \rightarrow \R^3$ of a material point $\xb \in \Omega_0$ to the so-called current configuration $\Omega(t)$ at time $t\in [0,T]$ is given as

\begin{equation}
\bm{\phi} (t, \xb) := \bm{id}(\xb) + \ub(t,\xb) = \xb(t, \xb)
\end{equation}

where $\ub : [0,T] \times \R^3 \rightarrow \R^3$ refers to the displacement

\begin{equation}
    \ub(t,\xb) := \xb(t,\xb) - \xb\text{.}
\end{equation}

The stretch $s : [0,T] \times \R^3 \times \R^3 \rightarrow \R^3$ between two material points $\xb$ and $\xb'$ in the reference configuration $\Omega_0$ is defined by

\begin{equation}
    s(t,\xb,\xb') := \bm{\phi}(t,\xb') - \bm{\phi}(t,\xb) = \xb(t,\xb) - \xb(t,\xb')\text{.}
\end{equation}

%%%%%%%%%%%%%%%%%%%%%%%%%%%%%%%%%%
\subsection{Partition of unity method}
%%%%%%%%%%%%%%%%%%%%%%%%%%%%%%%%%%
Partition of unity methods are a class of methods to numerically solve partial differential equations (PDEs).
They were introduced in~\cite{melenk1996partition, babuvska1997partition} to overcome limitations in the choice of basis functions of classical finite element methods.
The key concept of a PUM is the use of a compactly supported PU, that covers the computational domain \(\Omega\).
To each PU function \(\varphi_i\) a local approximation space \(\mathrm{V}_i\) is attached, which yields a global, finite-dimensional space \(\mathrm{V}^{PU} = \sum_i \varphi_i \mathrm{V}_i\), that is then used in a Galerkin approach.
An advantage over classical finite element methods is the ability to incorporate arbitrary basis functions.
Here, the intent is to use only a few problem specific basis functions, thereby requiring less degrees of freedom (DOF) to attain the desired global accuracy.
Well-known instances of FE based PUMs are the generalized finite element method by Duarte and Babu\v{s}ka~\cite{duarte2000generalized} and the extended finite element method by Mo\"es, Dolbow and Belytschko~\cite{moes1999finite}.
The particular meshfree PUM employed in this paper was introduced in~\cite{schweitzer2003phd} and is referred to as the PUM in the following.\@
All computations in this paper were carried out using the PUMA software framework~\cite{schweitzer2017rapid} developed at Fraunhofer SCAI\@.
In the following we present only a very short review of the PUM and refer the reader to~\cite{schweitzer2003phd} for details.

Given a computational domain \(\Omega\), we assume to have a partition of unity \(\{\varphi_i\}\) with \(\varphi_i \geq 0\) and

\begin{equation}
    \sum_i \varphi_i\big(\xb\big) = 1 \quad \forall \xb\in\Omega
\end{equation}

that covers the domain.
We call the support of a PU function \(\varphi_i\) a patch \(\omega_i := \mathrm{supp}(\varphi_i)\).
In the PUMA software framework, patches are constructed as follows.
First, we compute a cubic bounding-box \(\mathrm{C}\) of the domain \(\Omega\).
This bounding box corresponds to discretization level zero.
For level \(l\), we subdivide the bounding-box \(l\)-times and obtain the cells

\begin{equation}\label{eq:def_cells}
    C_i = \prod_{k=1}^{d} \Big( c_i^k - h, c_i^k + h \Big), \quad \mathrm{with} \quad h = \frac{width\,(\mathrm{C})}{2^{l+1}}
\end{equation}

where the \(c_i\) are the mid-points of the cells and \(h\) the discretization parameter usually reported for finite element method meshes.
Note that this sub-division can be done uniformly on all cells, or locally in an adaptive fashion.
We then obtain the patches \(\omega_i\) by scaling the cells via

\begin{equation}\label{eq:def_patches}
    \omega_i := \prod_{k=1}^{d} \Big( c_i^k - \alpha h, c_i^k + \alpha h \Big), \qquad 2 > \alpha > 1.
\end{equation}

On these patches \(\omega_i\) we construct a Shepard PU with the help of B-spline weight functions.
Note that by construction this PU has the flat-top property, i.e. on every patch \(\omega_i\) we can find an open ball on which it evaluates to one.
Details can be found in, \emph{e.g.}~\cite{schweitzer2003phd,schweitzer2009algebraic}.

To construct a higher order basis, each PU function \(\varphi_i\) is multiplied with a local approximation space

\begin{equation}\label{eq:def_structure_local_space}
    \mathrm{V}_i := \mathcal{P}_i \oplus \mathcal{E}_i = \Bspan{\bm{\psi}_i^s, \bm{\eta}_i^t},
\end{equation}

of dimension \(n_i\), where \(\mathcal{P}_i\) are spaces of polynomials of degree \(d^{\mathcal{P}_i}\).
The spaces \(\mathcal{E}_i\) denote so-called enrichment spaces of dimension \(d^{\mathcal{E}_i}\).
The latter are arbitrary functions locally incorporated into the simulation, which we can choose with respect to the problem at hand.
We can either obtain the \(\mathcal{E}_i\) from a-priori analytic knowledge about the structure of the solution to a problem.
This is \emph{e.g.}\ done in \(2D\) fracture mechanics problems, where we have an analytic expansion of the solution around a crack tip available.
Or the enrichments themselves are results of other simulations.
In the PUM, the global approximation space then reads as

\begin{equation}\label{eq:def_pu_space}
    \mathrm{V}^{PU} := \sum_i \varphi_i \mathrm{V}_i = \sum_i \varphi_i \mathcal{P}_i + \varphi_i \mathcal{E}_i.
\end{equation}

\textcolor{black}{For PDEs with smooth solution the PUM, just as standard finite element methods, converges~\cite{babuvska1997partition, schweitzer2013variational} with order \(\mathcal{O}(h^{p+1})\) in  $L^2$ and \(\mathcal{O}(h^{p})\) in \(H^1\)-norm if all local approximation spaces \(\mathcal{P}_i\) span polynomials of order $p$.
However, this rate can be recovered even in the presence of singularities with appropriate enrichment functions.}

Observe, that we do not assume the discrete functions,

\begin{equation}\label{eq:def_discrete_function}
    \ub^{PU} = \sum_i \varphi_i \left(
        \sum_{s=1}^{d^{\mathcal{P}_i}} \ub_i^s \bm{\psi}_i^s
        +
        \sum_{t=1}^{d^{\mathcal{E}_i}} \ub_i^{t + d^{\mathcal{P}_i}} \bm{\eta},
    \right)
\end{equation}

i.e.~the employed basis functions, to be interpolatory.
That is, the coefficients \(\ub_i^j\) do not necessarily correspond to function values at specific points.
We therefore use the direct splitting of the local spaces \(\mathrm{V}_i\) presented in~\cite{schweitzer2009algebraic} to enforce Dirichlet boundary conditions.
By the above-mentioned flat-top property of the PU, it is possible to construct a basis transformation that guarantees numerical stability and linear independence of the basis.
The latter is of special importance when using enrichment functions that are computed on-the-fly.
Other PUM approaches such as the finite element based generalized finite element method or extended finite element method cannot construct such a transformation.

%%%%%%%%%%%%%%%%%%%%%%%%%%%%%%%%%%%%%%%%%%%%%%%%%%%%%%%%%%%%%%%%%%
\subsubsection*{Equation of motion}

Let us first introduce our general model problem, the equation of (linear elastic) motion.
On the global domain~\(\Omega\) we consider

\begin{equation}\label{eq:lin_elas_equi}
    \varrho(\xb) \ddot{\ub}(t, \xb)= -\nabla \cdot \bm{\sigma} + \bb \qquad \mathrm{in} \; [0, T] \times \Omega_0,
\end{equation}

where \(\bb\) are the volume forces acting on the body, \emph{e.g.}~gravity, \(\bm{\sigma}\) is the Cauchy stress tensor and \(\varrho\) the mass density.
The latter is computed from the linear strain tensor \(\bm{\varepsilon}\) via Hooke's law

\begin{equation}\label{eq:Hookes_law}
    \bm{\sigma} = \mathbf{C} : \bm{\varepsilon} = 2\mu\bm{\varepsilon}(\ub) + \lambda \mathrm{tr}\left(\bm{\varepsilon}(\ub)\right) \bm{\mathbb{I}} ,
\end{equation}

with \(\mathbf{C}\) denoting Hooke's tensor, $\mu, \lambda$ the Lame parameters, $\mathrm{tr}(\cdot)$ the trace and $\mathbb{I}$ the identity.
The linear strain tensor is computed from the displacement field \(\ub\) by

\begin{equation}\label{eq:def_strain}
    \bm{\varepsilon}\big(\ub\big)(t, \xb) = \frac{1}{2}\left( \nabla\ub(t, \xb) + {\big(\nabla\ub(t, \xb)\big)}^T \right).
\end{equation}

To obtain a unique solution of~\eqref{eq:lin_elas_equi} we impose boundary conditions on \( \partial\Omega = \Gamma_N \mathbin{\dot{\cup}} \Gamma_D \)

\begin{align}\label{eq:def_bcs}
    \ub(0, \xb) &= \ub_0(\xb) \quad &\mathrm{in}\; &\Omega \\
    \ub(t, \xb &= \bar{\ub}(t, \xb) \quad &\mathrm{on}\; &(0, T] \times \Gamma^D\\
    \bm{\sigma}(t, \xb) \cdot \mathbf{n}(\xb) &= \traction(t, \xb) \quad &\mathrm{on}\; &(0, T] \times \Gamma^N,
\end{align}

where \(\mathbf{n}\) is the outward unit normal to \(\Gamma^N\) and \(\bar{\ub}\) and \(\traction\) are the prescribed displacement and traction conditions, respectively.

%%%%%%%%%%%%%%%%%%%%%%%%%%%%%%%%%%
\subsubsection*{Discretization in space}
%%%%%%%%%%%%%%%%%%%%%%%%%%%%%%%%%%

The weak formulation of \eqref{eq:lin_elas_equi} is given by:
For fixed $t \in [0, T]$ find \(\ub \in \mathrm{V}^{\mathrm{PU}}(\Omega)\) such that

\begin{equation}\label{eq:pumweakform}
    \int\limits_{\Omega} \varrho \ddot{\ub} \vb \Bdx
    =
    - \int\limits_{\Omega} \bm{\sigma}\big(\ub\big) : \bm{\varepsilon}\big(\vb\big) \Bdx
    + \int\limits_{\Gamma^N} \traction \vb \Bds
    + \int\limits_{\Omega} \bb \vb \Bdx
\end{equation}

for all test functions \(\vb \in \mathrm{V}^{\mathrm{PU}}\) that vanish on \(\Gamma^D\), see~\cite{schweitzer2009algebraic} for details.
With $\tilde{\bullet}$ denoting the coefficient vector of $\bullet$ with respect to $\mathrm{V}^{\mathrm{PU}}$, we can write \eqref{eq:pumweakform} as follows:

\begin{equation}\label{eq:pumvectorweakform}
    \varrho\Nb \tilde{\ab} = - \Kb \tilde{\ub} + \tilde{\traction} + \tilde{\bb},
\end{equation}

with $\Nb$ the mass matrix weighted by the mass density, $\ab = \ddot{\ub}$ the acceleration and $\Kb$ the stiffness matrix corresponding to linear elasticity.
%%%%%%%%%%%%%%%%%%%%%%%%%%%%%%%%%%
\subsubsection*{Discretization in time}
\label{subsubsec:pum:discret:time}
%%%%%%%%%%%%%%%%%%%%%%%%%%%%%%%%%%
We discretize \eqref{eq:pumvectorweakform} in time using discrete time steps \( t_n = t_{n-1} + \delta t_n \), where \(\delta t_n\) denotes the current time step size.
First we obtain the current acceleration from the current displacement at \(t_n\) via

\begin{align}
    \tilde{\ab}_n = {(\varrho \Nb)}^{-1} \left( - \Kb \tilde{\ub}_n + \tilde{\traction}_n + \tilde{\bb}_n\right).
\end{align}

Using central differences and half-time steps, we compute the $(n+\frac{1}{2})$-velocity coefficient $\tilde{\vb}$ by

\begin{align}
    \tilde{\vb}_{n+\frac{1}{2}} = \delta t_{n+\frac{1}{2}} \tilde{\ab}_n + \tilde{\vb}_{n-\frac{1}{2}}
\end{align}

and similarly the next displacement by

\begin{align}
    \tilde{\ub}_{n+1} = \delta t_{n+1} \tilde{\vb}_{n+\frac{1}{2}} + \tilde{\ub}_n.
\end{align}

As usual, central differences are second order in time.

%%%%%%%%%%%%%%%%%%%%%%%%%%%%%%%%%%
\subsubsection*{Quasi-static problem}
%%%%%%%%%%%%%%%%%%%%%%%%%%%%%%%%%%

In the numerical examples, we also solve the quasi-static version of \eqref{eq:pumweakform}.
There, the assumption is that the acceleration \(\ddot{\ub} = 0\) is zero.
The corresponding weak form thus is: find \(\ub \in \mathrm{V}^{\mathrm{PU}}(\Omega)\) such that

\begin{equation}\label{eq:pumstaticweakform}
    \int\limits_{\Omega} \bm{\sigma}\big(\ub\big) : \bm{\varepsilon}\big(\vb\big) \Bdx
    = \int\limits_{\Gamma^N} \traction \vb \Bds
    + \int\limits_{\Omega} \bb \vb \Bdx
\end{equation}

for all test functions \(\vb \in \mathrm{V}^{\mathrm{PU}}\) that vanish on \(\Gamma^D\).

%%%%%%%%%%%%%%%%%%%%%%%%%%%%%%%%%%
\subsection{Peridynamic theory}
\begin{figure}[htb]%{.45\linewidth}
    \centering
        \begin{tikzpicture}[xscale=0.6,yscale=0.6]
		    \draw [<-,thick] (0,3) -- (0,-3);
			\draw [->,thick] (-5,0) -- (3.5,0);
			\draw [-,thick] (0,0) to [out=60,in=140] (1.5,1.5) to [out=-45,in=180] (3,0.0);

			\draw [-,thick] (-3.0,-0.0) to [out=0,in=130] (-1.5,-1.5) to [out=-50, in=245] (0,0);

			\draw (3.0,-0.2) -- (3.0, 0.2);
			\draw (-3.0,-0.2) -- (-3.0, 0.2);
			\draw (1.2,-0.2) -- (1.2, 0.2);
			\draw (-1.2,-0.2) -- (-1.2, 0.2);
			\node [below] at (1.2,-0.2) {${r}^c$};
			\node [below] at (-1.2,-0.2) {$-{r}^c$};
			\node [below] at (3.0,-0.2) {${r}^+$};
			\node [below] at (-3.0,-0.2) {$-{r}^+$};
			\node [right] at (3.5,0) {${r}$};
			\node [right] at (0,2.2) {$g'(r)$};
		  \end{tikzpicture}
		   \caption{Plot of the derivative $g'(r)$ of the potential function $g(r)$ used in the cohesive force in Equation~\ref{eq:derivbond}. The force goes smoothly to zero at $\pm r^+$.}
		   \label{fig:ConvexConcaveb}
\end{figure}
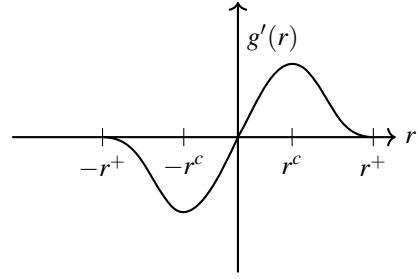

%%%%%%%%%%%%%%%%%%%%%%%%%%%%%%%%%%
Peridynamics (PD)~\cite{silling2000reformulation,silling2005meshfree}, a non-local generalization of classical continuum mechanics (CCM) which allows for discontinuous
displacement fields and provides an attractive framework for simulating cracks and fractures. PD was employed in many validations against experimental data~\cite{diehl2019review}. The governing equation for bond-based peridynamics reads as

\begin{align}
    \varrho(\xb) & \ddot{\ub}(\xb) (\xb,t) = \notag \\
    & \int\limits_{B_\delta(\xb)} \fb(t,\xb'-\xb,\ub(\xb',t)-\ub(\xb,t)) \,\mathrm{d}\xb' + \bb(\xb,t)
    \label{eq:pd:motion}
\end{align}

where $\varrho(\xb)$ is the material's density at the material point $\xb$, $\fb:[0,T]\times \R^3 \times \R^3 \rightarrow \R^3$ is the pair-wise force density, and $\bb : [0,T] \times \R^3 \rightarrow \R^3$ is the external force density. For more details about PD, we refer to~\cite{javili2019peridynamics}. The material model is included in the pair-wise force function $\fb$ and for this paper the bond-based softening model~\cite{lipton2014dynamic,lipton2016cohesive} is chosen as the constitutive law. Figure~\ref{fig:ConvexConcaveb} shows the derivative $g'(r)$ of the potential function $g(r)$ of this constitutive model. In the double well model, the force acting between material points $\xb$ and $\xb'$ is initially elastic and then softens and decays to zero as the bond stretch between points increases. The critical bond stretch $S_c>0$ for which the force begins to soften is given by

\begin{equation}
S_c=\frac{{r}^c}{\sqrt{|\xb'-\xb|}}\text{.}
\label{eq:crittensileplus}
\end{equation}

The bond stretch $S:[0,T] \times \R^3 \times \R^3 \rightarrow \R$ between two points $\xb$ and $\xb'$ in $\Omega_0$ is defined as

\begin{align}\label{eq:strain}
s(t,\xb,\xb')=\frac{\ub(\xb',t)-\ub(\xb,t)}{|\xb'-\xb|} \circ \mathbf{e}_{\xb'-\xb},
\end{align}

where $ \mathbf{e}_{\xb'-\xb}=\frac{\xb'-\xb}{\vert\xb'-\xb\vert}$ is a unit vector and ``$\circ$'' is the dot product. The non-local force is defined in terms of a double well potential.
The potential is a function of the bond stretch and is defined for all $\xb'$ and $\xb$ in $\Omega_0$ by

\begin{align}\label{eq:tensilepot}
&\mathcal{W}^\epsilon s(t,\xb',\xb)= \notag\\
&J^\epsilon(|\xb'-\xb|)\frac{1}{\delta^{d+1}\omega_d|\xb'-\xb|}g(\sqrt{\vert \xb'-\xb\vert}s(t,\xb',\xb))
\end{align}

where $\mathcal{W}^\epsilon (S(t,\xb',\xb))$ is the pairwise force potential per unit length between two points $\xb$ and $\xb'$. It is described in terms of its potential function $g:\R \rightarrow \R$, given by

\begin{equation}\label{eq:pot}
g(r)=h(r^2)
\end{equation}

 where $h$ is concave. Here $\omega_d$ is the volume of the unit ball in dimension $d$, and $\delta^{d}\omega_d$ is the volume of the horizon $B_{\delta}(\xb)$. The influence function $J^\delta(|\xb'-\xb|)$ is a measure of the influence that the point $\xb'$ has on $\xb$. Note that we chose a constant influence function $J^\delta(|\xb'-\xb|)=1$ for all simulations in this paper. Only points inside the horizon can influence $\xb$ so $J^\delta(|\xb'-\xb|)$ is nonzero for $|\xb' - \xb| < \delta$ and zero otherwise. One common example of double well potential $g$ out of the family of potentials is:
 
 \begin{align}
     g(r) = C(1 - \exp[-\beta r])
    \label{eq:potential}
 \end{align}
 
 where $C, \beta$ are material dependent parameters. Thus, the pair-wise force $\fb: [0,T]\times \R^3 \times \R^3 \rightarrow \R^3$ is given by
 
\begin{align}\label{eq:nonlocforcetensite}
\fb(t,\xb',\xb)&=2\partial_s\mathcal{W}^\delta(S(t,\xb;\xb))\mathbf{e}_{\xb'-\xb},
\end{align}

where

\begin{align}\label{eq:derivbond}
\partial_s\mathcal{W}^\delta &(s(t,\xb'\xb))= \notag \\
&\frac{1}{\delta^{d+1} \omega_d}\frac{J^\delta(|\xb'-\xb|)}{|\xb'-\xb|}\partial_s g(\sqrt{|\xb'-\xb|}S(t,\xb',\xb)).
\end{align}

%%%%%%%%%%%%%%%%%%%%%%%%%%%%%%%%%

\subsubsection*{Elastic Properties}
%%%%%%%%%%%%%%%%%%%%%%%%%%%%%%%
The PD model parameters $C$ and $\beta$ in Equation~\eqref{eq:potential} can be estimated via energy equivalence using classical material properties.
The first PD material constant is obtained as

\begin{align}
    C := \pi \frac{G_c}{4}
\end{align}

using the classical energy release rate $G_c$ as the reference material property.
The second PD material constant $\beta$ is obtained as

\begin{align}\label{eq:pd:beta}
\beta := \frac{4E\nu}{C(1-\nu)(1-2\nu)}
\end{align}

using the Poisson ratio $\nu$ and the Young's modulus $E$ from the classical theory. Note that in two dimensions, a bond-based PD model has a fixed Poisson ratio of $\sfrac{1}{3}$ for plain strain due to the pair-wise force $\fb$~\cite{kunin1975theory}. Therefore, only $G_c$ and $E$ are needed to calibrate the softening model. For a detailed discussion of the energy equivalence with linear elasticity, we refer to~\cite{https://doi.org/10.1002/nme.7005}. The PUM's classical linear elastic material model was calibrated using the Poisson ratio and the Young's modulus used in~\ref{eq:pd:beta}. The values used in the simulations are provided in Section~\ref{sec:numerical:results}.

%%%%%%%%%%%%%%%%%%%%%%%%%%%%%%%%%%
\subsubsection*{Discretization in time}
%%%%%%%%%%%%%%%%%%%%%%%%%%%%%%%%%%
The discretization in time of Equation~\eqref{eq:pd:motion} is done via a central difference scheme

\begin{align}
    \ub(t^{k+1},\xb) = 2 \ub(t^k,\xb) &- \ub(t^{k-1},\xb) \notag\\
    &+ \frac{\Delta t_s^2}{\varrho(\xb)} \left( \bb(t^k,\xb) + \int\limits_{B_\delta(\xb)} \fb(t,\xb',\xb)  \right)\text{.}
\end{align}

%%%%%%%%%%%%%%%%%%%%%%%%%%%%%%%%%%
\subsubsection*{Discretization in space}
\begin{figure}[htb]
    \centering
    \begin{tikzpicture}
\draw (-2,-2) -- (2,-2) -- (2,2) -- (-2,2) -- (-2,-2);
\node at (-1.75,2.25) {$\Omega_0$};
\draw (0,0) circle(1);
\node [above] at (0,0.9) {$B_\delta(\xb_i)$};
\node [below] at (0,-0.1) {$\xb_i$};
\draw [->] (0,0.075) -- (0,1);
\node [right] at (0,0.5) {$\delta$};
\shade [ball color = azure] (0,0) circle (0.075);
\foreach \x in {0,...,3}
	\foreach \y in {0,...,3}
		\shade [ball color = cadetgrey] (-1.5 + \x,-1.5 + \y) circle (0.075);
\end{tikzpicture}
    \caption{The material points at discrete positions $\xb_i$ in the domain $\Omega_0$ in the reference configuration at time $t= 0$. For each discrete material point $\xb_i$ the neighborhood $B_\delta(\xb_i):=\{j \vert \; \vert \xb_j - \xb_i \vert < \delta \}$ is computed. As an example, the neighborhood for the discrete node $\xb_i$ is shown. The material point $\xb_i$ exchanges force with all other discrete nodes within its neighborhood. Adapted from~\cite{diehl_2020_emu}.}
    \label{fig:discretization}
\end{figure}
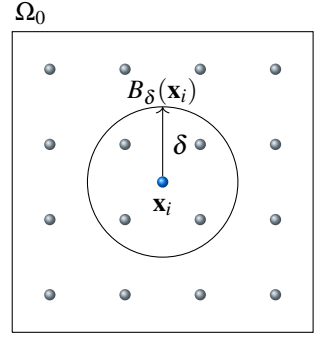

%%%%%%%%%%%%%%%%%%%%%%%%%%%%%%%%%%
The governing PD equation of motion~\eqref{eq:pd:motion} is discretized in space using a collocation approach, the so-called EMU nodal discretization~\cite{silling2005meshfree}. Figure~\ref{fig:discretization} sketches the reference configuration $\Omega_0$ with the discrete PD nodes $X = \{ \xb_i \vert i=1,\ldots n  \}$. For all discrete PD nodes, a surrounding volume $V = \{ V_i \in \R \vert i=1,\ldots n  \}$ is associated such that the volumes are non-overlapping $V_i \cap V_j = \emptyset, \text{ for } i \neq j$ and the sum of the surrounding volumes are approximately the volume of the reference configuration $\sum_{i=1}^n V_i(t) \approx \vert \Omega_0 \vert, \forall t \in [0,T]$. Thus, the space-discrete governing equation of motion reads as

\begin{align}
        \varrho(\xb_i) &\ddot{\ub}(\xb_i,t) = \notag\\
        &\sum\limits_{j \in B_\delta(\xb_i)} \fb(t,\xb_j-\xb_i,\ub(\xb_j,t)-\ub(\xb_i,t)) V_j + \bb(\xb_i,t)\text{.}
\end{align}

Combining the discretization in time and space thus yields

\begin{align}\label{eq:pd:discretized}
    \ub(t^{k+1},\xb_i) = & 2 \ub(t^k,\xb_i) - \ub(t^{k-1},\xb_i) \notag\\
    &+ \frac{\Delta t_s^2}{\varrho(\xb_i)} \left( \bb(t^k,\xb_i) + \sum\limits_{j \in B_\delta(\xb_i)} \fb(t,\xb_i,\xb) \right) \text{.}
\end{align}

\textcolor{black}{Note that other discretization techniques, e.g. continuous and discontinuous finite element methods~\cite{chen2011continuous,macek2007peridynamics}, Gauss quadrature~\cite{weckner2005numerical}, and spatial discretization~\cite{parks2008implementing,emmrich2007peridynamic}, are available for PD. Fore more details, we refer to~\cite[Section 2.1.2 ]{diehl_lipton_wick_tyagi_2021}.}
%%%%%%%%%%%%%%%%%%%%%%%%%%%%%%%%%%
\paragraph{Boundary conditions}
%%%%%%%%%%%%%%%%%%%%%%%%%%%%%%%%%%
One challenge for non-local models is the equivalent application of local boundary conditions. Many publications related to the application of local boundary conditions in non-local models are available~\cite{prudhomme2020treatment,jin2021coupling}, however, since we apply local traction conditions in our numerical examples, we focus on the conversion of local traction conditions within peridynamics. The external force $\bb(\xb,t)$ is usually applied in a region of horizon size $\delta$ times the length of the boundary and not on a single line of nodes close to the boundary to mimic a traction condition. Thus, the horizon can scale to zero so that in its limit the body forces applied to the area becomes a boundary traction \cite{LiptonJhaBdry}. The peridynamic forces on the crack faces also converge to zero normal traction on the crack faces as the horizon goes to zero, see \cite{LiptonJhaBdry}.

%%%%%%%%%%%%%%%%%%%%%%%%%%%%%%%%%%
\paragraph{Convergence}
%%%%%%%%%%%%%%%%%%%%%%%%%%%%%%%%%%
One has a-priori convergence of peridynamic models under space and time grid refinement that is shown for both finite difference and finite element discretizetions, see \cite{CMPer-JhaLipton}, \cite{CMPer-JhaLipton8}, and \cite{CMPer-JhaLipton3}. In addition, numerical simulations show convergence of solutions in the $L^2$ norm and the convergence of the PD energy to the Griffith fracture energy for fracture simulations in \cite{CMPer-JhaLipton8}, and \cite{CMPer-JhaLipton3}. Here convergence is seen under refinement of peridynamic horizon and mesh discritization.

%%%%%%%%%%%%%%%%%%%%%%%%%%%%%%%%%%
\subsubsection*{Damage and crack identification}
%%%%%%%%%%%%%%%%%%%%%%%%%%%%%%%%%%
Since we do not break bonds as in many other PD models and soften bonds instead, see Figure~\ref{fig:ConvexConcaveb}, we use the critical value $r^c=\sqrt{\sfrac{0.5}{\beta}}$ where the material model transitions from linear behavior to softening as an indication of damage.
To define a damage field $D$, we define for each discrete PD node \(\xb\)

\begin{align}
    d(\xb) = \frac{\max(r)}{r^c}
\end{align}

where $\max(r)$ is the maximal value over all bonds within the neighborhood $B_\delta(\xb)$.
In the case that $d(\xb) < 1 $ there is no damage and once,  $d(\xb) \geq 1$ we assume damage at the node $\xb$.
Now, the damage field for all nodes is used to determine the crack path, which is not directly encoded in the model.
Algorithm~\ref{algo:crack:path} sketches the extraction of the crack path out of a series of damage fields $D_{1,\ldots,n}$ for $n$ time steps.
First, we iterate over a subset of damage fields $d\in D_{1,250,500,\ldots,n}$, see Line~\ref{algo:crack:path:for1}, where we take only every 250th time step to extract the position of the crack tip.
Note that this value was chosen for the inclined crack example in Section~\ref{sec:numerical:results:inclined:crack}.
However, for faster crack growth, this value needs to be adjusted.
Second, we iterate over all the damage values $d_i$ at each PD node $\xb_i$, see Line~\ref{algo:crack:path:for2}.
In Line~\ref{algo:crack:path:if1} the maximal damage value is searched where we assume the current position of the crack path.
In Line~\ref{algo:crack:path:if2} we collect the crack positions of each time step and construct a linear interpolation to generate the crack path over all time steps, see Line~\ref{algo:crack:path:interpolation}.
For the inclined crack simulation in Section~\ref{sec:numerical:results:inclined:crack} the algorithm was used twice.
Once on the left side and once on the right side, only to extract the left and right crack branches separately, see Figure~\ref{fig:crack:path}.
Note that the limitation of this algorithm is that the crack path aligns with the PD nodes, and we do not take into account that the crack path is typically between two PD nodes. To clarify, the Algorithm~\ref{algo:crack:path} was run as post-processing once the PD simulation was done. For the coupled approach, the algorithm would run until the current time step, wait for the next time step to be processed, and so on. There is no need to run the algorithm from the beginning for each new crack position.

More sophisticated methods to extract a crack surface were proposed by the authors in~\cite{diehl2017extraction,bussler2017visualization} which are approximations of the crack path between PD nodes, however, these numerical models introduce an additional uncertainty and will be investigated in this context in the future.

\begin{algorithm}[tb]
\begin{algorithmic}[1]
\State crack\_pos = []
\State \textcolor{asparagus}{// Loop over every 250th time steps}
\ForAll{$d \in D_{1,250,500,\ldots,n}$} \label{algo:crack:path:for1}
\State \textcolor{asparagus}{// Loop over the damage value at each node}
\State $max=0$, $index=-1$
\ForAll{$d_i \in D$} \label{algo:crack:path:for2}
\State \textcolor{asparagus}{// Check if there is damage} \label{algo:crack:path:if1}
\State \textcolor{asparagus}{// and find the max damage}
\If{$d_i > 1$ and $d_i > max$ }
\State $max = d_i$
\State $index=i$
\EndIf
\EndFor
\State \textcolor{asparagus}{Check if we have damage and add the position of $\xb_{index}$}  \label{algo:crack:path:if2}
\If{$max > 0$ }
\State crack\_pos.append($\xb_{index}$)
\EndIf
\EndFor
\State linear\_interpolation(crack\_pos) \label{algo:crack:path:interpolation}
\end{algorithmic}
\caption{}
\label{algo:crack:path}
\end{algorithm}

%%%%%%%%%%%%%%%%%%%%%%%%%%%%%%%%%%
\section{Peridynamics as Multiscale Enrichments in the Partition of Unity Method}
\label{sec:enrichments}
%%%%%%%%%%%%%%%%%%%%%%%%%%%%%%%%%%
Solving fracture mechanics with the PD model allows for naturally developing and growing cracks with the downside of high computational costs independent of the actual deformation, damage and crack evolution, i.e. in regions where the material response is essentially linearly elastic the PD model does not provide any approximation benefit while being much more expensive to simulate.

On the other hand, generalizations of the finite element method like PUM can solve large linear elastic crack evolution problems rather efficiently when a crack path evolution is available, since PUM employs only a few basis functions to resolve the known crack path.
In these methods, however, the initiation of cracks and their growth over time requires additional modelling and is typically the most challenging task.

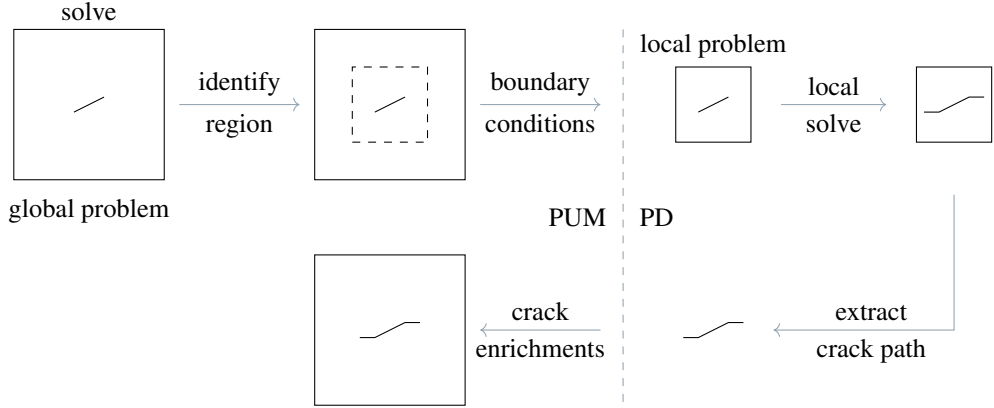
\begin{figure*}[tb]
    \centering
    \begin{tikzpicture}
        \draw (-4,3) -- (-2,3) -- (-2,5) -- (-4,5) -- cycle;
        \draw (-3.2,3.9) -- (-2.8,4.1);
        \node[above] at (-3,2.3) {global problem};
        \node[above] at (-3,5) {solve};
        % --------------------
        \node[above] at (-1,4) {identify};
        \draw[thin,cadetgrey,->] (-1.8,4) -- (-0.2,4) ;
        \node[below] at (-1,4) {region};
        % --------------------
        \draw (0,3) -- (2,3) -- (2,5) -- (0,5) -- cycle;
        \draw[dashed] (0.5,3.5) -- (1.5,3.5) -- (1.5,4.5) -- (0.5,4.5) -- cycle;
        \draw (0.8,3.9) -- (1.2,4.1);
        % --------------------
        \node[above] at (3,4) {boundary};
        \draw[thin,cadetgrey,->] (2.2,4) -- (3.8,4) ;
        \node[below] at (3,4) {conditions};
        % ---------------------------------------------------------------
        \node[right] at (4.2,2.5) {PD};
        % --------------------
        \node[above] at (5.3,4.5) {local problem};
        \draw (4.8,3.5) -- (5.8,3.5) -- (5.8,4.5) -- (4.8,4.5) -- cycle;
        \draw (5.1,3.9) -- (5.5,4.1);
        % --------------------
        \node[above] at (6.9,4) {local};
        \draw[thin,cadetgrey,->] (6.2,4) -- (7.6,4) ;
        \node[below] at (6.9,4) {solve};
        % --------------------
        \draw (8,3.5) -- (9,3.5) -- (9,4.5) -- (8,4.5) -- cycle;
        \draw (8.1,3.9) -- (8.3,3.9) -- (8.7,4.1) -- (8.9,4.1);
        % --------------------
        \node[above] at (7.4,1) {extract};
        \draw[thin,cadetgrey,->] (8.5,2.8) -- (8.5,1) -- (6.1,1) ;
        \node[below] at (7.4,1) {crack path};
        % --------------------
        \draw (4.9,0.9) -- (5.1,0.9) -- (5.5,1.1) -- (5.7,1.1);
        % ---------------------------------------------------------------
        \node[left] at (4,2.5) {PUM};
        % --------------------
        \node[above] at (3,1) {crack};
        \draw[thin,cadetgrey,<-] (2.2,1) -- (3.8,1) ;
        \node[below] at (3,1) {enrichments};
        % --------------------
        \draw (0,0) -- (2,0) -- (2,2) -- (0,2) -- cycle;
        \draw (0.6,0.9) -- (0.8,0.9) -- (1.2,1.1) -- (1.4,1.1);
        % --------------------
        \draw[thin, dashed, cadetgrey] (4.1,5) -- (4.1,0);
    \end{tikzpicture}
    \caption{Enriching PUM with PD using a global-local approach.
        We are interested in the global problem, which is solved twice using the PUM.
        First it is solved using the initial or old crack configuration to obtain boundary conditions for the local problem.
        Once PD provides a new crack path through the local problem, the updated global problem is solved again.}
    \label{fig:coupling:glcycle}
\end{figure*}

Thus, the benefits and weaknesses of PD and PUM are completely complementary to each other and their merger promises to allow for the efficient and reliable treatment of arbitrary crack growth by combining the best of both worlds.
The aim of this work is thus to construct such a combined PD/PUM simulator which automatically determines the region where PD should be employed so that the local PD approximation can be utilized to enrich the global PUM approximation to capture the true material response with high accuracy efficiently.
To this end, we need to identify dynamically within a global PUM simulation those regions in which cracks will initiate or grow in the next time step, \emph{e.g.}\ by some strain or stress based damage model.
Then, we extract the respective subdomains together with respective boundary conditions on the subdomain boundaries obtained from the global PUM approximation.
These subdomain problems are then passed to the PD simulator to approximate the local material response, including crack initiation and growth up to the next global PUM time step.
Then, we need to post-process these local PD approximations to define appropriate enrichment functions for the global PUM simulation that encode the computed crack path and the respective deformation in the vicinity of the crack.
With the help of these enrichment functions, we can then define a PUM approximation space that is appropriate for the resolution of the global material response in the next global time step and thus can advance the global PUM simulation in time.
Thus, we envision a so-called global-local enrichment strategy, see \cite{duarte2007global, birner2017global} and Figure~\ref{fig:coupling:glcycle}.

As a first step towards this research goal we actually employ PD and PUM both globally on the complete domain to identify under which conditions the deformations computed by PD and PUM agree well locally, i.e. we identify the complement of the local regions on which the PD model should ultimately be employed only.
This study ensures that away from damaged regions the global PUM approximation can safely replace the PD approximation and thus can be used to prescribe boundary data for the local PD simulations in the future.
Then, we study how information from the PD approximation can be passed back into the global PUM approximation, i.e. how to construct enrichments for PUM from PD, compare ~\cite{schweitzer2014moving}.
Here, we again start with a rather simple strategy to validate the overall approach. To this end, we extract the crack path predicted by PD, see Algorithm~\ref{algo:crack:path}, and pass only this geometric information to PUM.
On this crack geometry, we then define the typical crack enrichments~\cite{schweitzer2009adaptive} based on the Heaviside and Westergaard functions and finally compare the global PD evolution with the respective global PUM solution to validate our approach.

\begin{algorithm}[tb]
    \begin{algorithmic}[1]
        \State \textcolor{asparagus}{// Divide time interval T into $N$ synchronization time steps}
        \State $n$ = 0
        \State $C_0$ = initial crack patch
        \State \textcolor{asparagus}{// Solve global PUM problem until first sync time step $t_0$}
        \State $u_{PUM, 0}$ = Solution of \eqref{eq:pumvectorweakform} for $t_0 = 0$
        \State Identify damaged area around crack $\rightarrow$ local domain \label{algo:coupling:identify}
        \State Set up PD problem on local domain \label{algo:coupling:setup:pd}
        \State \textcolor{asparagus}{// Loop over synchronization time steps}
        \While{$n < N$} \label{algo:coupling:1}
            \State Extract PD problem boundary data from $u_{PUM, n}$ \label{algo:coupling:write:bc}
            \State Solve PD problem \eqref{eq:pd:discretized} until time $t_{n}$ for $u_{PD, n}$ \label{algo:coupling:read:bc}
            \State \textcolor{asparagus}{// Extract crack path as in Algorithm~\ref{algo:crack:path}}
            \State $C_{n+1}$ = crack path from $u_{PD, n}$
            \State Update $V_{PU}$ crack enrichments corresponding to $C_{n+1}$
            \State \textcolor{asparagus}{// Solve updated PUM problem until $t_{n+1}$}
            \State $u_{PUM, n+1}$ = Solution of \eqref{eq:pumvectorweakform} for $t_{n+1}$
            \State Update local domain
            \State $n = n+1$
        \EndWhile
    \end{algorithmic}
    \caption{}
    \label{algo:coupling}
\end{algorithm}

The coupling then works as follows.
We use global-local enrichments (GL) as introduced in \cite{duarte2007global, birner2017global}
as a coupling framework, see Figure~\ref{fig:coupling:glcycle} for an overview.
Assume the problem, which we refer to as the global problem, is given on a simple domain with a single crack~\cite{duarte2007global,birner2017global}.
By \emph{e.g.}~some strain or stress based damage model, we identify a region of interest around the crack.
This is not performed in this paper, but left for future work.
On that region we set up a local problem, using a buffer zone, such that PD has access to a volumetric boundary strip to prescribe boundary conditions.
Initially, we solve the global problem with the PUM, i.e.~\eqref{eq:pumvectorweakform} as described in Section~\ref{subsubsec:pum:discret:time}, disregarding the crack.
Thereby, we have a displacement field around the local problem that captures the overall linear elastic response of the body.
The solution is evaluated to set boundary conditions on the local problem, where we then solve~\eqref{eq:pd:discretized} using PD.
From this PD solution on the local domain we extract the crack path as described in Algorithm~\ref{algo:crack:path}.
In the PUM this crack path is for now modeled using a Heaviside function as an enrichment along the crack and the Westergaard enrichment functions at the tips.
With the updated crack path, the global problem is solved by the PUM again to arrive at the coupled solution.
In dynamic simulations, this has to be done in every synchronization time step.
Here, we assume that the crack does not grow outside the local problem domain.

Using the full PD solution, and not only the crack path, as an enrichment has already been studied in~\cite{schweitzer2014moving}.
The main difficulty here, is that the PD solution is piece-wise constant, thus has zero derivative.
Further questions that will arise in coupling the two methods is how to handle time step sizes between them and the initial identification of the area of interest around the crack.
These issues however need to be addressed in future work.

\textcolor{black}{In the literature, there are several approaches similar in nature to the one proposed here.
For instance, the global-local method is coupled~\cite{geelen2020extended} with a phase-field method~\cite{wu2020phase} to compute the crack geometry. A detailed comparison of peridynamics with phase-field models can be found in~\cite{diehl_lipton_wick_tyagi_2021}. Peridynamics have also been already successfully coupled~\cite{giannakeas2020coupling1} with FEM based partition of unity methods as the XFEM/GFEM, however to our knowledge the peridynamics solution has not been used as an enrichment so far, but rather constitutes the near tip material response. This requires changing the global FEM mesh in crack growth~\cite{giannakeas2020coupling2}, rather than moving enrichments. Moreover, the approach in~\cite{talebi2014computational} is similar in spirit, however operates on the micro scale.}

%%%%%%%%%%%%%%%%%%%%%%%%%%%%%%%%%%
\FloatBarrier
\section{Verification of the proposed approach}
\label{sec:numerical:validation}
%%%%%%%%%%%%%%%%%%%%%%%%%%%%%%%%%%
Before presenting numerical results in the next section, we do some verification of the proposed approached. First, we look into a two-dimensional bar and simulating pure linear elasticity without any initial crack, see Section~\ref{sec:numerical:result:bar}. Here, we focus on the agreement of the local and non-local model in the linear elastic case, where both models are compatible due to the absence of damage. Second, we compare for a stationary crack the results of the explicit PD and explicit PUM simulation with the implicit PUM, see Section~\ref{sec:numerical:results:modei}.
%Third, we show that the crack will not grow after the loads drops to zero and the simulation continues, see Section~\ref{sec:numerical:stability}.

For all simulations the following material properties are considered: material density $\varrho=$\num{1200}\si{\kilo\gram\per\cubic\meter},  Young's modulus $E=$\num{3.25}\si{\giga\pascal}, bulk modulus $K=$\num{2.16}\si{\giga\pascal}, a Poisson ratio $\nu=\nicefrac{1}{3}$ and a critical energy release rate $G=$\num{500}\si{\joule\per\square\meter}.
All peridynamic simulations were done with the author's NLMech/PeriHPX code~\cite{Diehl2020hpx}, all PUM simulations with PUMA~\cite{schweitzer2017rapid}.
For the comparison, we define the maximal magnitude of displacement as

\begin{align}
    \mathbf{U}_\text{max} = \max_{i=1,\ldots,n}(\vert\ub(t,\xb_i)\vert)
\end{align}

to compare the simulation results of both methods.
For the PUM results, the locations \(\{\xb_i\}\) refer to the grid points of the mesh the solution is visualized on.
All PUM simulations were run with a discretization on \(\#\varphi_i = 4096\) patches, as further refinement did not change the results in any meaningful digit.
We report the respective discretization parameter $\hPum$ for each experiment as calculated by~\eqref{eq:def_cells}.
The local approximation spaces~\eqref{eq:def_structure_local_space} are given by first order polynomials and the given crack enrichments, if a crack is present in the problem.
There, we enrich patches intersecting the crack, but not the crack tips with a Heaviside function and patches in an area around the tips with the Westergaard functions.
Note that though we ultimately want to couple an explicit dynamic PUM with an explicit dynamic PD simulation, we primarily compare PD results with quasi-static PUM solutions here.
This should work due to the undynamic nature of the problems.
To justify this, we show dynamic results for all verification experiments as well.

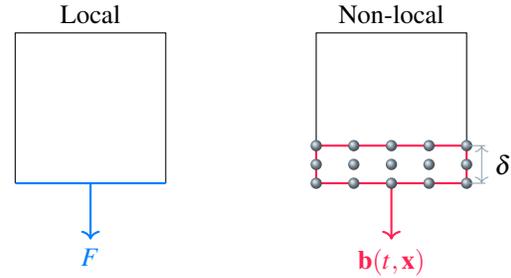
\begin{figure}[htb]
    \centering
    \begin{tikzpicture}
    \draw (-3,-1.5) -- (-3,0.5) -- (-1,0.5) -- (-1,-1.5) ;
    \draw[azure,thick] (-3,-1.5) -- (-1,-1.5);
    \node[above] at (-2,0.5) {Local};
    \draw[azure,thick,->] (-2,-1.5) -- (-2,-2.25);
    \node[below,azure] at (-2,-2.25) {$F$};
    \draw (3,-1.5) -- (3,0.5) -- (1,0.5) -- (1,-1.5) ;
    \draw[awesome,thick] (3,-1.5) -- (1,-1.5) -- (1, -1) -- (3,-1) -- cycle;
    \node[above] at (2,0.5) {Non-local};
    \draw[awesome,thick,->] (2,-1.5) -- (2,-2.25);
    \node[below,awesome] at (2,-2.25) {$\bb(t,\xb)$};
    \shade [ball color = cadetgrey] (1.0, -1.5) circle (0.075);
    \shade [ball color = cadetgrey] (1.5, -1.5) circle (0.075);
    \shade [ball color = cadetgrey] (2.0, -1.5) circle (0.075);
    \shade [ball color = cadetgrey] (2.5, -1.5) circle (0.075);
    \shade [ball color = cadetgrey] (3, -1.5) circle (0.075);
	\shade [ball color = cadetgrey] (1., -1.25) circle (0.075);
    \shade [ball color = cadetgrey] (1.5, -1.25) circle (0.075);
    \shade [ball color = cadetgrey] (2.0, -1.25) circle (0.075);
    \shade [ball color = cadetgrey] (2.5, -1.25) circle (0.075);
    \shade [ball color = cadetgrey] (3, -1.25) circle (0.075);
    \shade [ball color = cadetgrey] (1., -1.) circle (0.075);
    \shade [ball color = cadetgrey] (1.5, -1.) circle (0.075);
    \shade [ball color = cadetgrey] (2.0, -1.) circle (0.075);
    \shade [ball color = cadetgrey] (2.5, -1.) circle (0.075);
    \shade [ball color = cadetgrey] (3, -1.) circle (0.075);

    \draw[thin,cadetgrey] (3,-1.5) -- (3.25,-1.5) ;
    \draw[thin,cadetgrey] (3,-1.) -- (3.25,-1.) ;
    \draw[thin,cadetgrey,<->] (3.2,-1.5) -- (3.2,-1) ;
    \node[right] at (3.25,-1.25) {$\delta$};
    \end{tikzpicture}
    \caption{Left: traction condition in the local model on the lower boundary in \textcolor{azure}{blue}. Right: Application of the non-local traction conditions in the layer of horizon size $\delta$. If the horizon can scale to zero so that in its limit the body forces applied to the layer of horizon size is equivalent to the boundary traction on the right-hand side~\cite{LiptonJhaBdry}.}
    \label{fig:sketch:boundary}
\end{figure}

For the non-local model, we apply the local traction condition highlighted in \textcolor{azure}{blue} in Figure~\ref{fig:sketch:boundary} using the external force density $\bb(\xb_i,t)$ in a layer of horizon-size $\delta$ along the local traction condition (which corresponds to the \textcolor{awesome}{red} rectangle).
The external force density at a node $\xb_i$ within the layer of horizon-size $\delta$ yields

\begin{align}
    \bb(\xb_i,t) =  \frac{F}{Vm}
    \label{eq:externalforce:traction}
\end{align}

where $F$ is the force in Newton applied at the local traction condition, $V$ is the sum of all volumes $V_i$ associated to nodes $\xb_i$ in the layer of horizon-size $\delta$, and $m$ is the amount of nodes within the horizon $\delta$.
More theoretical details are shown in~\cite{LiptonJhaBdry}.
These non-local boundary conditions are applied in all remaining examples.
Future work will incorporate recently developed quadrature weights \cite{parksyueyoutrask} delivering asymptotically optimal truncation error.

%%%%%%%%%%%%%%%%%%%%%%%%%%%%%%%%%%
% \FloatBarrier
\subsection{Two-dimensional bar (pure linear elasticity)}
\label{sec:numerical:result:bar}
%%%%%%%%%%%%%%%%%%%%%%%%%%%%%%%%%%

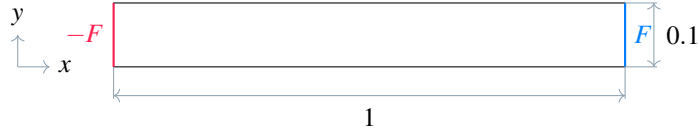
\begin{figure*}[htb]
    \centering
    \begin{tikzpicture}[scale=0.85]
        \draw (0,0) -- (0,1) -- (8,1) -- (8,0) -- cycle;
        \draw[awesome,thick] (0,0) -- (0,1);
        \draw[azure,thick] (8,0) -- (8,1);
        \node[awesome,left] at (0,0.5) {$-F$};
        \node[azure,right] at (8,0.5) {$F$};
        \draw[thin,cadetgrey] (0,0) -- (0,-0.5) ;
        \draw[thin,cadetgrey] (8,0) -- (8,-0.5) ;
        \draw[thin,cadetgrey,<->] (0,-0.45) -- (8,-0.45) ;
        \node[below] at (4,-0.5) {\num{1}};
        \draw[thin,cadetgrey] (8,0) -- (8.5,0) ;
        \draw[thin,cadetgrey] (8,1) -- (8.5,1) ;
        \draw[thin,cadetgrey,<->] (8.45,0) -- (8.45,1) ;
        \node[right] at (8.5,.5) {\num{0.1}};
        \draw[->,cadetgrey](-1.5,0) -- (-1,0);
        \draw[->,cadetgrey](-1.5,0) -- (-1.5,0.5);
        \node[right] at (-1,0) {$x$};
        \node[above] at (-1.5,0.5) {$y$};
    \end{tikzpicture}
    \caption{Sketch of the two-dimensional bar which is used to study the influence of bond softening on the displacement. Initially, a force of $F=\pm$\num{9e5}\si{\newton} is applied on the left-hand side and right-hand side of the bar. The force increases linearly to $2F$, $4F$, and $8F$ which results in a bond damages of \num{7.3}\si{\percent} up to \num{106.1}\si{\percent}.}
    \label{fig:bar:geometry}
\end{figure*}

\begin{figure*}
    \centering
    \includegraphics[trim=0 250 0 300, clip,width=0.495\textwidth]{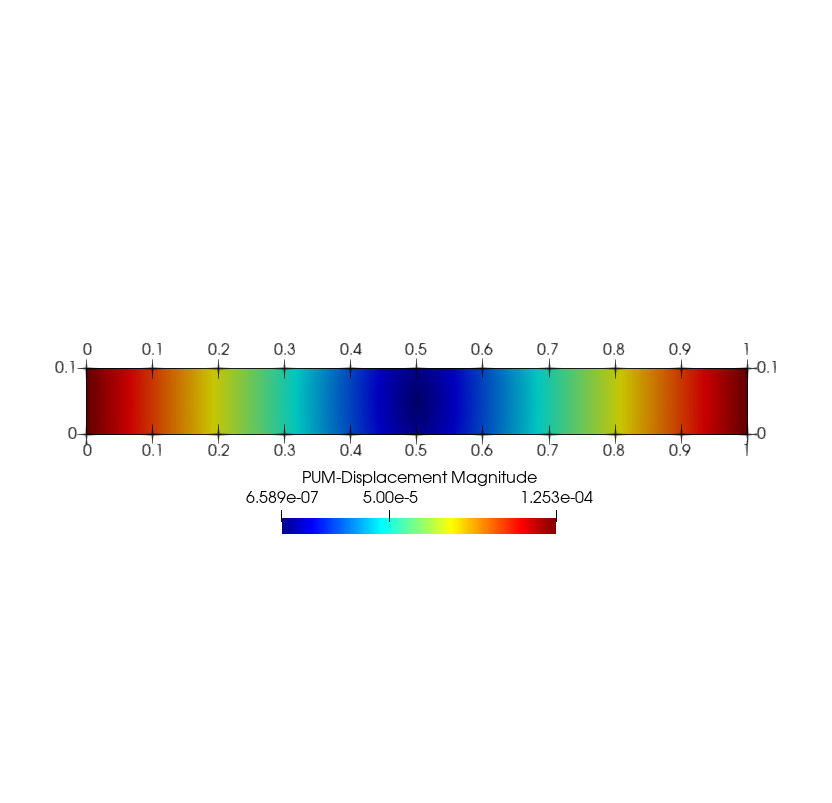}
    \caption{Obtained displacement field by the explicit dynamic PUM simulation for the load F=\num{9e5}\si{\newton}. For comparison, the displacement field obtained by the quasi-static PUM simulation is shown in Figure~\ref{fig:bar:load:1:pum}.}
    \label{fig:bar:load:1:pum:dynamic}
\end{figure*}

\begin{figure*}[p]
    \centering
    \begin{tikzpicture}
    \node at (-4,0) {\bf Force = \num{9e5}\si{\newton}};
    \node at (4,0) {\bf Force = \num{8}$\times$\num{9e5}\si{\newton}};
    \end{tikzpicture}
    \vspace{-0.5cm}
    \subfloat[PUM (Displacement)\label{fig:bar:load:1:pum}]{
    \includegraphics[trim=0 250 0 300, clip,width=0.495\linewidth]{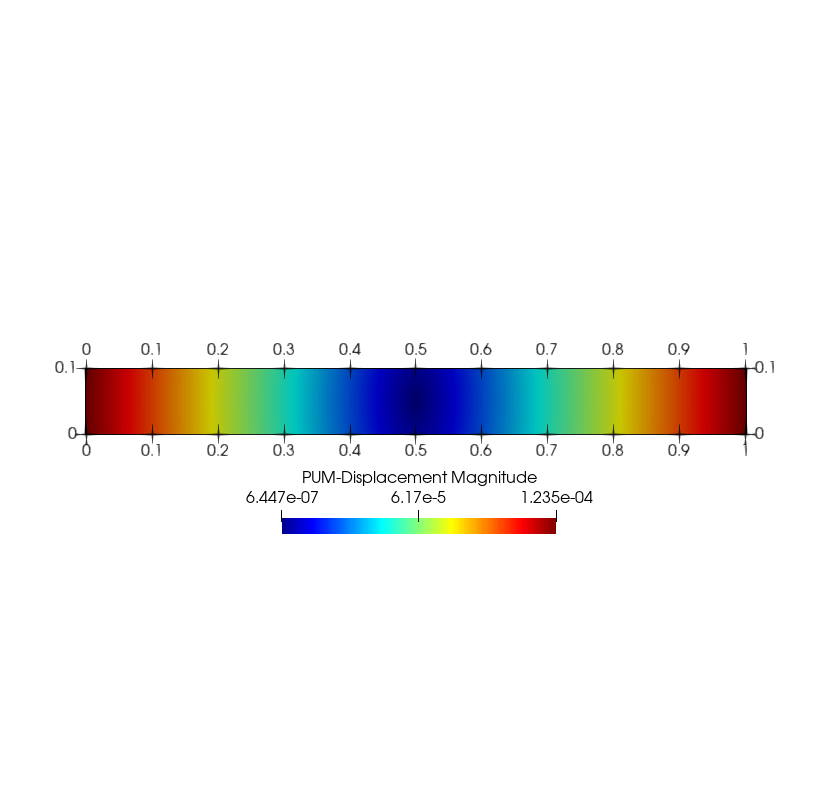}
    }
    \subfloat[PUM (Displacement)\label{fig:bar:load:8:pum}]{
    \includegraphics[trim=0 250 0 300, clip,width=0.495\linewidth]{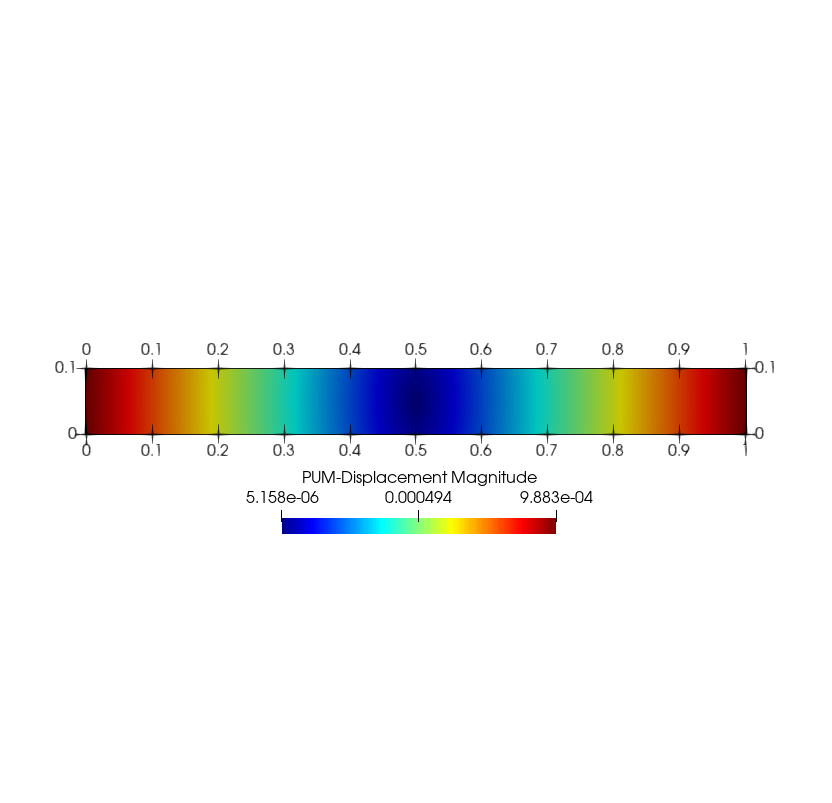}
    }
    \vspace{0.1cm}
    \subfloat[PD (Displacement)\label{fig:bar:load:1:pd}]{
    \includegraphics[trim=22 250 22 300, clip,width=0.495\linewidth]{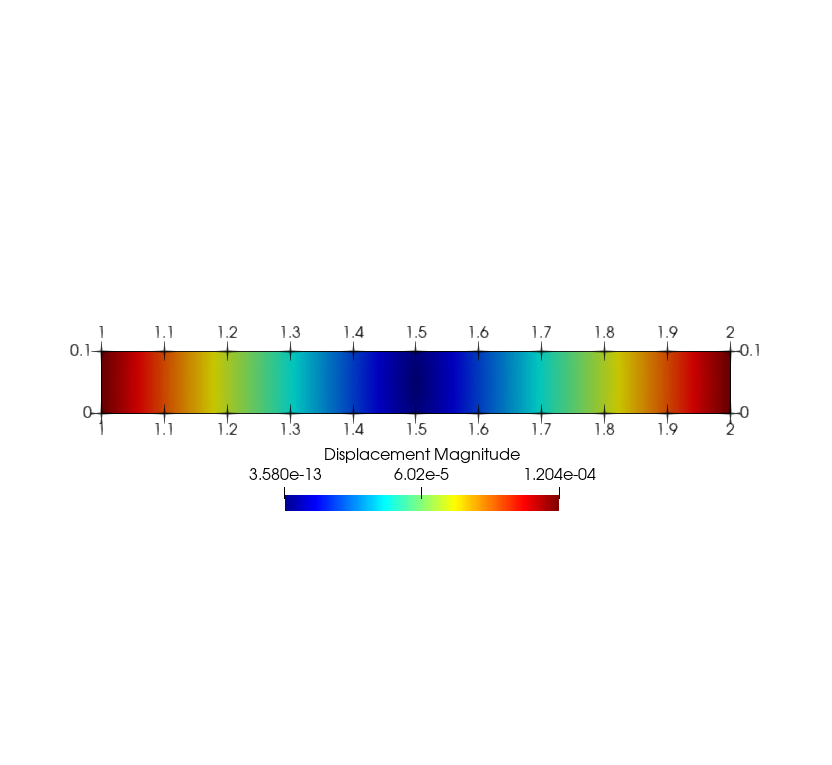}
    }
    \subfloat[PD (Displacement)\label{fig:bar:load:8:pd}]{
    \includegraphics[trim=22 250 22 300, clip,width=0.495\linewidth]{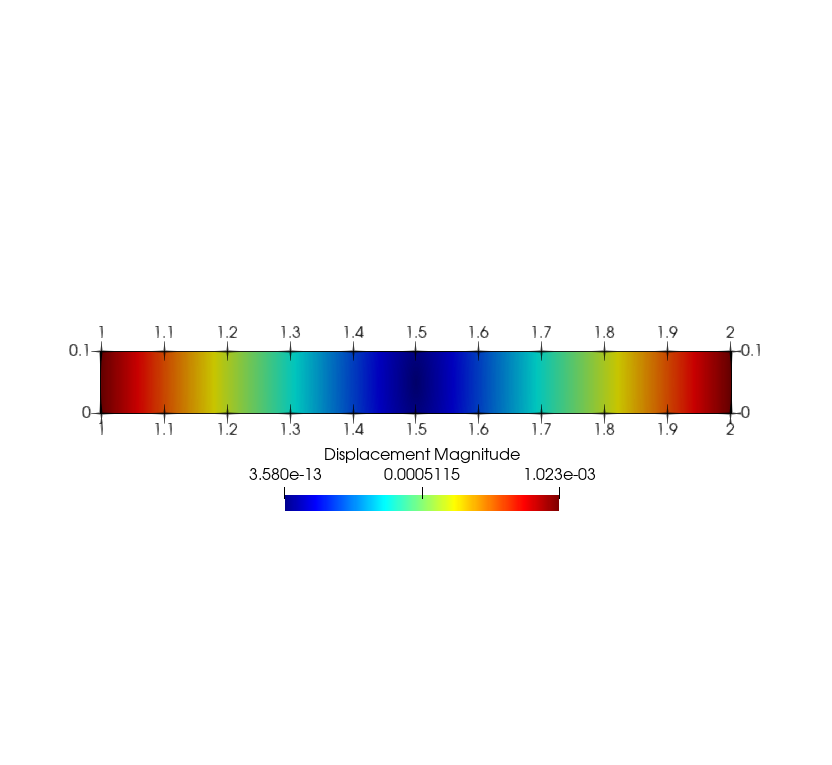}
    }

    \subfloat[PD (Damage)\label{fig:bar:load:1:pd:damage}]{
    \includegraphics[trim=22 250 22 300, clip,width=0.495\linewidth]{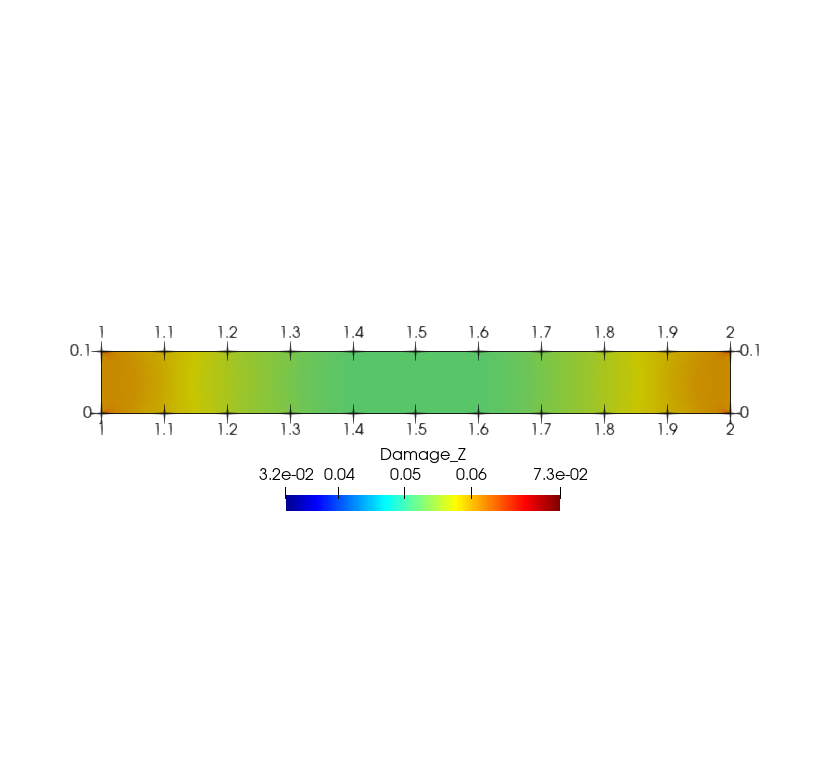}
    }
    \subfloat[PD (Damage)\label{fig:bar:load:8:pd:damage}]{
    \includegraphics[trim=22 250 22 300, clip,width=0.495\linewidth]{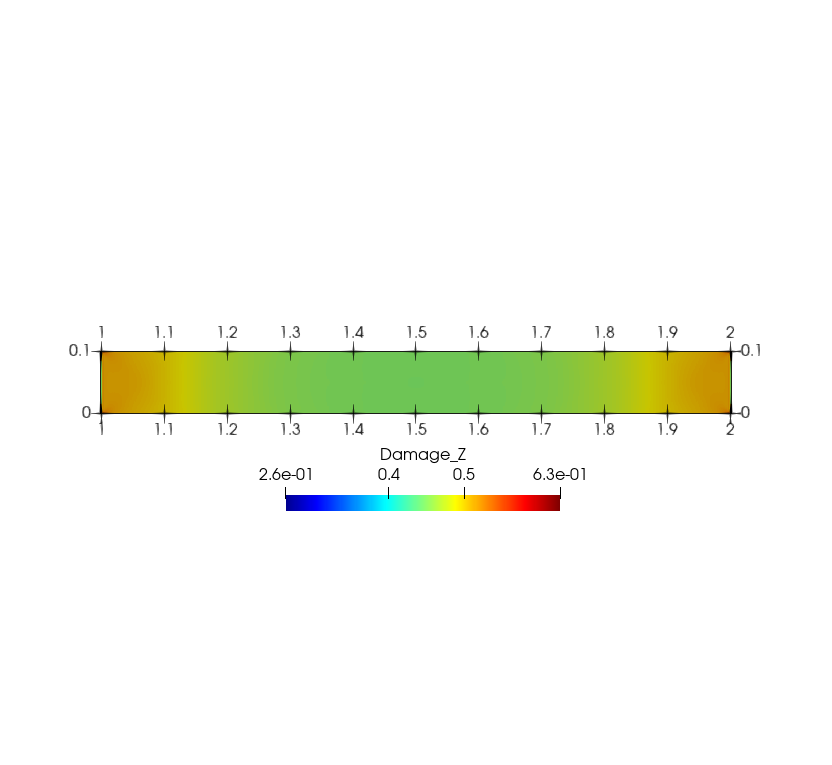}
    }
    \vspace{0.5cm}
    \begin{tikzpicture}
    \node at (-4,0) {\bf Force = \num{4}$\times$\num{9e5}\si{\newton}};
    \node at (4,0) {\bf Force = \num{12}$\times$\num{9e5}\si{\newton}};
    \end{tikzpicture}

    \subfloat[PUM (Displacement)\label{fig:bar:load:4:pum}]{
    \includegraphics[trim=0 250 0 300, clip,width=0.495\linewidth]{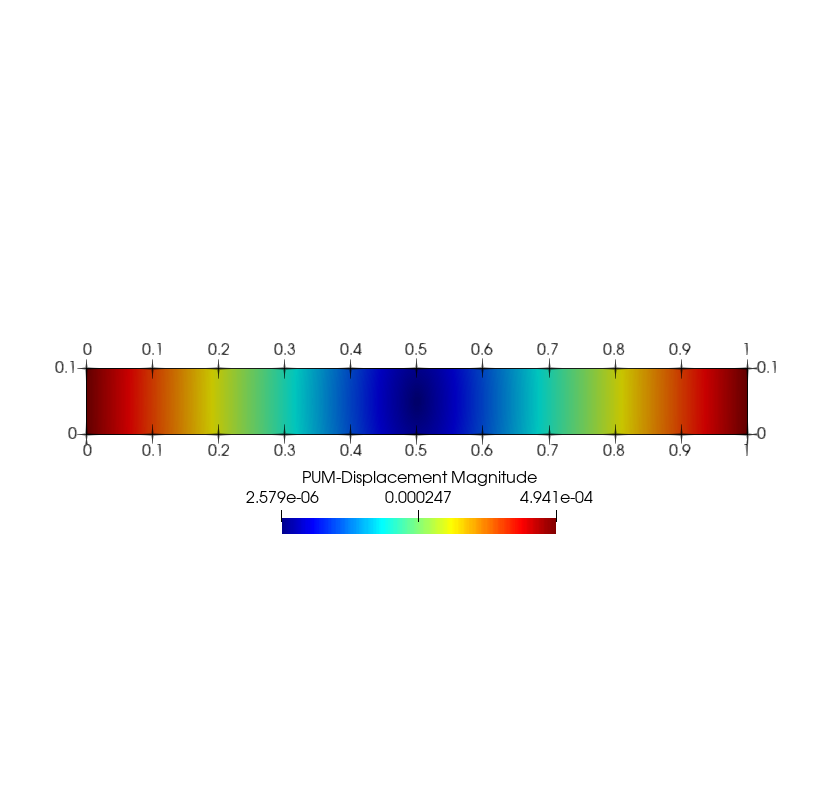}
    }
    \subfloat[PUM (Displacement)\label{fig:bar:load:12:pum}]{
    \includegraphics[trim=0 250 0 300, clip,width=0.495\linewidth]{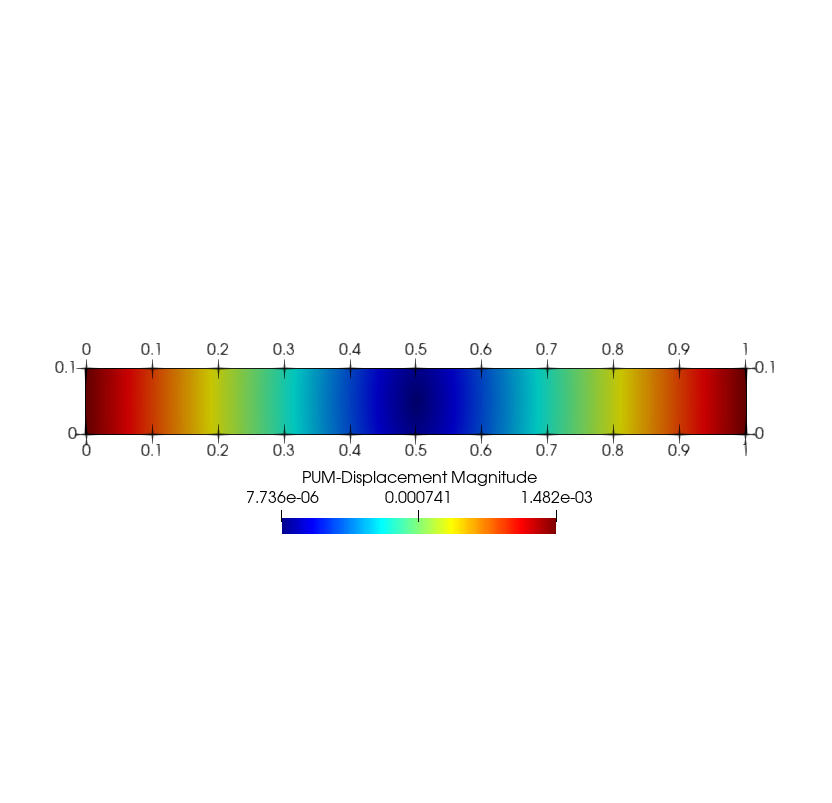}
    }

    \subfloat[PD (Displacement)\label{fig:bar:load:4:pd}]{
    \includegraphics[trim=22 250 22 300, clip,width=0.495\linewidth]{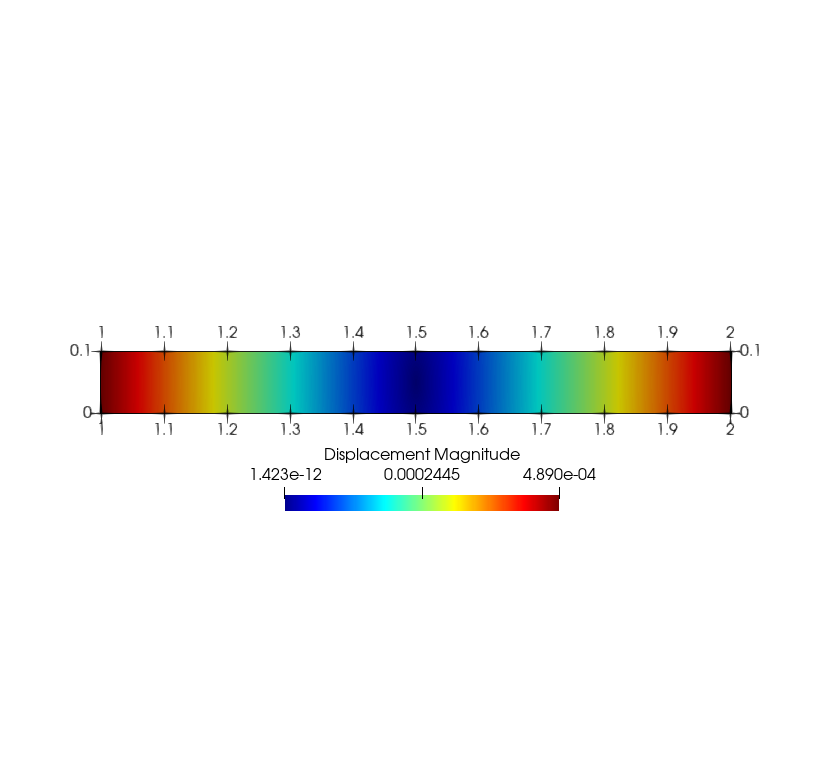}
    }
    \subfloat[PD (Displacement)\label{fig:bar:load:12:pd}]{
    \includegraphics[trim=22 250 22 300, clip,width=0.495\linewidth]{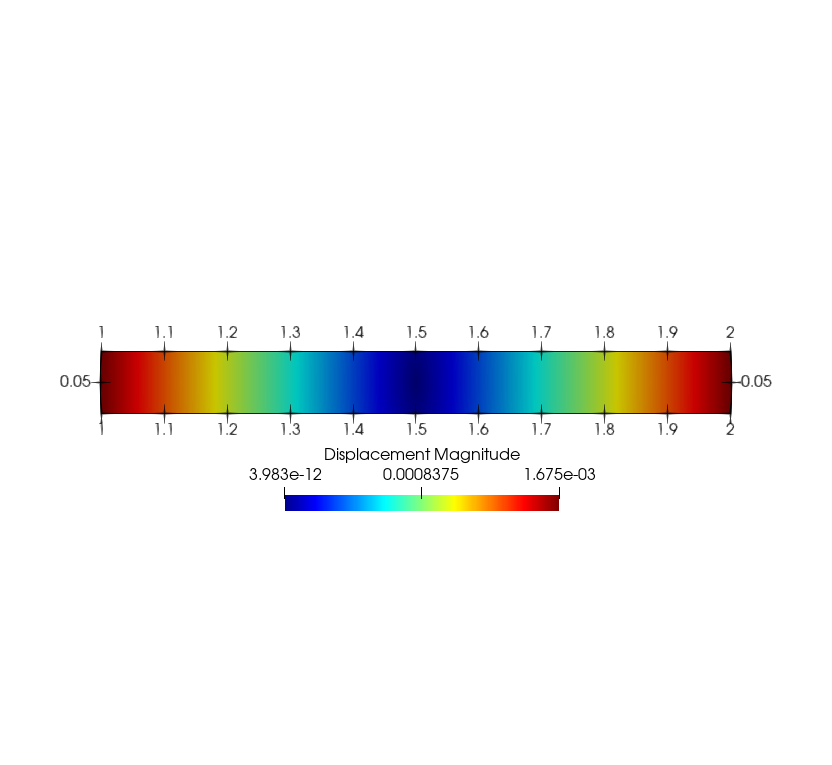}
    }

    \subfloat[PD (Damage)\label{fig:bar:load:4:pd:damage}]{
    \includegraphics[trim=22 250 22 300, clip,width=0.495\linewidth]{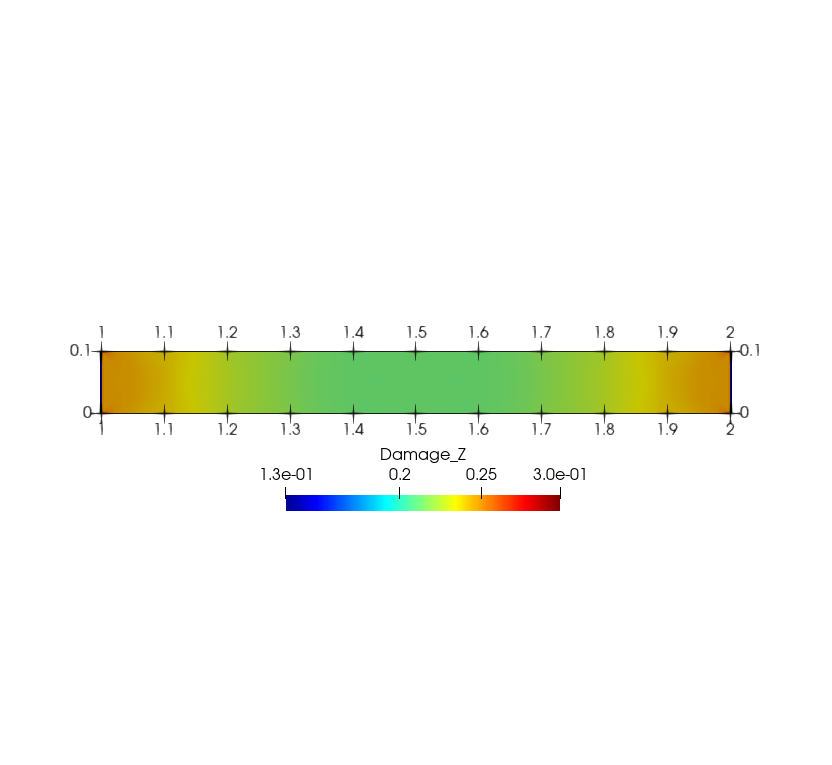}
    }
    \subfloat[PD (Damage)\label{fig:bar:load:12:pd:damage}]{
    \includegraphics[trim=22 250 22 300, clip,width=0.495\linewidth]{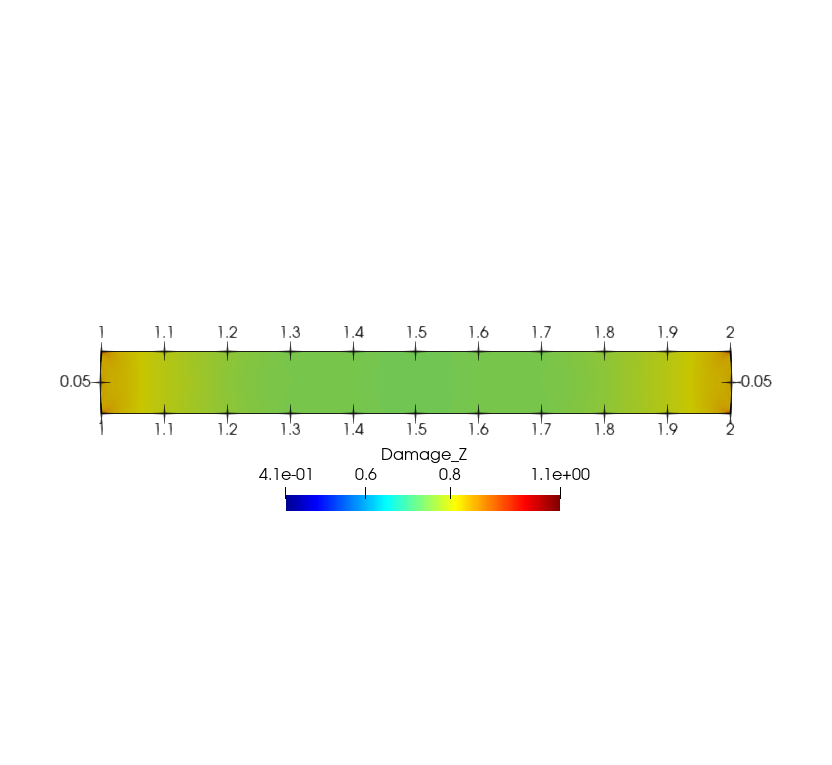}
    }
    \caption{Magnitude of the displacement for the two-dimensional bar, see Figure~\ref{fig:bar:geometry}. The traction condition's load various from \num{9e5}\si{\newton} up to \num{12}$\times$\num{9e5}\si{\newton} to study the effect of softened bonds. As long as the PD model stays in the linear regime, the displacements agree. Once the softening starts the results differ since the PUM still simulates linear elasticity.}
    %\label{fig:my_label}
\end{figure*}

In this first example, we use a two-dimensional bar, see Figure~\ref{fig:bar:geometry}, to investigate the effect of softening bonds in the PD model. The PD potential in Figure~\ref{fig:ConvexConcaveb} stays in the linear regime until the stretch $r$ exceeds its critical value $r^c$ which means that the material behaves linearly elastic.
In that case, the same material behavior as in the PUM model is simulated.
Therefore, this example can validate that both methods and codes simulate the same problem.
However, after the bonds start to soften, i.e. when the deformation is large enough to impose a bond stretch larger than \(r^c\), the PD model simulates a different material behavior, while the PUM model stays in the linear elastic regime.
To investigate the difference in the displacement fields, we apply force $F=\num{9e5}\si{\newton}$ stretching the bonds to $\approx 7.3$\% to the critical stretch $r^c$.
Table~\ref{tab:bar:parameters} lists all simulation parameters.
\begin{table*}[tb]
    \centering
    \caption{Simulation parameters for the discretization in time and space for the two-dimensional bar problem.}
    \label{tab:bar:parameters}
    \begin{tabular}{l|l}
    \toprule
     Force $F=$\num{9e5}\si{\newton}  & Time steps $t_n$=\num{50000} \\
     Node spacing $\hPd=$\num{0.0005}\si{\meter} & Time step size $t_s=$\num{2e-8}\si{\second} \\
     and $\hPum=$\num{0.0078125}\si{\meter} & \\
     Horizon $\delta=4\hPd=$\num{0.002}\si{\meter} & Final time $T$=\num{0.001}\si{\second}  \\\bottomrule
    \end{tabular}
\end{table*}
Here, the maximal displacement magnitude $\mathbf{U}_\text{max}$ obtained by the PUM method is \num{\pm 1.235e-4}\si{\meter} for the quasi-static simulation and \num{\pm 1.253 e-4}\si{\meter} for the dynamic simulation, respectively.
The displacement field for the quasi-static case is shown in Figure~\ref{fig:bar:load:1:pum} and the displacement field for the explicit dynamic case in Figure~\ref{fig:bar:load:1:pum:dynamic}.
The maximal displacement magnitude $\mathbf{U}_\text{max}$ obtained by the explicit PD simulations was \num{1.20335e-4}\si{\meter}, see Figure~\ref{fig:bar:load:1:pd}, and the obtained damage is around \num{7.3}\si{\percent}, see Figure~\ref{fig:bar:load:1:pd:damage}.
Note that we add the external force density, see~\eqref{eq:externalforce:traction}, such that the force increases linearly over time such that we have the final force \(F\) at the final time $T$.

Next, the force is increased by a factor of 4, which results in a bond damage of \num{29.8}\si{\percent}. The maximal displacement magnitude $\mathbf{U}_\text{max}$ obtained by PUM is \num{4.941e-4}\si{\meter} and obtained by PD is \num{4.890e-4}\si{\meter}, respectively. The PUM displacement is shown in Figure~\ref{fig:bar:load:4:pum} and the PD displacement and PD damage in Figure~\ref{fig:bar:load:4:pd} and Figure~\ref{fig:bar:load:4:pd:damage}, respectively.

Next, the force is increased by a factor of 8, which results in a bond damage of \num{62.7}\si{\percent}. The maximal displacement magnitude $\mathbf{U}_\text{max}$ obtained by PUM is \num{9.883e-4}\si{\meter} and obtained by PD is \num{1.020e-3}\si{\meter}, respectively. The PUM displacement is shown in Figure~\ref{fig:bar:load:8:pum} and the PD displacement and PD damage in Figure~\ref{fig:bar:load:8:pd} and Figure~\ref{fig:bar:load:8:pd:damage}, respectively.

Next, the force is increased by a factor of 12, which results in a bond damage of \num{106.1}\si{\percent}. The maximal displacement magnitude $\mathbf{U}_\text{max}$ obtained by PUM is \num{1.675e-3}\si{\meter} and obtained by PD is \num{1.482e-3}\si{\meter}, respectively. The PUM displacement is shown in Figure~\ref{fig:bar:load:12:pum} and the PD displacement and PD damage in Figure~\ref{fig:bar:load:12:pd} and Figure~\ref{fig:bar:load:12:pd:damage}, respectively.

\begin{table*}[tb]
    \centering
    \caption{The maximal displacement magnitude $\mathbf{U}_\text{max}$ obtained by the quasi-static PUM simulation and by the explicit PD simulation for the two-dimensional bar, see Figure~\ref{fig:bar:geometry}. The traction condition's load is increased up to twelve times to showcase the influence of softened bond to the displacement field. As long as the PD model stays in the linear regime, both methods result in a similar displacement and once the softening starts the results diverges as expected.}
    \begin{tabular}{l|cll}
        Load [\si{\newton}] & \multicolumn{2}{c}{$\mathbf{U}_\text{max}$ [\si{\meter}]} & Damage [\si{\percent}] \\\toprule
          & PUM & PD &    \\\midrule
        \num{9e5} & \num{1.235e-4} & \num{1.203e-4} & \num{7.3} \\
        \num{4}$\times$\num{9e5} & \num{4.941e-4} & \num{4.890e-4}  & \num{29.8} \\
        \num{8}$\times$\num{9e5} & \num{9.883e-4} & \num{1.020e-3}  & \num{62.7} \\
        \num{12}$\times$\num{9e5} & \num{1.482e-3} & \num{1.675e-3}  & \num{106.1} \\\bottomrule
    \end{tabular}
    \label{tab:bar:overview}
\end{table*}

Table~\ref{tab:bar:overview} lists all results for the different bond damage values.
One clearly sees that if the bond damage stays below one, the obtained displacements agree well, since the PD model is in the linear regime and both models simulate linear elasticity.
However, if the bond damage is greater than one, the observed displacements begin to differ, since the PD model starts to soften and the PUM model still simulates linear elasticity.
Thus, we have shown that for the linear behavior, both models results in a similar maximal displacement magnitude $\mathbf{U}_\text{max}$.

Note that for the explicit dynamic PUM and the explicit PD simulations, one could do a more detailed convergence study. Especially for PD the suitable choice of the horizon and the grid size is important for convergence.
Another aspect is that the traction condition only converges to the local traction condition while the horizon scales to zero, see \cite{LiptonJhaBdry}.
For the sake of this work, we use this discretization in time and space to show the proof of concept of the proposed multiscale method only.

%%%%%%%%%%%%%%%%%%%%%%%%%%%%%%%%%%
% \FloatBarrier

\subsection{Stationary Mode I crack}
\label{sec:numerical:results:modei}
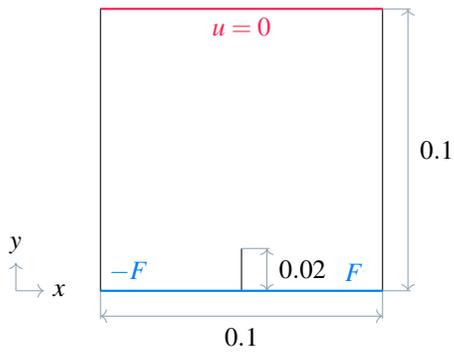
\begin{figure}[htb]
    \centering
        \begin{tikzpicture}[scale=0.75]
            \draw (0,0) -- (5,0) -- (5,5) -- (0,5) -- cycle;
            \draw (2.5,0) -- (2.5,0.75);
            \draw[thin,cadetgrey] (0,0) -- (0,-0.5) ;
            \draw[thin,cadetgrey] (5,0) -- (5,-0.5) ;
            \draw[thin,cadetgrey,<->] (0,-0.45) -- (5,-0.45) ;
            \node[below] at (2.5,-0.5) {\num{0.1}};
            \draw[thin,cadetgrey] (5,0) -- (5.5,0) ;
            \draw[thin,cadetgrey] (5,5) -- (5.5,5) ;
            \draw[thin,cadetgrey,<->] (5.45,0) -- (5.45,5) ;
            \node[right] at (5.5,2.5) {\num{0.1}};
            \draw[thin,cadetgrey] (2.5,0.75) -- (3,0.75) ;
            \draw[thin,cadetgrey,<->] (2.95,0) -- (2.95,0.75) ;
            \node[right] at (3,0.375) {\num{0.02}};
            \draw[->,cadetgrey](-1.5,0) -- (-1,0);
            \draw[->,cadetgrey](-1.5,0) -- (-1.5,0.5);
            \node[right] at (-1,0) {$x$};
            \node[above] at (-1.5,0.5) {$y$};
            \draw[thick,azure] (0,0) -- (2.5,0);
            \node[above,azure] at (0.5,0) {$-F$};
            \draw[thick,azure] (5,0) -- (2.5,0);
            \node[above,azure] at (4.5,0) {$F$};
            \draw[thick,awesome] (0,5) -- (5,5);
            \node[below,awesome] at (2.5,5) {$u = 0$};
        \end{tikzpicture}
    \caption{Sketch of the square plate $(\num{0.1}\si{\meter}\times\num{0.1}\si{\meter})$ with an initial crack of length \num{0.02}\si{\meter}.
        For the local PUM model a constant traction boundary condition of $\pm \num{1e3}\si{\newton}$ is applied to the left-hand side and right-hand side of the initial crack.
        A Dirichlet boundary condition is applied on the top of the plate and the displacement is fixed in both directions.}
    \label{fig:modelproblem:scrack}
\end{figure}

%%%%%%%%%%%%%%%%%%%%%%%%%%%%%%%%%%
Figure~\ref{fig:modelproblem:scrack} shows a square plate $(\num{0.1}\si{\meter}\times\num{0.1}\si{\meter})$ with an initial crack of length $l=\num{0.02}$\si{\meter}.
For the PDE-based PUM model, a traction boundary condition of $\pm \num{1e3}\si{\newton}$ is applied to the bottom of the plate on the left-hand side and right-hand side of the initial crack.
A Dirichlet boundary condition is applied to the top of the plate and the displacement there is fixed. However, in the non-local PD model boundary conditions are applied differently.
For the Dirichlet boundary condition, all nodes within a layer of horizon size at the top of the square plate are fixed in displacement.
The force is applied using the external force density in~\eqref{eq:externalforce:traction}.
Note that we add the external force such that the force increases linearly over time, such that we have the final force \(F\) at the final time $T$.
Table~\ref{tab:stationary:crack:parameters} lists all simulation parameters.
\begin{table*}[tb]
    \centering
    \caption{Simulation parameters for the discretization in time and space for the stationary crack problem.}
    \label{tab:stationary:crack:parameters}
    \begin{tabular}{l|l}
    \toprule
    Force $F=$\num{1e3}\si{\newton}  & Time steps $t_n$=\num{50000}   \\
    Node spacing $\hPd=$\num{0.0005}\si{\meter}  & Time step size $t_s=$\num{2e-8}\si{\second}  \\
    and $\hPum=$\num{0.00078125}\si{\meter} & \\
    Horizon $\delta=4\hPd=$\num{0.002}\si{\meter}  & Final time $T$=\num{0.001}\si{\second}  \\\bottomrule
    \end{tabular}
\end{table*}

\begin{figure*}[htb]
    \centering
    \subfloat[Quasi-static\label{fig:stationary:crack:pum:static}]{\includegraphics[width=0.5\linewidth]{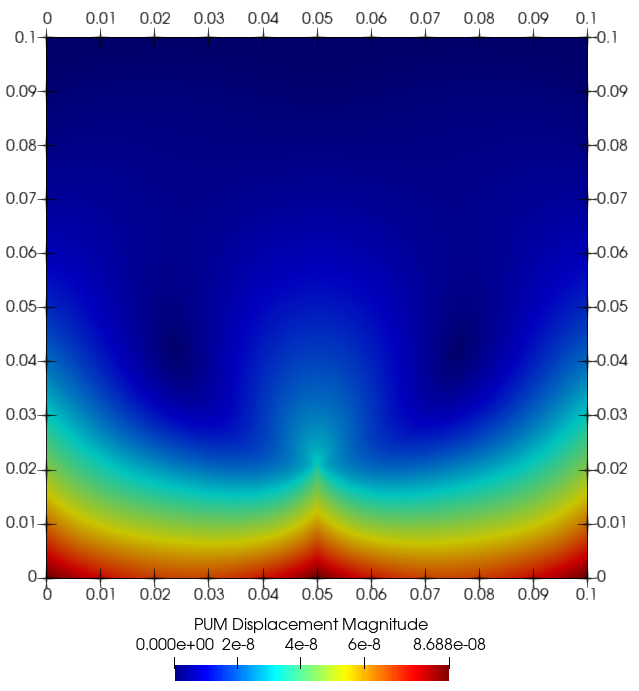}}
    \subfloat[Dynamic\label{fig:stationary:crack:pum:implicit}]{\includegraphics[width=0.5\linewidth]{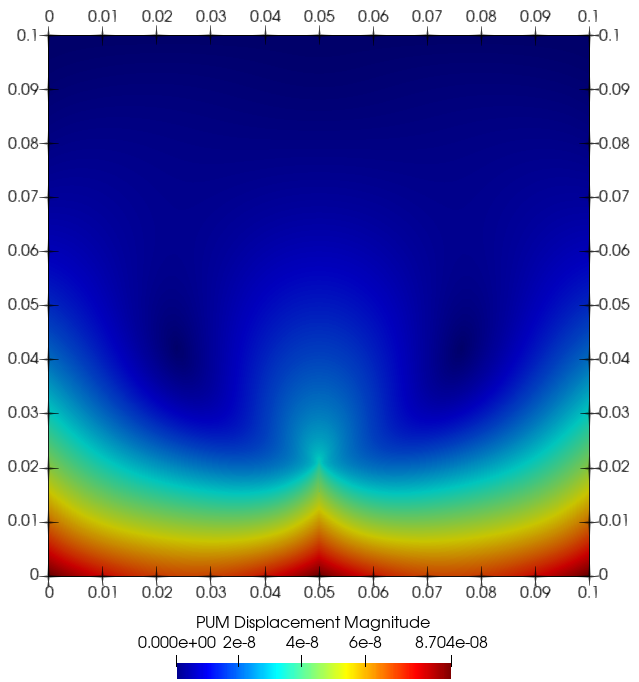}}
    \caption{Obtained displacement magnitudes by PUM for the \protect\subref{fig:stationary:crack:pum:static} quasi-static case and \protect\subref{fig:stationary:crack:pum:implicit} for the explicit dynamic case, which shows good agreement.}
\end{figure*}

First, we compare the displacement field of the quasi-static and explicit dynamic PUM simulation with an initial crack.
Figure~\ref{fig:stationary:crack:pum:static} shows the displacement magnitude for the quasi-static PUM and Figure~\ref{fig:stationary:crack:pum:implicit} shows the displacement magnitude for the dynamic PUM, respectively.
In both cases, the maximal displacement magnitude is \num{8.688e-8}\si{\meter} for the quasi-static simulation and \num{8.704e-8}\si{\meter} for the dynamic simulation.
Again, as for the previous example, both values are sufficiently close.

\begin{figure*}[tbp]
    \centering

    \begin{tikzpicture}
    \node at (0,0) {$\deltaCoarse=0.002$ and $\hCoarse=\sfrac{\deltaCoarse}{4}$};
    \end{tikzpicture}

    \subfloat[Displacement\label{fig:stationary:crack:pd:const}]{\includegraphics[width=0.5\linewidth]{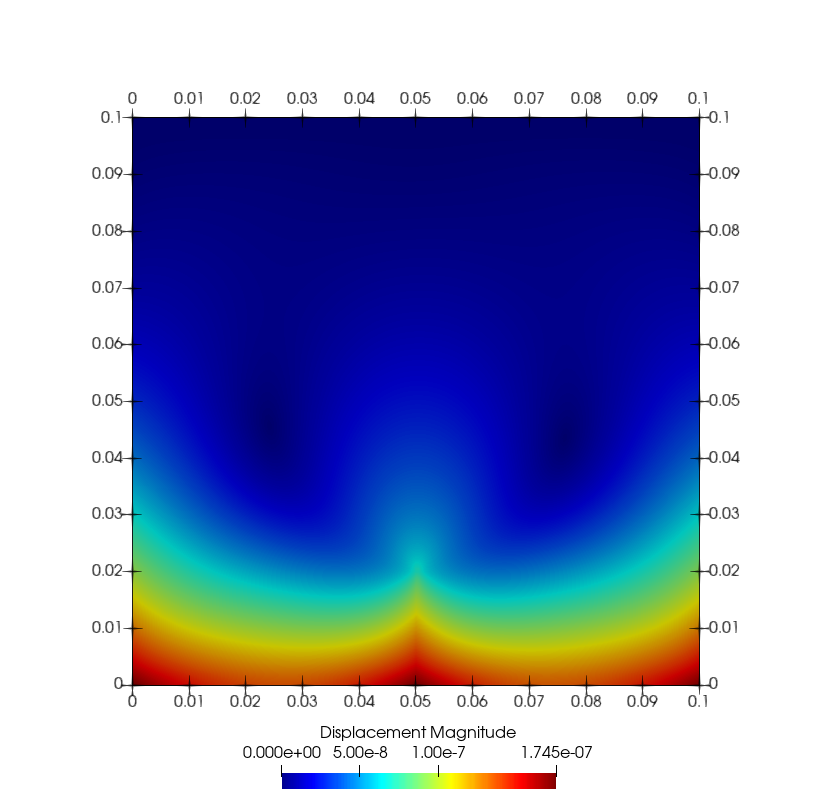}}
    \subfloat[Damage\label{fig:stationary:crack:pd:const:damage}]{\includegraphics[width=0.5\linewidth]{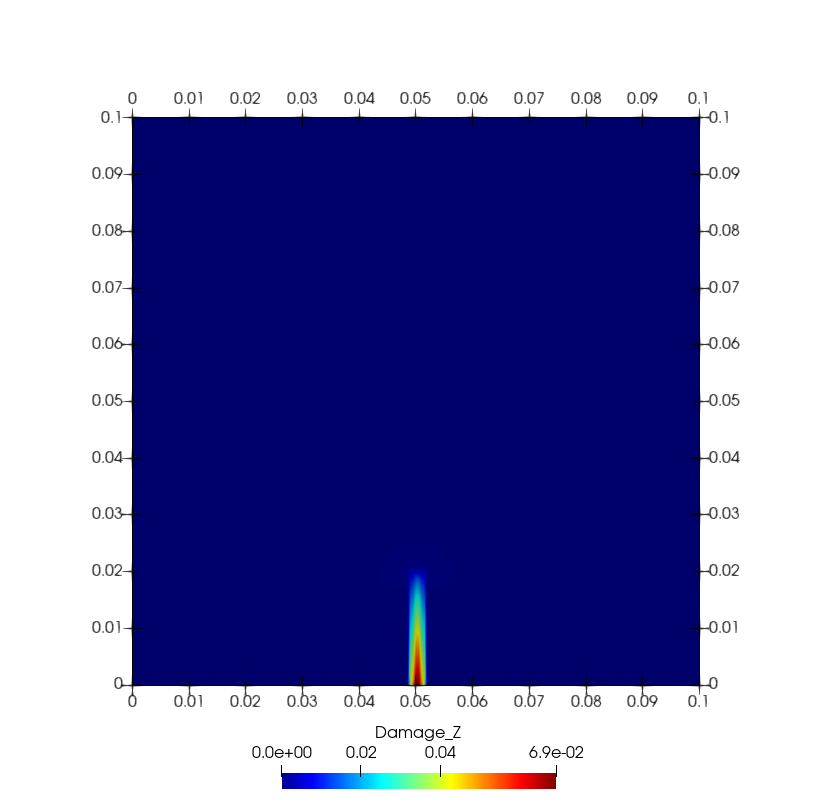}}

    \begin{tikzpicture}
    \node at (0,0) {$\deltaFine=0.001$ and $ \hFine=\sfrac{\deltaFine}{8}$};
    \end{tikzpicture}

    \subfloat[Displacement\label{fig:stationary:crack:pd:const:fine}]{\includegraphics[width=0.5\linewidth]{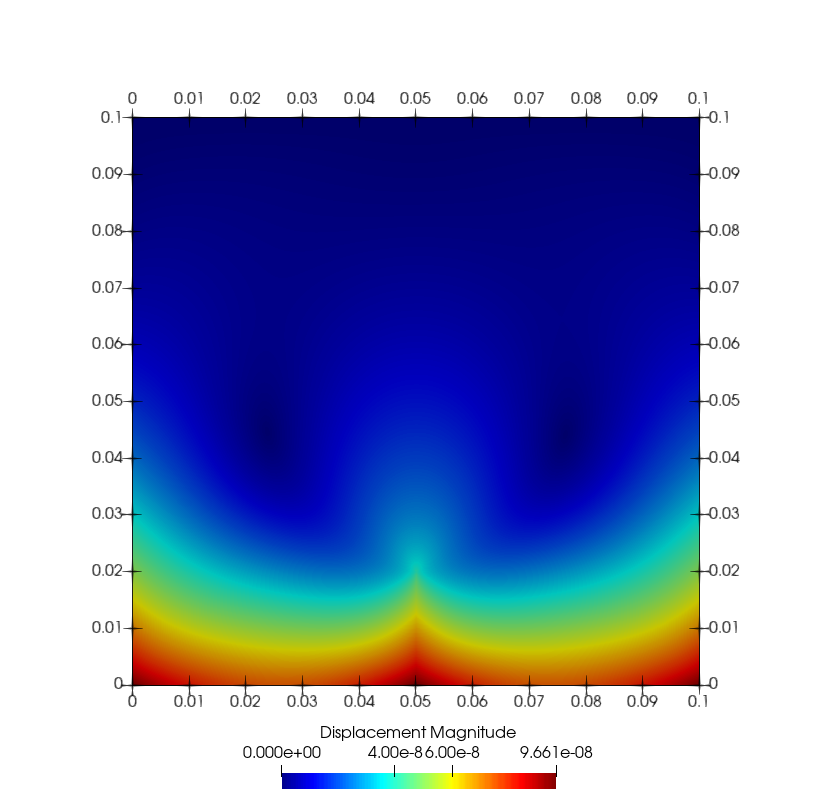}}
    \subfloat[Damage\label{fig:stationary:crack:pd:const:fine:damage}]{\includegraphics[width=0.5\linewidth]{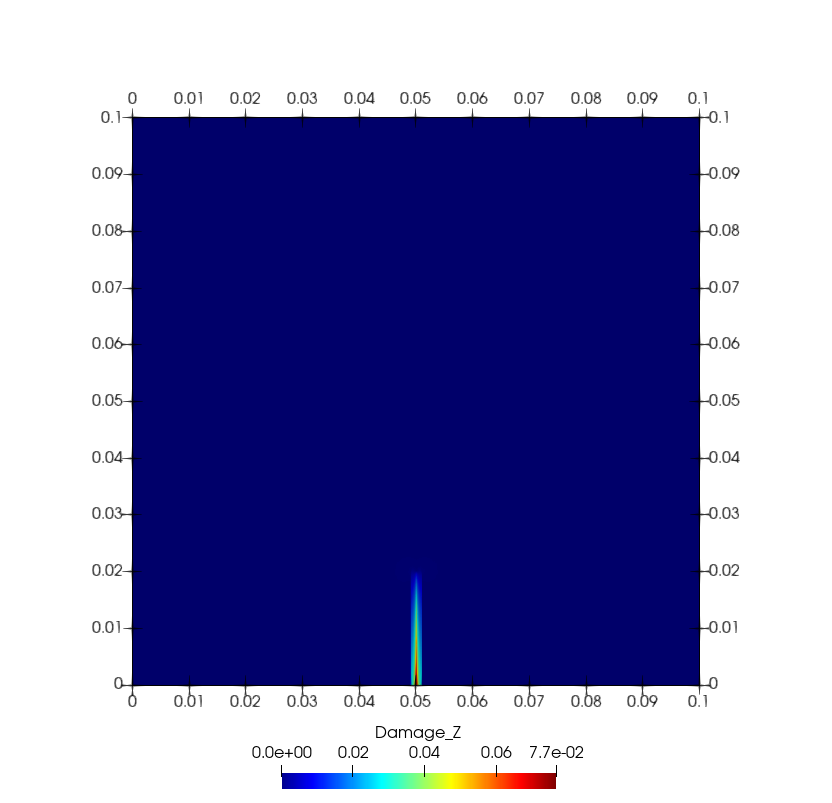}}

    \caption{The displacement magnitude and the damage for the stationary crack example~\ref{sec:numerical:results:modei} as computed by PD on two different nodal spacing.}
    \label{fig:stationary:crack:pd}
\end{figure*}

Second, the displacement field for two different horizons and nodal spacing are compared. For the first PD simulation, the same $\delta$ and the same $\hPd$ as for the two-dimensional bar in Section~\ref{sec:numerical:result:bar} was used. The magnitude of the displacement field is shown in Figure~\ref{fig:stationary:crack:pd:const} and the corresponding damage in Figure~\ref{fig:stationary:crack:pd:const:damage}. Here, the maximal displacement magnitude is \num{1.745e-7}\si{\meter} and the damage is around \num{6.9}\si{\percent}.
Note that for a similar damage percentage in the two-dimensional bar example without any pre-crack both methods predicted a much closer maximum displacement magnitude, see first row of Table~\ref{tab:bar:overview}.
We attribute this observation to a preasymptotic PD approximation.
In~\cite{du2015robust,tian2014asymptotically} various limits are shown and in the case that the horizon $\delta \rightarrow 0$ and the nodal spacing $\hPd\rightarrow 0$ the non-local models converges to the continuum PDE model.
Furthermore, the nodal spacing $\hPd$ has to shrink faster than the horizon $\delta$ to obtain convergence~\cite{diehl2016bond}. Therefore, we decrease the horizon by $\sfrac{1}{2}$ and the nodal spacing $\hPd$ by $\sfrac{1}{4}$. The magnitude of the displacement field is shown in Figure~\ref{fig:stationary:crack:pd:const:fine} and the corresponding damage in Figure~\ref{fig:stationary:crack:pd:const:fine:damage}. Here, the maximal displacement magnitude is \num{9.661e-8}\si{\meter} and the damage is around \num{7.7}\si{\percent}. In that case, the order of magnitude is the same and the maximal displacement magnitude differs by \num{0.993e-8}. Table~\ref{tab:modei:overview} gives an overview of all results. 

One can see that PD approaches the continuum PDE model rather slowly, and a small horizon and nodal spacing is needed to attain acceptable accuracy in the PD results.
In fact, a very small horizon is needed for the convergence of the traction condition, as shown in \cite{LiptonJhaBdry}. The convergence of this model has been discussed in \cite{Diehl2020hpx} and here we wanted to find a suitable horizon and nodal spacing. Therefore, it is beneficial to use PD solely where crack and fractures arise and have the PUM deal with the bulk of the undamaged material and provide appropriate boundary data for the local PD simulations.

\begin{table*}[tb]
    \centering
        \caption{The maximal obtained displacement magnitude for the stationary crack, obtained by a quasi-static and explicit dynamic PUM simulation on the left two columns. On the right two columns an explicit PD simulation on a coarse nodal spacing and finer nodal spacing.}
    \begin{tabular}{ll|ll}
    \multicolumn{4}{c}{ $\mathbf{U}_\text{max}$ [\si{\meter}]} \\\toprule
       \multicolumn{2}{c|}{PUM} &  \multicolumn{2}{c}{PD}
         \\\midrule
        quasi-static & dynamic & Coarse & Fine  \\\midrule
        \num{8.688e-8} & \num{8.704e-8} & \num{1.745e-7} & \num{9.661e-8} \\\bottomrule
    \end{tabular}
    \label{tab:modei:overview}
\end{table*}

%%%%%%%%%%%%%%%%%%%%%%%%%%%%%%%%%%
\FloatBarrier
\section{Numerical results}
\label{sec:numerical:results}
%%%%%%%%%%%%%%%%%%%%%%%%%%%%%%%%%%
After the previous verification of the proposed global-local approach, we now investigate parts of the scheme, where we transfer information between the two methods.
Specifically, we first consider the upper part of Figure~\ref{fig:coupling:glcycle} by running a PD simulation with boundary data generated by a PUM simulation.
Here, the question is whether the piece wise constant PD basis functions can capture the smooth displacement field of the PUM.
In a second experiment, we focus on the lower part by computing a crack path in a PD simulation and using that in a PUM simulation.
Both substantial crack growth and different crack geometry representation in the two methods introduce some technical challenges here.
Unless stated otherwise, we use the same material and discretization parameters as in Section~\ref{sec:numerical:validation}.

%%%%%%%%%%%%%%%%%%%%%%%%%%%%%%%%%%
% \FloatBarrier
\subsection{Verification of coupling PUM and PD using displacement}
\label{sec:numerical:results:coupling:pum}
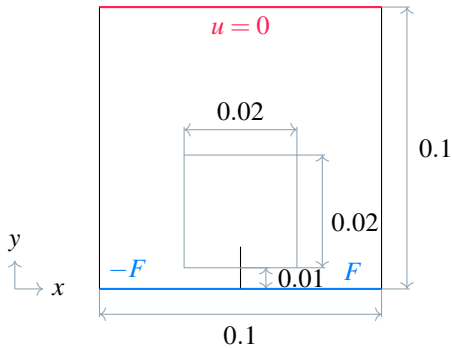
\begin{figure}[htb]
    \centering
    \begin{tikzpicture}[scale=0.75]
        \draw (0,0) -- (5,0) -- (5,5) -- (0,5) -- cycle;
        \draw (2.5,0) -- (2.5,0.75);
        \draw[thin,cadetgrey] (0,0) -- (0,-0.5) ;
        \draw[thin,cadetgrey] (5,0) -- (5,-0.5) ;
        \draw[thin,cadetgrey,<->] (0,-0.45) -- (5,-0.45) ;
        \node[below] at (2.5,-0.5) {\num{0.1}};
        \draw[thin,cadetgrey] (5,0) -- (5.5,0) ;
        \draw[thin,cadetgrey] (5,5) -- (5.5,5) ;
        \draw[thin,cadetgrey,<->] (5.45,0) -- (5.45,5) ;
        \node[right] at (5.5,2.5) {\num{0.1}};
        \draw[thin,cadetgrey] (2.5,0.375) -- (3,0.375) ;
        \draw[thin,cadetgrey,<->] (2.95,0) -- (2.95,0.375) ;
        \node[right] at (3,0.1875) {\num{0.01}};
        \draw[thin,cadetgrey] (1.5,0.375) -- (3.5,0.375) -- (3.5,2.375) -- (1.5,2.375) -- cycle  ;
        \draw[thin,cadetgrey] (3.5,2.375) -- (4,2.375) ;
        \draw[thin,cadetgrey] (3.5,0.375) -- (4,0.375) ;
        \draw[thin,cadetgrey,<->] (3.95,2.375) -- (3.95,0.375) ;
        \node[right] at (3.95,1.1875) {\num{0.02}};
        \draw[thin,cadetgrey] (1.5,2.375) -- (1.5,2.875) ;
        \draw[thin,cadetgrey] (3.5,2.375) -- (3.5,2.875) ;
        \draw[thin,cadetgrey,<->] (1.5,2.825) -- (3.5,2.825) ;
        \node[above] at (2.5,2.825) {\num{0.02}};
        \draw[->,cadetgrey](-1.5,0) -- (-1,0);
        \draw[->,cadetgrey](-1.5,0) -- (-1.5,0.5);
        \node[right] at (-1,0) {$x$};
        \node[above] at (-1.5,0.5) {$y$};
        \draw[thick,azure] (0,0) -- (2.5,0);
        \node[above,azure] at (0.5,0) {$-F$};
        \draw[thick,azure] (5,0) -- (2.5,0);
        \node[above,azure] at (4.5,0) {$F$};
        \draw[thick,awesome] (0,5) -- (5,5);
        \node[below,awesome] at (2.5,5) {$u = 0$};
    \end{tikzpicture}
    \caption{Sketch of the local problem within the global problem. The small square plate (\num{0.02} $\times$ \num{0.02}) is the area of the local problem, including half of the length of the initial crack. For all PD nodes within the layer of horizon size $\delta$ the displacement is prescribed by the displacement obtained by the global PUM simulation.}
    \label{fig:modelproblem:scrack:coupling}
\end{figure}

%%%%%%%%%%%%%%%%%%%%%%%%%%%%%%%%%%
In this example, the focus is on the transfer of information from the PUM to PD, see the upper part of Figure~\ref{fig:coupling:glcycle}.
That is, displacement values computed by the PUM in the global problem are used within the local PD region.
The question we want to answer here is whether the piece wise constant PD basis functions are able to capture the smooth PUM displacement field such that the local end result still fits into the global solution.
To this end, we revisit the stationary Mode I example in Section~\ref{sec:numerical:results:modei}.
Figure~\ref{fig:modelproblem:scrack} sketches the geometry for the global problem.
For the local problem, a square area (\num{0.02} $\times$ \num{0.02}) including half the length of the initial crack is chosen, compare Figure~\ref{fig:modelproblem:scrack:coupling}.

Within the square plate we define a layer of horizon size $\delta$ highlighted in green.
To test local and global coupling, a simplified version of Algorithm~\ref{algo:coupling} is used.
In Line~\ref{algo:coupling:identify} the local domain around the crack which becomes the local PD domain is prescribed.
Note that later a criterion, \emph{e.g.}\ high stress concentration, will be used to determine this region.
In Line~\ref{algo:coupling:setup:pd} the local domain is discretized using the PD code and a VTK file with the discrete PD nodes is written to hard disk.
% Now let us do on iteration of the loop in Line~\ref{algo:coupling:1}.
In Line~\ref{algo:coupling:write:bc} the PUM code reads the VTK file, adds displacement values for all nodes within the horizon layer, and write the file back to hard disk.
In Line~\ref{algo:coupling:read:bc} the PD code reads this file, uses the PUM displacement values as boundary conditions and all other initial displacement values are set to zero, and runs until time $t_{n+1}$.
Note that we linearly increase the displacement at each node within the horizon layer such that the prescribed value by PUM is reached at time $t_{n+1}$.
For this example, the algorithm terminates here, since the goal was to show the displacement obtained in the local region using PD matches the one in the same area of the global region obtained by a pure PUM simulation.
Recall that the open challenges here are to find a criterion to identify the local region, find the synchronization time step size, and time step size for the local problem.
However, these will be investigated in a forthcoming second part of the paper. \\

\begin{figure*}
    \centering
 \includegraphics[width=0.75\linewidth,trim=20 40 20 40, clip]{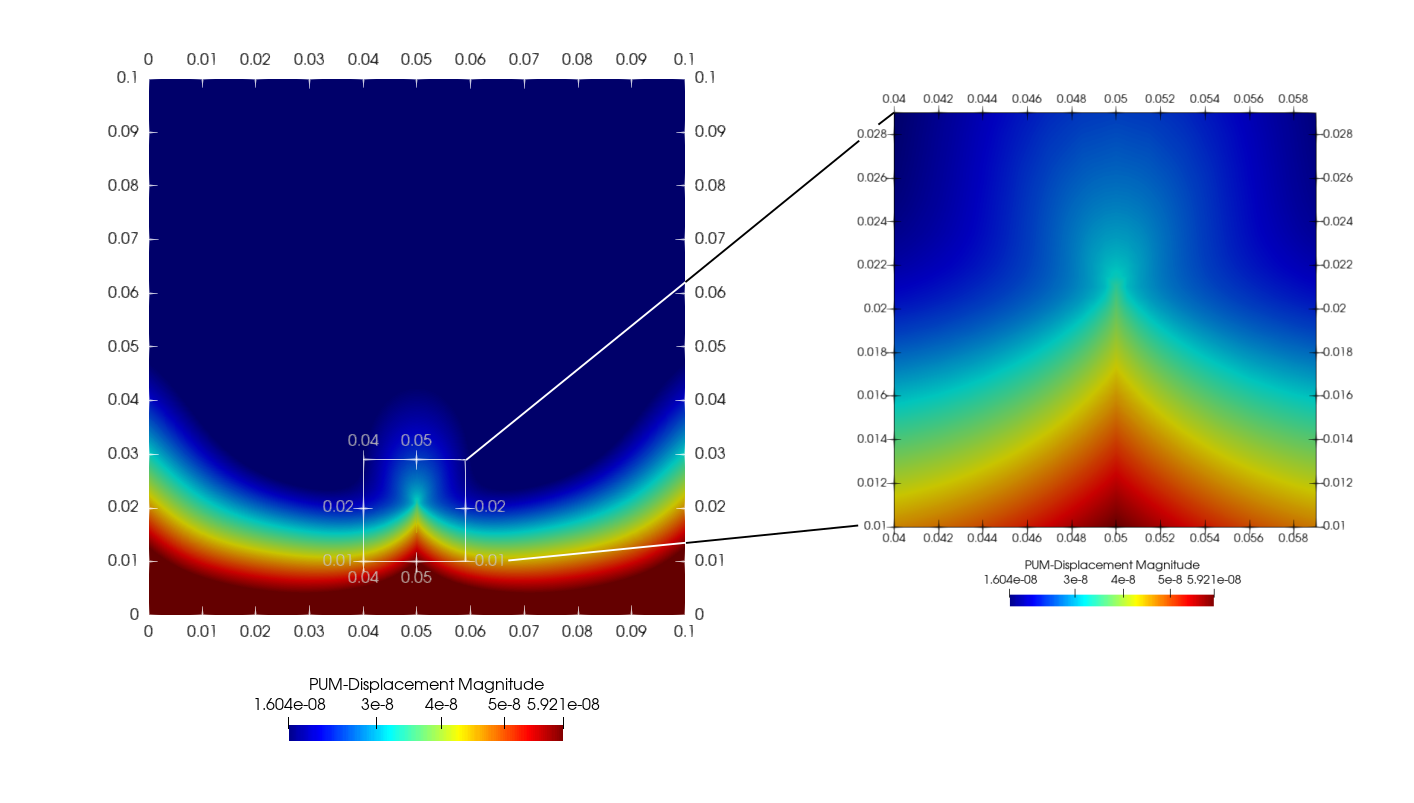}
    \caption{\textbf{Left}: displacement magnitude ($\mathbf{U}_\text{max}$) of the global solution. \textbf{Right}: $\mathbf{U}_\text{max}$ in the local region obtained by the PUM.}
    \label{fig:cutoff:pum}
\end{figure*}

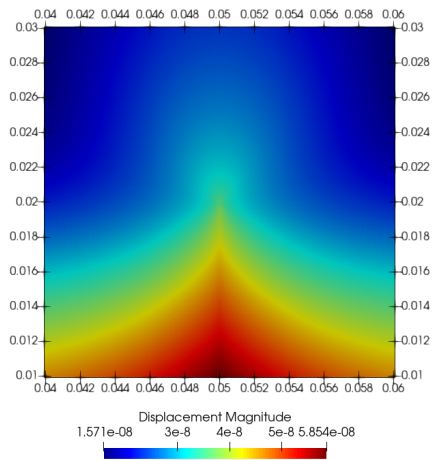
\begin{figure*}
    \centering
    \begin{tikzpicture}
    \node at (-3.5,0) {$\delta_1=0.00031$ and $h_\text{PD}^1=\sfrac{\delta_1}{2}$};
    \node at (3.5,0) {$\delta_2=0.000311$ and $h_\text{PD}^2=\sfrac{\delta_2}{4}$};
    \end{tikzpicture}
    \begin{center}
    \subfloat[\label{fig:cutoff:pd}]{
    \includegraphics[width=0.35\linewidth,trim=0 0 0 0, clip]{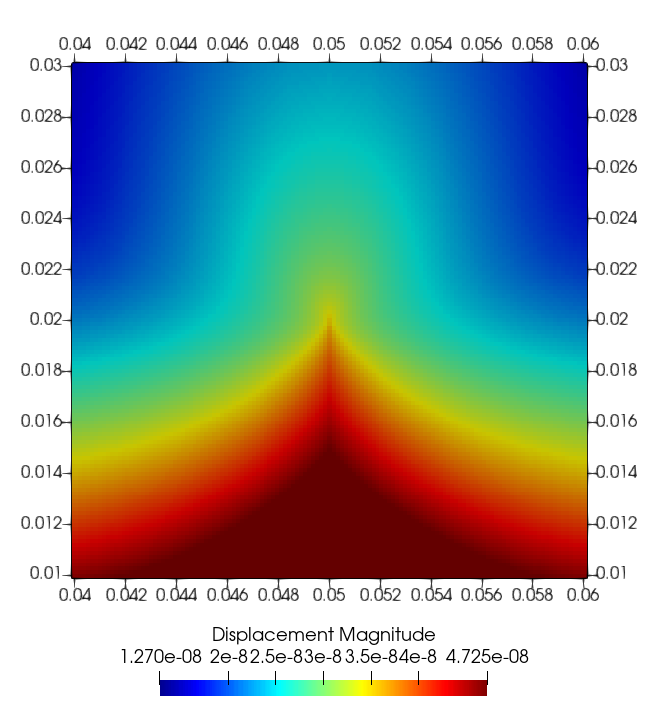}
    }
    \subfloat[\label{fig:cutoff:pd:1}]{
    \includegraphics[width=0.35\linewidth,trim=0 0 0 0, clip]{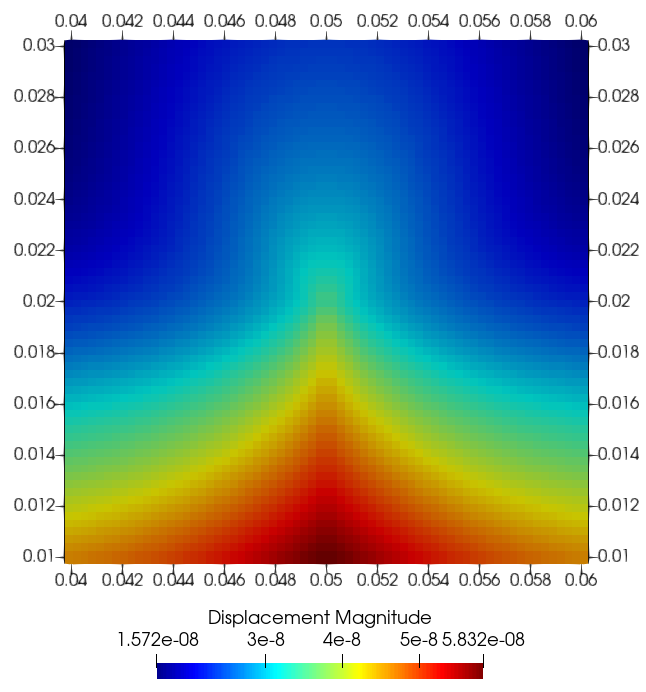}
    }
    \end{center}
    \vspace{0.35cm}
    \begin{tikzpicture}
    \node at (-3.5,0) {$\delta_3=0.00062$ and $h_\text{PD}^3=\sfrac{\delta_3}{8}$};
    %\node at (3.5,0) {$\delta^3=0.00047$ and $h_\text{PD}^3=\sfrac{\delta^3}{12}$};
    \end{tikzpicture}
    \\
    \subfloat[\label{fig:cutoff:pd:2}]{
    \includegraphics[width=0.35\linewidth,trim=0 0 0 0, clip]{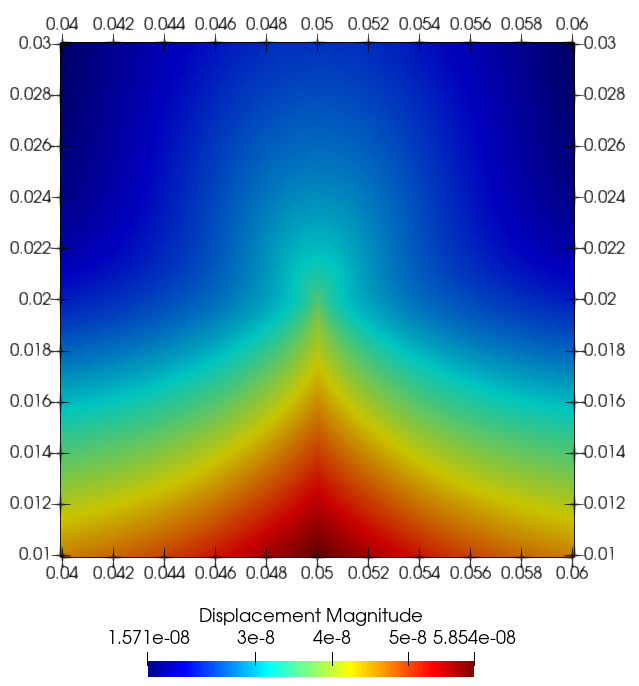}
    }
    %\subfloat[\label{fig:cutoff:pd:3}]{
    %\includegraphics[width=0.35\linewidth,trim=0 0 0 0, clip]{pd_cuttoff-new-25-slice.png}
    %}
    \caption{Displacement magnitude on the local region as obtained by PD on two different nodal spacing, coarse on the left and fine on the right.}
\end{figure*}

\begin{figure*}
    \centering
    \begin{tikzpicture}
    \node at (-3.5,0) {$\delta_1=0.00031$ and $h_text{PD}^1=\sfrac{\delta_1}{2}$};
    \node at (3.5,0) {$\delta_2=0.000311$ and $h_\text{PD}^2=\sfrac{\delta_2}{4}$};
    \end{tikzpicture}
    \begin{center}
    \subfloat[Damage\label{fig:cutoff:damage}]{
    \includegraphics[width=0.35\linewidth,trim=0 0 0 0, clip]{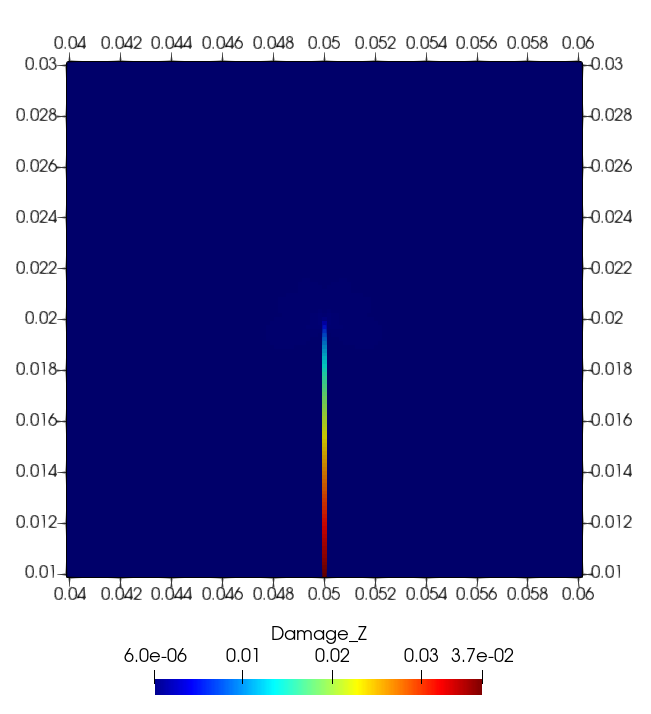}
    }
    \subfloat[Damage \label{fig:cutoff:damage:1}]{
    \includegraphics[width=0.35\linewidth,trim=0 0 0 0, clip]{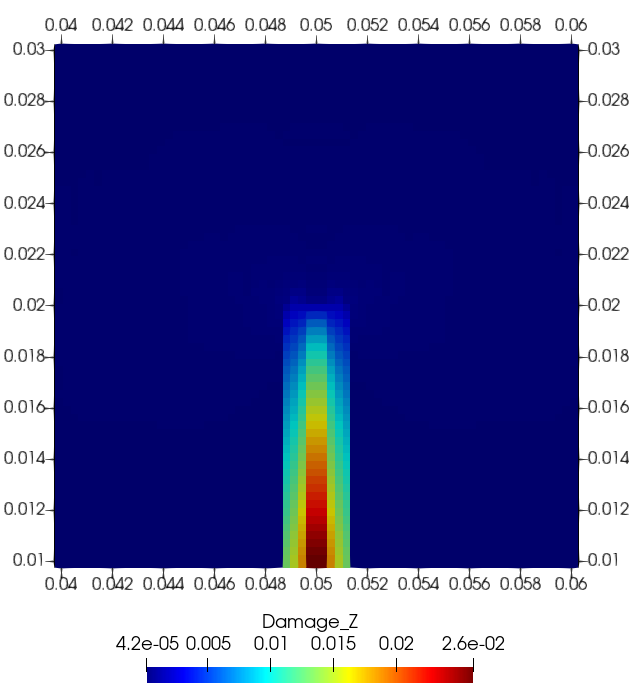}
    }
    \end{center}
    \vspace{0.35cm}
    \begin{tikzpicture}
    \node at (-3.5,0) {$\delta_3=0.00062$ and $h_\text{PD}^3=\sfrac{\delta_3}{8}$};
    %\node at (3.5,0) {$\delta^3=0.00047$ and $h_\text{PD}^3=\sfrac{\delta^3}{12}$};
    \end{tikzpicture}
    \\
    \subfloat[Damage\label{fig:cutoff:damage:2}]{
    \includegraphics[width=0.35\linewidth,trim=0 0 0 0, clip]{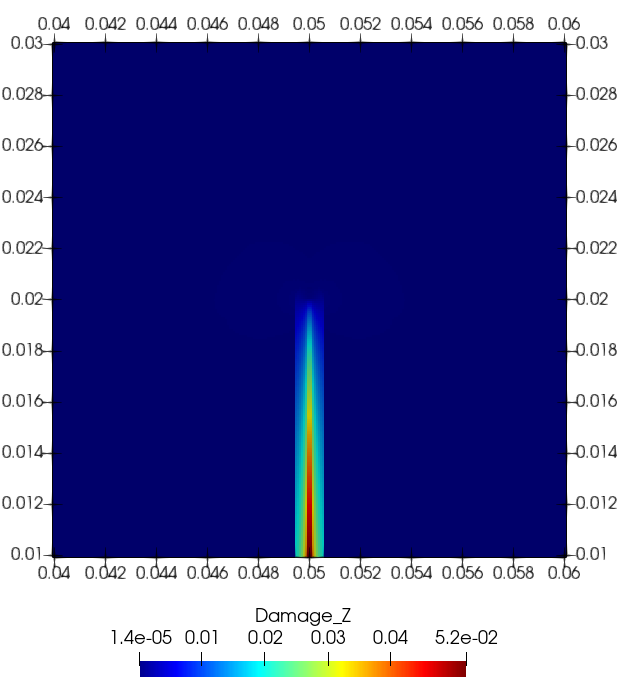}
    }
    %\subfloat[Damage \label{fig:cutoff:damage:3}]{
    %\includegraphics[width=0.35\linewidth,trim=0 0 0 0, clip]{pd_cuttoff_damage-new-245-slice.png}
    %}
    \caption{Damage on the local region as obtained by PD on two different nodal spacing, coarse on the left and fine on the right.}
    \label{fig:cutoff:damage:all}
\end{figure*}

Figure~\ref{fig:cutoff:pum} shows the global displacement magnitude within the local area at the synchronization time step $0.001$\si{\second}. Figure~\ref{fig:cutoff:pd} shows the displacement magnitude of the local problem at the final time $T=0.001$\si{\second} with \num{50000} explicit time steps for the following discretization: \subref{fig:cutoff:pd} $\delta_1=0.00031$ and $h_\text{PD}^1=\sfrac{\delta_1}{2}$; \subref{fig:cutoff:pd:1} $\delta_2=0.000311$ and $h_\text{PD}^2=\sfrac{\delta_2}{4}$; and \subref{fig:cutoff:pd:2} $\delta_3=0.00062$ and $h_\text{PD}^3=\sfrac{\delta_3}{8}$. Figure~\ref{fig:cutoff:damage:all} shows the damage plot of the local problem at the final time $T=0.001$\si{\second} with \num{50000} explicit time steps for the following discretization: \subref{fig:cutoff:damage} $\delta_1=0.00031$ and $h_\text{PD}^1=\sfrac{\delta_1}{2}$; \subref{fig:cutoff:damage:1} $\delta_2=0.000311$ and $h_\text{PD}^2=\sfrac{\delta_2}{4}$; and  \subref{fig:cutoff:damage:2} $\delta_3=0.00062$ and $h_\text{PD}^3=\sfrac{\delta_3}{8}$. \\

Looking at the results from a bird's eye view, one could say that the displacement field matches the shape and has the same order of magnitude.
This suggests that transferring information from the PUM to PD is possible, as the PD basis functions capture the smooth PUM displacement field accurately. To get a straight crack at the boundary between the PUM region and the PD region, the discretization of the PD region needs to be aligned with the PUM region.
Here, experience from adaptive mesh refinement with PD~\cite{bobaru2011adaptive,dipasquale2014crack,bobaru2009convergence,xu2016multiscale} or coupling FE with PD~\cite{zaccariotto2017enhanced} can be used to construct a suitable PD discretization in the local problem's domain employing the prescribed displacement, especially around the crack or discontinuity
This will be investigated in the second part of the paper where the focus is on transferring information from PUM to PD.

%%%%%%%%%%%%%%%%%%%%%%%%%%%%%%%%%%
% \FloatBarrier
\subsubsection{Computational costs}
%%%%%%%%%%%%%%%%%%%%%%%%%%%%%%%%%%
Let us compare the computational costs for running the full PD and full PUM simulations in Section~\ref{sec:numerical:results:modei} and the computational costs of the sub grid in this example. The computational times and number of nodes/elements are shown in Table~\ref{tab:stationary:times}. Note that for the PD simulations, we report the values for the coarse grid first and the fine grid second. Once can see that we save computational time by simulation only a sub grid of the PD region. The exchange of the boundary conditions took \num{0.05}\si{\second} for the coarse grid and \num{0.74}\si{\second} for the fine grid. This includes reading, extraction the condition, and writing the files. The PD simulations were executed on a single node with two 20-core Intel Xeon Gold 6148 Skylake CPUs and 96 GB of memory. So in total 40 cores were used.
The PUM simulations were run sequentially on the author's laptop, parallel execution is possible though.
For the explicit dynamic PUM simulation, the time step size was set to \num{4e-8}\si{\second}, which is half the critical time step size for this problem.
Details about the used software and dependencies are provided in the appendix.

\begin{table*}[tb]
    \centering
    \begin{tabular}{l|ll}
     Method    & Time [\si{\second}] & \# nodes/DOF \\\midrule 
     PUM quasi-static & \num{1.5} & \num{25148} \\
     PUM dynamic  & \num{237} & \num{25148}\\
     Full PD  & \num{4388.33} / \num{17263.5} & \num{641601} / \num{2563201} \\
     Small PD  & \num{522.324} / -- & \num{4356} / \num{66564} \\\bottomrule
    \end{tabular}
    \caption{Simulation details for the PUM simulation using dynamic and quasi-static time integration. For the PD simulations, we show first the details for the full simulations. Later, the details for the small sub grid is shown. Note that we report the values for the coarse grid first and the fine grid second. The timing for the PD discretization with $\delta_1$ and $\delta_3$ is not available, since the cluster was on maintenance and the simulation was run on a work station with different hardware.}
    \label{tab:stationary:times}
\end{table*}

%%%%%%%%%%%%%%%%%%%%%%%%%%%%%%%%%%
% \FloatBarrier
\subsection{Inclined crack}
\label{sec:numerical:results:inclined:crack}
%%%%%%%%%%%%%%%%%%%%%%%%%%%%%%%%%%
In this example, we focus on the transfer of information from PD to the PUM, so on the lower part of Figure~\ref{fig:coupling:glcycle}.
To this end, consider a problem that exhibits substantial crack growth, thus also amplifying the dynamic nature of this simulation compared to the previous ones.
This introduces several new challenges and further stresses the agreement of PUM and PD simulations.
Note that again we compare PD results with quasi-static PUM solutions here, but plan to use explicit dynamics in the PUM too, for coupling both methods.
To justify this, we show dynamic PUM results of the inclined crack problem as well.
Even for this problem, there is only a small difference between the two.

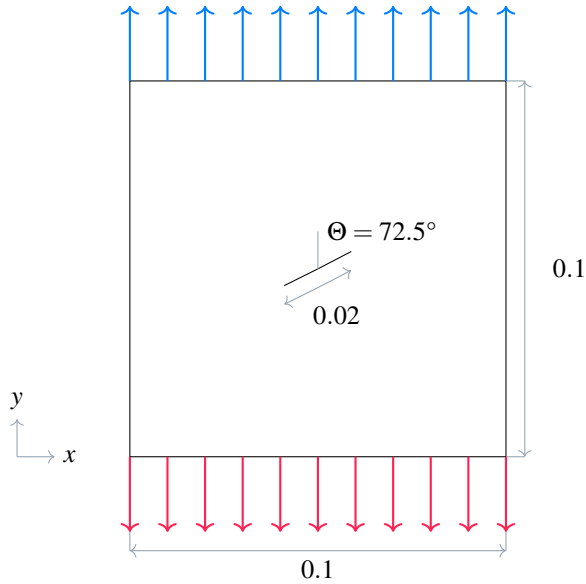
\begin{figure}[htb]
    \centering
    \begin{tikzpicture}
        \draw (0,0) -- (5,0) -- (5,5) -- (0,5) -- cycle;
        \draw[thin,cadetgrey] (5,0) -- (5,-1.25) ;
        \draw[thin,cadetgrey] (0,0) -- (0,-1.25) ;
        \draw[thin,cadetgrey,<->] (0,-1.25) -- (5,-1.25) ;
        \node[below] at (2.5,-1.25) {\num{0.1}};
        \draw[thin,cadetgrey] (5,0) -- (5.25,0) ;
        \draw[thin,cadetgrey] (5,5) -- (5.25,5) ;
        \draw[thin,cadetgrey,<->] (5.25,0) -- (5.25,5) ;
        \node[right] at (5.5,2.5) {\num{0.1}};
        \draw(2.5,2.5) -- ++(27.5:0.5);
        \draw[thin,cadetgrey,->] (2.5,2.25) -- ++(27.5:0.5);
        \draw (2.5,2.5) -- ++(-153.5:0.5);
        \draw[thin,cadetgrey,->] (2.5,2.25) -- ++(-153.5:0.5);
        \node[below] at (2.75,2.125) {\num{0.02}};
        \draw[thin,cadetgrey] (2.5,2.5) -- (2.5,3);
        \node[right] at (2.5,3) {$\Theta=72.5$\si{\degree}};
        \draw[->,cadetgrey](-1.5,0) -- (-1,0);
        \draw[->,cadetgrey](-1.5,0) -- (-1.5,0.5);
        \node[right] at (-1,0) {$x$};
        \node[above] at (-1.5,0.5) {$y$};
        \foreach \i in {1,...,11}
        \draw[->,azure,thick] (5.5-0.5*\i,5) -- (5.5-0.5*\i,6.0);
        \foreach \i in {0,...,10}
        \draw[->,awesome,thick] (5-0.5*\i,0) -- (5-0.5*\i,-1.0);
    \end{tikzpicture}
    \caption{Sketch of the square plate $(\num{0.1}\si{\meter}\times\num{0.1}\si{\meter})$ with an initial inclined crack of length \num{0.02}\si{\meter}.
        For the local PUM model a constant traction boundary condition of $\pm \num{1e3}\si{\newton}$ was applied to the top and bottom of the square plate.}
    \label{fig:modelproblem:incline}
\end{figure}

Figure~\ref{fig:modelproblem:incline} shows a square plate of length $(\num{0.1}\si{\meter}\times\num{0.1}\si{\meter})$ with an initial inclined crack of length \num{0.02}\si{\meter} inclined by $\Theta=72.5$\si{\degree}.
A force $F$ is applied in normal direction at the top and bottom of the square.
In the dynamic PD simulations, the force is zero initially and scales linear to the maximal force $F$ at the final time $T$.

\begin{table*}[tb]
    \centering
    \caption{Simulation parameters for the discretization in time and space for the inclined crack problem.}
    \label{tab:inclined:crack:parameters}
    \begin{tabular}{l|l}
        \toprule
        Force $F_{\mathrm{coarse}}=$\num{4.25e6}\si{\newton}  & Final time $T$=\num{0.001}\si{\second}  \\
        Force $F_{\mathrm{fine}}=$\num{9e6}\si{\newton} & Time steps $t_n$=\num{50000}  \\
        Node spacing:  &  Time step size $t_s=$\num{2e-8}\si{\second}  \\
        $\hPum=$\num{0.00078125}\si{\meter} & Horizon:  \\
        $\hCoarse=$\num{0.0005}\si{\meter}  &  $\deltaCoarse=4\hPd^\text{coarse}$ \\
        $\hFine=$\num{0.000125}\si{\meter} &  $\deltaFine=8\hPd^\text{fine}$\\
        \bottomrule
    \end{tabular}
\end{table*}

Table~\ref{tab:inclined:crack:parameters} lists the simulation parameters for the discretization in time and space.
In PD the boundary condition is applied as described in~\eqref{eq:externalforce:traction}.
As in the previous example, we run two PD simulations with a coarse and a fine nodal spacing.

\begin{figure*}
    \centering
    \subfloat[$\deltaCoarse=0.002$ and $\hCoarse=\sfrac{\deltaCoarse}{4}$]{
    \begin{tikzpicture}[scale=0.9]
    \begin{axis}[xmin=0,xmax=0.1,ymin=0,ymax=0.1]
    \draw [thick, draw=black]
        (axis cs: 0.04023505748323,0.04653142352928) -- (axis cs: 0.059644249830343,0.052772541916514);
    \addplot[azure,mark=*] table [x=x1, y=y1, col sep=comma] {crack_pos_inclined.csv};
    \addplot[awesome,mark=*] table [x=x2, y=y2, col sep=comma] {crack_pos_inclined.csv};
    \end{axis}
    \end{tikzpicture}
    }
    \subfloat[$\deltaFine=0.001$ and $\hFine=\sfrac{\deltaFine}{8}$]{
    \begin{tikzpicture}[scale=0.9]
    \begin{axis}[xmin=0,xmax=0.1,ymin=0,ymax=0.1]
    \draw [thick, draw=black]
        (axis cs: 0.04023505748323,0.04653142352928) -- (axis cs: 0.059644249830343,0.052772541916514);
    \addplot[azure,mark=*] table [x=x1, y=y1, col sep=comma] {crack_pos_inclined-2.csv};
    \addplot[awesome,mark=*] table [x=x2, y=y2, col sep=comma] {crack_pos_inclined-2.csv};
    \end{axis}
    \end{tikzpicture}
    }
    \caption{Extracted crack tip positions obtained by the PD simulation. The black line is the initial inclined crack. The red and blue lines are the crack branches and each dot is the crack position extracted between the explicit time step \num{47500} to \num{50000} with a resolution of 500 explicit time steps in between. The crack tip positions were extracted using Algorithm~\ref{algo:crack:path}.}
    \label{fig:crack:path}
\end{figure*}
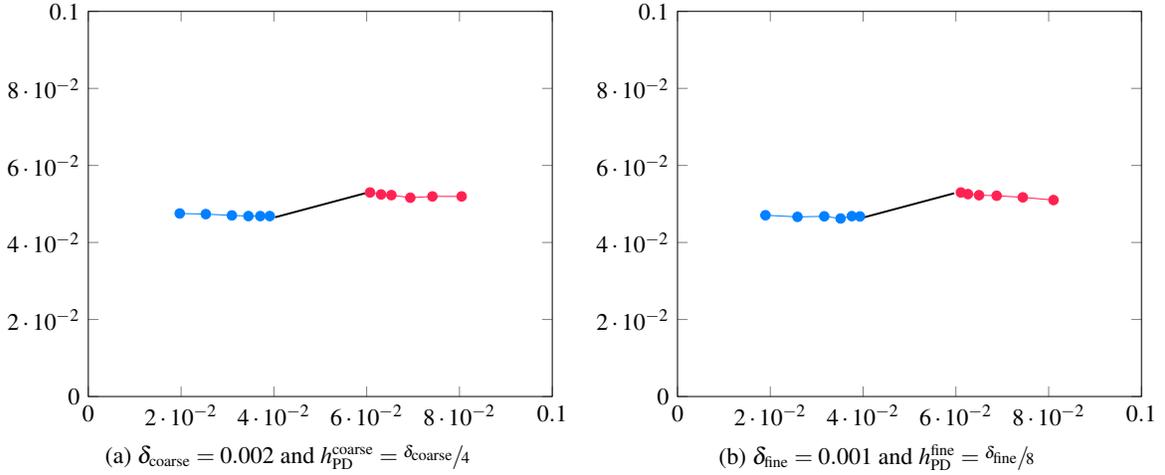

Figure~\ref{fig:crack:path} shows the extracted crack path for both nodal spacing.
The crack path was extracted in a post-processing step using the damage field $D$ and the current positions of the PD nodes, see Algorithm~\ref{algo:crack:path}.
Note that PD has the notion of damage included in the model, however, the crack tip or crack surface is not encoded in the model and needs to be extracted~\cite{diehl_lipton_wick_tyagi_2021}.
More sophisticated methods to extract the crack surface were proposed by the authors~\cite{diehl2017extraction,bussler2017visualization} which are approximations of the crack path between PD nodes, however, these numerical models introduce an additional uncertainty and will be investigated in the future.\\

\begin{figure*}[tbp]
    \centering

    \subfloat[PUM \label{fig:inclined:crack:pum}]{
    \includegraphics[width=0.4\linewidth,trim=20 40 20 40, clip]{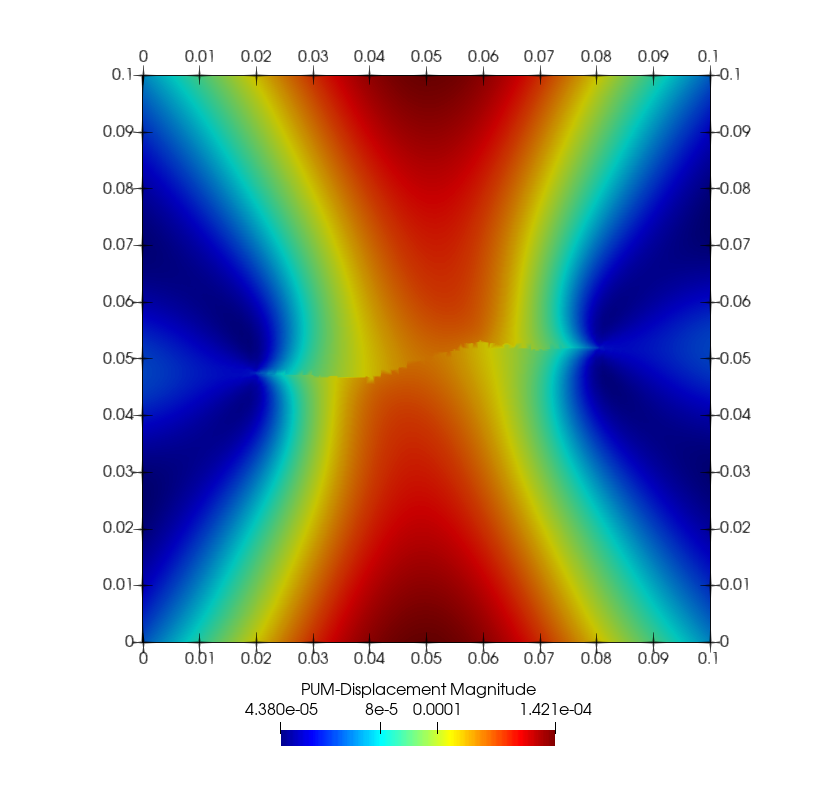}
    }
    \subfloat[PUM \label{fig:inclined:crack:pum:fine}]{
    \includegraphics[width=0.4\linewidth,trim=20 40 20 40, clip]{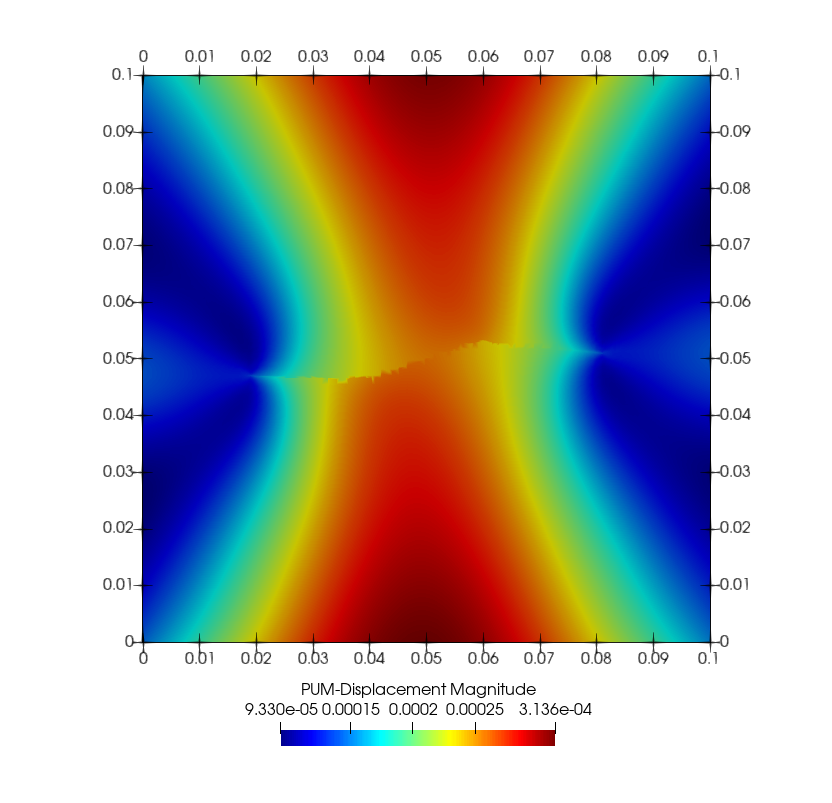}
    }
    \vspace{0.125cm}
    \begin{tikzpicture}
    \node at (-3.5,0) {$\deltaCoarse=0.002$ and $\hCoarse=\sfrac{\deltaCoarse}{4}$};
    \node at (3.5,0) {$\deltaFine=0.001$ and $\hFine=\sfrac{\deltaFine}{8}$};
    \end{tikzpicture}

    \vspace{-0.35cm}

    \subfloat[PD \label{fig:inclined:crack:pd}]{
    \includegraphics[width=0.4\linewidth,trim=20 40 20 40, clip]{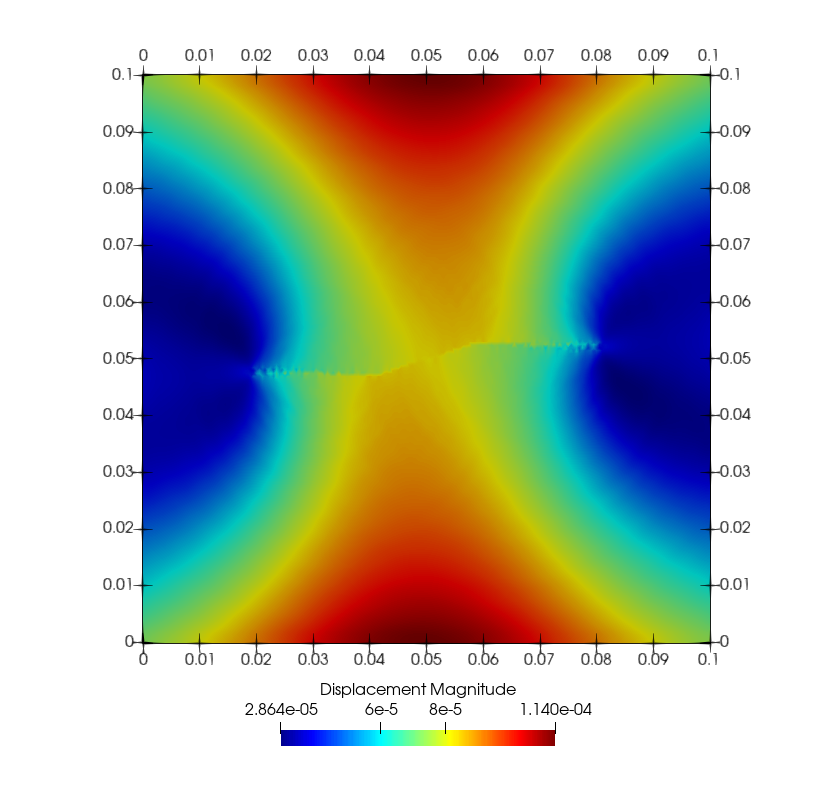}
    }
    \subfloat[PD \label{fig:inclined:crack:pd:fine}]{
    \includegraphics[width=0.4\linewidth,trim=20 40 20 40, clip]{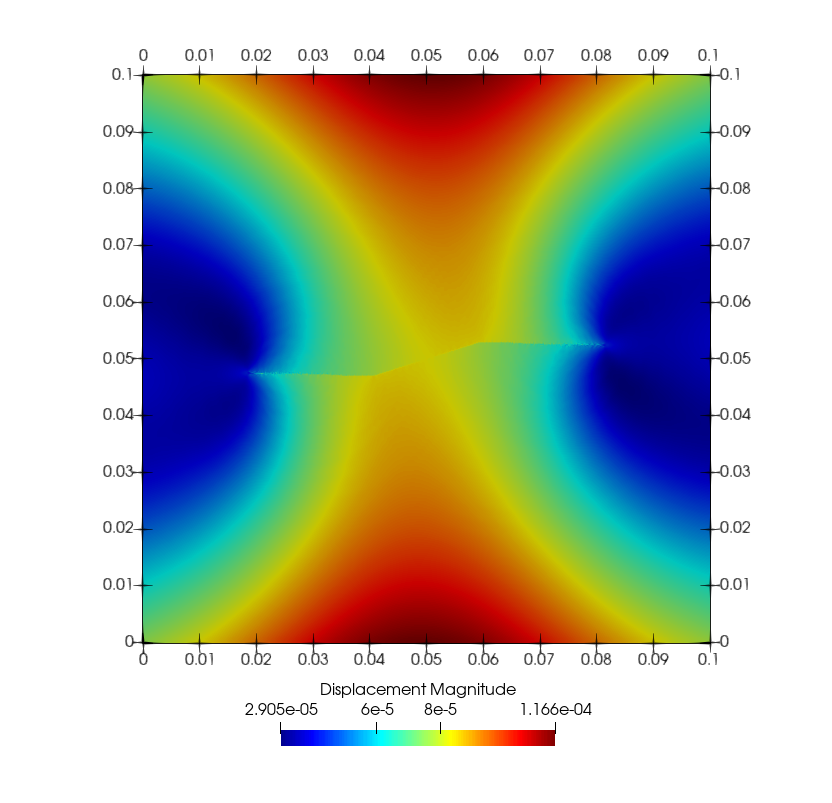}
    }

    \vspace{-0.35cm}

    \subfloat[Damage\label{fig:inclined:crack:pd:damage}]{
    \includegraphics[width=0.4\linewidth,trim=20 40 20 40, clip]{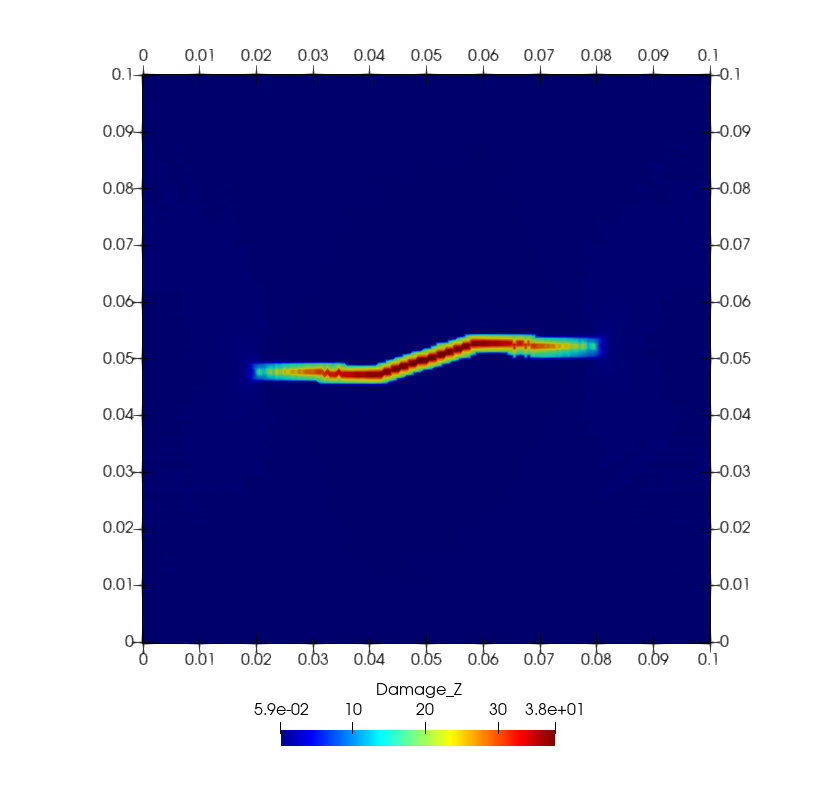}
    }
    \subfloat[Damage \label{fig:inclined:crack:pd:damage:fine}]{
    \includegraphics[width=0.4\linewidth,trim=20 40 20 40, clip]{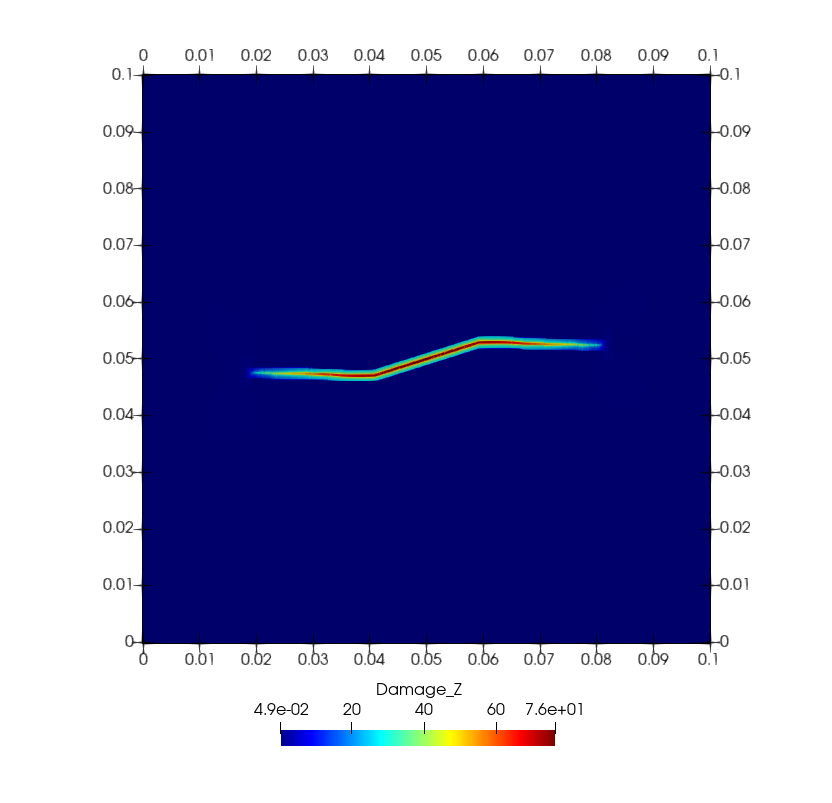}
    }
    \caption{Magnitude of the displacement in the last time step of the inclined crack for a coarse nodal spacing (left) and a fine nodal spacing (right).}
    \label{fig:inclined:crack:u}
\end{figure*}

Figure~\ref{fig:inclined:crack:pum} shows the displacement magnitude obtained by the PUM method using the crack path from the coarse nodal spacing. Figure~\ref{fig:inclined:crack:pd} shows the displacement magnitude obtained by PD and Figure~\ref{fig:inclined:crack:pd:damage} the corresponding damage.
Figure~\ref{fig:inclined:crack:pum:fine} shows the displacement magnitude obtained by the PUM method using the crack path from the fine nodal spacing. Figure~\ref{fig:inclined:crack:pd:fine} shows the displacement magnitude obtained by PD and Figure~\ref{fig:inclined:crack:pd:damage:fine} the corresponding damage.\\

In the coarse case, there is only a small gap between the two solutions, as PD computes a maximal displacement magnitude of \num{1.140e-4}\si{\meter} and the PUM \num{1.421e-4}\si{\meter}.
In the fine case, with a greater force applied, the gap widens as PD computes a maximal displacement magnitude of \num{1.166e-4}\si{\meter} whereas with \num{3.136e-4}\si{\meter} the PDE model of the PUM reacts linearly to the increased load.
These are comparisons with quasi-static PUM solutions.

\begin{figure*}[htb]
    \centering
    \subfloat[PUM solution]{
    \includegraphics[width=0.54\linewidth]{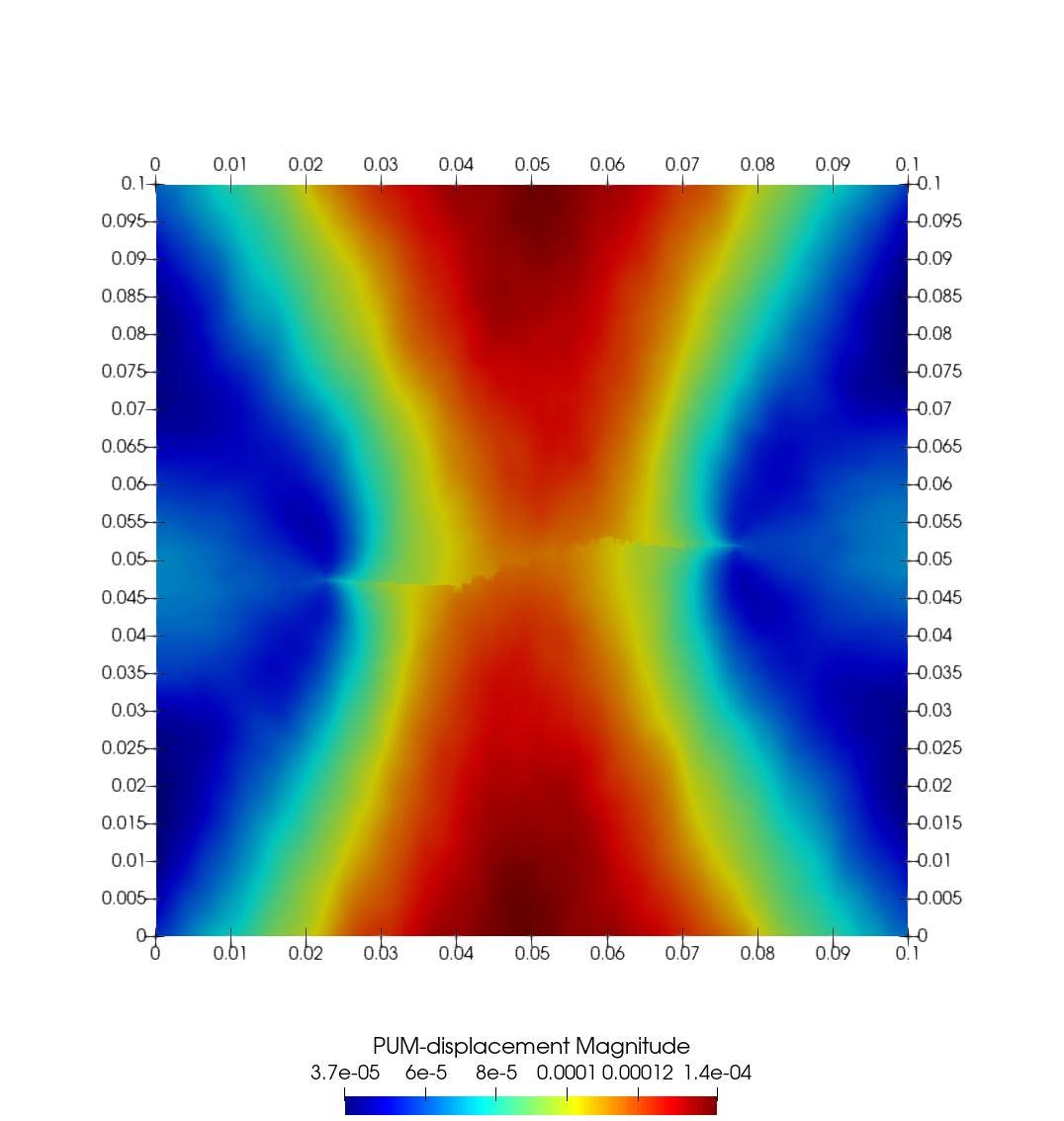}}
    \subfloat[Point-wise difference]{
    \includegraphics[width=0.5\linewidth,trim=250 40 250 40, clip]{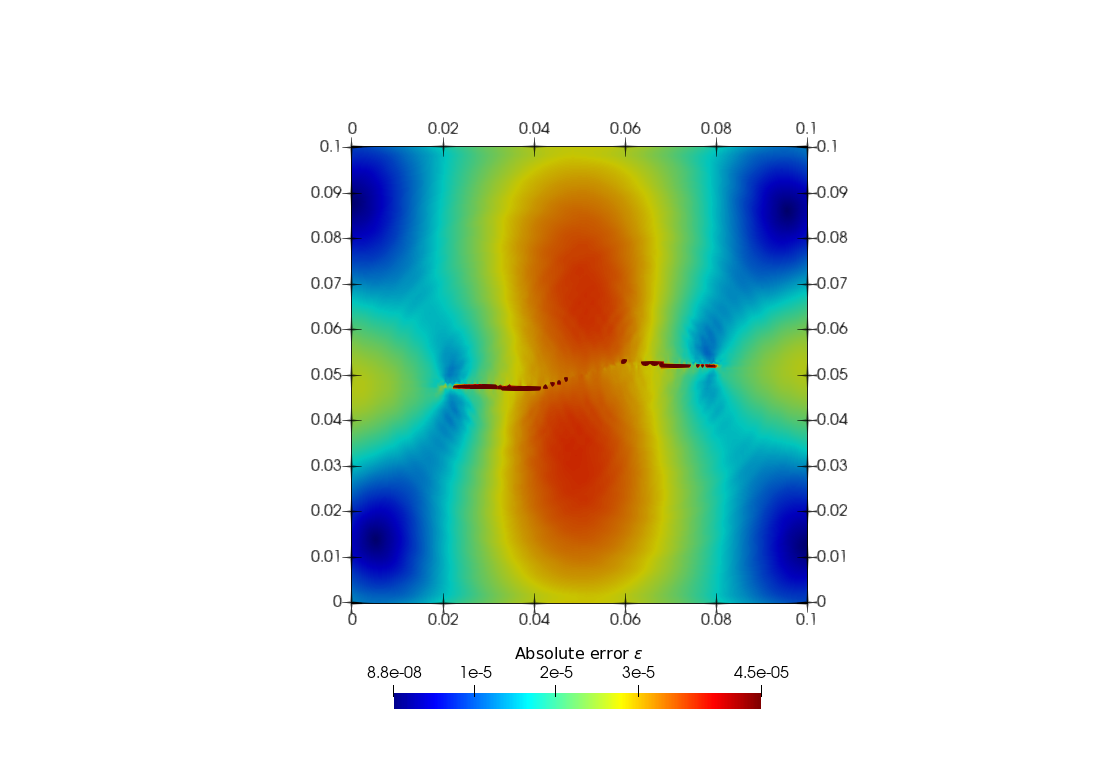}}
    \caption{(a) Explicit dynamic PUM solution of the coarse inclined crack problem. Maximum displacement and overall shape match the quasi-static solution, but small issues remain.
    (b) Point-wise absolute error $\epsilon_i:=\vert \mathbf{U}_\text{PD}(\xb_i)-\mathbf{U}_\text{PUM}(\xb_i) \vert$  between the PUM and the PD solution at the final time $T$. Note that the largest error is at the crack and dominating. Thus, we use a different color scale excluding values close to the crack.}
    \label{fig:inclined:dynamic:pum}
\end{figure*}

For the coarse problem, we provide an explicit dynamic PUM solution in Figure~\ref{fig:inclined:dynamic:pum}.
While maximum displacement and overall shape again match the quasi-static solution, some wavy artifacts are visible.
We believe this to have two reasons:
First, due to the involved crack enrichments, the ansatz space evolves over time, which can introduce stability problems in a naive explicit time stepping scheme used here.
Stable time stepping schemes adapted to this problem are available in the literature, see e.g.~\cite{chang2002explicit, chang2007enhanced, gravouil2009explicit}.
Second, we currently advance the crack front in rather big steps, which probably introduces dynamic effects into the simulation.
Both problems will be addressed in the future parts, when we actually couple the methods.

Previously, the magnitude of the displacement was in a good agreement for the example where the PD model stayed in the linear regime, see Section~\ref{sec:numerical:result:bar}, and for stationary Mode I crack, see Section~\ref{sec:numerical:results:modei}. To get a better understanding of the difference in the displacement field, we compute the point-wise absolute error $\epsilon_i:=\vert \mathbf{U}_\text{PD}(\xb_i)-\mathbf{U}_\text{PUM}(\xb_i) \vert$, see Figure~\ref{fig:inclined:dynamic:pum}. Note that the relative error was avoided, since the displacement could be zero, which results in a division by zero. One clearly identifies that the absolute error $\epsilon_i$ is large on the left-hand side and right-hand side at the crack tip and there is the band going from the initial crack to the top and bottom of the plate.\\

The obvious issues would be that the PD model left the linear regime and the softening started.
However, from the damage fields (Figure~\ref{fig:inclined:crack:pd:damage} and Figure~\ref{fig:inclined:crack:pd:damage:fine}) we can observe that in both cases most of the domain is in the linear regime of the PD material model, as the damage is way below \num{20}\si{\percent} except for the crack.
We currently assert the following reasons might explain the differences in the displacement fields once crack growth is involved in the PD model:
\begin{itemize}
    \item \textbf{Sharp crack vs crack zone}:\\
    Using the extracted crack path within the PUM yields in a sharp crack, however in the PD model the crack is instead a narrow zone of failed bonds of finite width equal to twice the peridynamic horizon $\delta$. In addition, at the crack tip there is a process zone on the order of the horizon size $\delta$ where the bonds are in the process of softening to failure. Because of this, PUM exhibits a strain singularity where PD does not. This difference between the strain fields surrounding the cohesive zone in PD and the strain field around the sharp crack in PUM contributes to the difference in displacement around the crack seen for these two models.
    \item \textbf{Boundary conditions:}\\
    In the PUM simulation, a traction condition is applied. However, in the PD simulation the traction condition is applied within a layer of horizon size, see Equation~\ref{eq:externalforce:traction}. It has shown that if the horizon goes to zero, the same traction condition as in the local PUM model is applied~\cite{LiptonJhaBdry}. In the future, the application of the non-local traction condition can be improved as described in~\cite{parksyueyoutrask}.
    \item  \textbf{Local interactions vs nonlocal interactions}\\
    The both models only agree in the linear regime of the potential in Figure~\ref{fig:ConvexConcaveb}. Once the inflection point $r^C$ is approached and bonds start to soften, the two models do not agree anymore. Hence, the displacement field around the crack looks different due to the non-local effects in the PD model. Note that in the area where bonds soften in the PD model, we use still linear elasticity in the PUM model.
\end{itemize}
We want to emphasize that in the first part of the series, we wanted to showcase a proof of concept for the proposed methods and show its validity and potential.
Some effects seen in the figure will be addressed in the part two of the paper series.

% \begin{figure}[htbp]
%     \centering
%     \includegraphics[width=0.5\linewidth,trim=250 40 250 40, clip]{absolute_error.png}

%     \caption{Point-wise absolute error $\epsilon_i:=\vert \mathbf{U}_\text{PD}(\xb_i)-\mathbf{U}_\text{PUM}(\xb_i) \vert$  between the PUM and the PD solution at the final time $T$. Note that the largest error is at the crack and dominating. Thus, we use a different color scale excluding values close to the crack.}
%     \label{fig:inclined:crack:error}
% \end{figure}

\begin{table}[tb]
    \centering
    \caption{The maximal displacement magnitude $\mathbf{U}_\text{max}$ obtained by the quasi-static PUM simulation and by the explicit PD simulation for the inclined crack problem, see Figure~\ref{fig:modelproblem:incline}. Even for this example, dynamic and quasi-static PUM solutions are almost equal, compare Figure~\ref{fig:inclined:dynamic:pum}.}
    \begin{tabular}{l|ccl}
       $\hPd$ & \multicolumn{2}{c}{$\mathbf{U}_\text{max}$ [\si{\meter}]} &  Load [\si{\newton}] \\\toprule
        &PD & PUM &  \\\midrule
       Coarse & \num{1.140e-4} & \num{1.421e-4} &  \num{4.25e6} \\
     Fine & \num{1.166e-4} & \num{3.136e-4} & \num{9e6}
    \end{tabular}
    \label{tab:my_label}
\end{table}

\subsection{Square plate with a local PD region}
\begin{figure}[tb]
    \centering
\begin{tikzpicture}
\draw (0,0) -- (5,0) -- (5,5) -- (0,5) -- cycle;
\draw[thick] (0,2.5) -- (2.5,2.5);
\draw[thin,gray] (2,1.5) -- (5,1.5);
\draw[thin,gray] (2,3.5) -- (5,3.5);
\draw[thin,gray] (2,1.5) -- (2,2.5);
\draw[thin,gray] (2,3.5) -- (2,2.5);
\node[gray] at (2.5+2.5/2,2.5) {PD};
\draw [<->] (2,1.3) -- (5,1.3);
\node[below] at (3.5,1.3) {$6$};
\draw (2,1.5) -- (2,1.3);
\draw [<->] (0,-0.5) -- (5,-0.5);
\node[below] at (2.5,-0.5) {$10$};
\draw (0,0) -- (0,-0.5);
\draw (5,0) -- (5,-0.5);
\draw[<->] (5.3,1.5) -- (5.3,3.5);
\node[right] at (5.3,2.5) {4};
\draw (5,1.5) -- (5.3,1.5);
\draw (5,3.5) -- (5.3,3.5);
\draw[<->] (5.8,0) -- (5.8,5);
\node[right] at (5.8,2.5) {10};
\draw (5,0) -- (5.8,0);
\draw (5,5) -- (5.8,5);
\draw[<->] (-0.3,0) -- (-0.3,2.5);
\node[left] at (-0.3,2.5/2) {5};
\draw (-0.3,0) -- (0,0);
\draw (-0.3,2.5) -- (0,2.5);
\draw[->,cadetgrey](-1.5,0) -- (-1,0);
\draw[->,cadetgrey](-1.5,0) -- (-1.5,0.5);
\node[right] at (-1,0) {$x$};
\node[above] at (-1.5,0.5) {$y$};
\foreach \i in {1,...,11}
    \draw[->,azure,thick] (5.5-0.5*\i,5) -- (5.5-0.5*\i,5.25);
\foreach \i in {0,...,10}
    \draw[->,awesome,thick] (5-0.5*\i,0) -- (5-0.5*\i,-.25);
\end{tikzpicture}
    \caption{Square plate ($10\times 10$) which is the domain of the partition of unity with an initial crack of length $5$. Around the crack existing crack tip, the gray region is the domain of peridynamics.
    On the top and bottom, a local Dirichlet boundary condition with $\pm u = \num{6.5e-6}$ is applied for the PUM. In the PD region, the displacement obtained by PU< is prescribed in a collar of size of the horizon on the top and bottom.}
    \label{fig:global:local:sketch}
\end{figure}
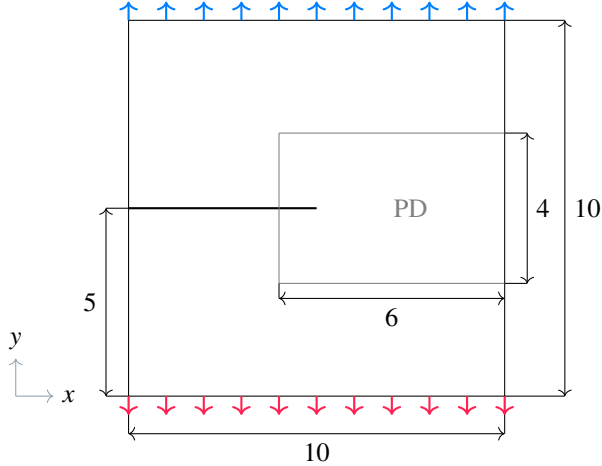

In this section, we do few cycles including crack growth of the global-local approach shown in Figure~\ref{fig:coupling:glcycle}. Figure~\ref{fig:global:local:sketch} sketches the global problem, which is a square plate ($10 \times 10$) with an initial crack of length $5$. A local Dirichlet condition is applied on the top and bottom of the plate. The local problem is the rectangle ($6 \times 4$) including a part of the initial crack. All PD bonds crossing the initial crack were softened to zero at the beginning of each PD solution. Table~\ref{tab:global:local:crack:parameters} shows the simulation parameters for the discretization in time and space for the global and local problem.
We keep the mesh width $h_\text{PUM}$ fixed and refine the PD discretization three times. 

\begin{table}[tb]
    \centering
    \caption{Simulation parameters for the discretization in time and space for the global-local problem with crack growth.}
    \label{tab:global:local:crack:parameters}
    \begin{tabular}{l|l}
        \toprule
        E = \num{2e11} & Final time $T$=\num{0.001}\si{\second}  \\
        Node spacing:  &  Time steps $t_n$=\num{50000}  \\
        $\hPum=$\num{0.078125}\si{\meter} & Horizon:  \\
        $\hPd^1=$\num{0.25}\si{\meter}  &  $\delta_1=2\hPd^\text{1}$ \\
        $\hPd^2=$\num{0.05}\si{\meter} &  $\delta_2=4\hPd^2$\\
        $\hPd^3=$\num{0.015625}\si{\meter} &  $\delta_3=8\hPd^\text{3}$\\
        \bottomrule
    \end{tabular}
\end{table}

\begin{figure}
    \centering
    \includegraphics[width=0.75\linewidth]{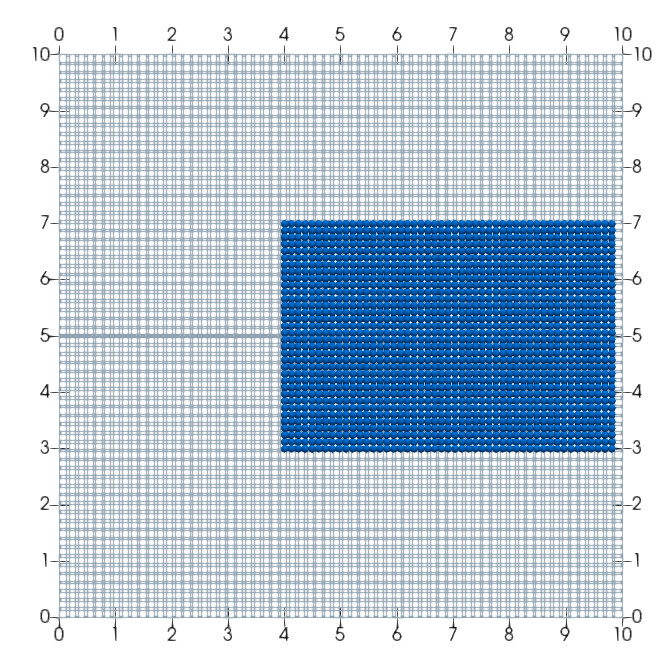}
    \caption{Discretization of the global-local approach. The gray mesh is used for the global PUM simulation. The blue dots represent the discrete PD nodes within the PD region. Note that the PUM mesh is constant and only the PD region is refined. For simplicity, we plotted the second refinement of the PD discretization.}
    \label{fig:mesh:global:local}
\end{figure}

As one example, we show the two discretizations in Figure~\ref{fig:mesh:global:local}. The gray mesh is the PUM discretization of the global problem. The blue dots represent the PD nodes in the local region. 
For the other two PD discretizations, we do not show the figures, since just more discrete PD nodes would be shown. We first start the global PUM simulation and apply the load in small load steps, which accumulate to the final load at the last time step. For each load step, we communicate the PUM displacement within a layer of horizon size $\delta$ on the top and bottom of the local region to the PD simulation. We stop the global PUM simulation and run a local PD simulation. We extract the crack path using Algorithm~\ref{algo:crack:path} from the local PD simulation and feed this information back to the global PUM simulation. For the first 85 steps, the crack was not growing in the local simulations. However, with the 86th load step, the crack started to grow. Figure~\ref{fig:square:crack:path} shows the extracted crack path until the final load for all three PD discretizations. For the coarse discretizaiton, the extracted crack and displacement differs with respect to the two other discretizations. Mostly due to the fact that the PD mesh was too coarse. However, for the two finer meshes the extracted crack tip position is very close for each of the load steps, see Figure~\ref{fig:square:crack:path:r1} and Figure~\ref{fig:square:crack:path:r2}. 
% shows the global PUM solution after the third exchange of the crack geometry and the final step. 

\begin{figure*}
    \centering
    \includegraphics[trim=350 0 350 0, width=0.495\linewidth]{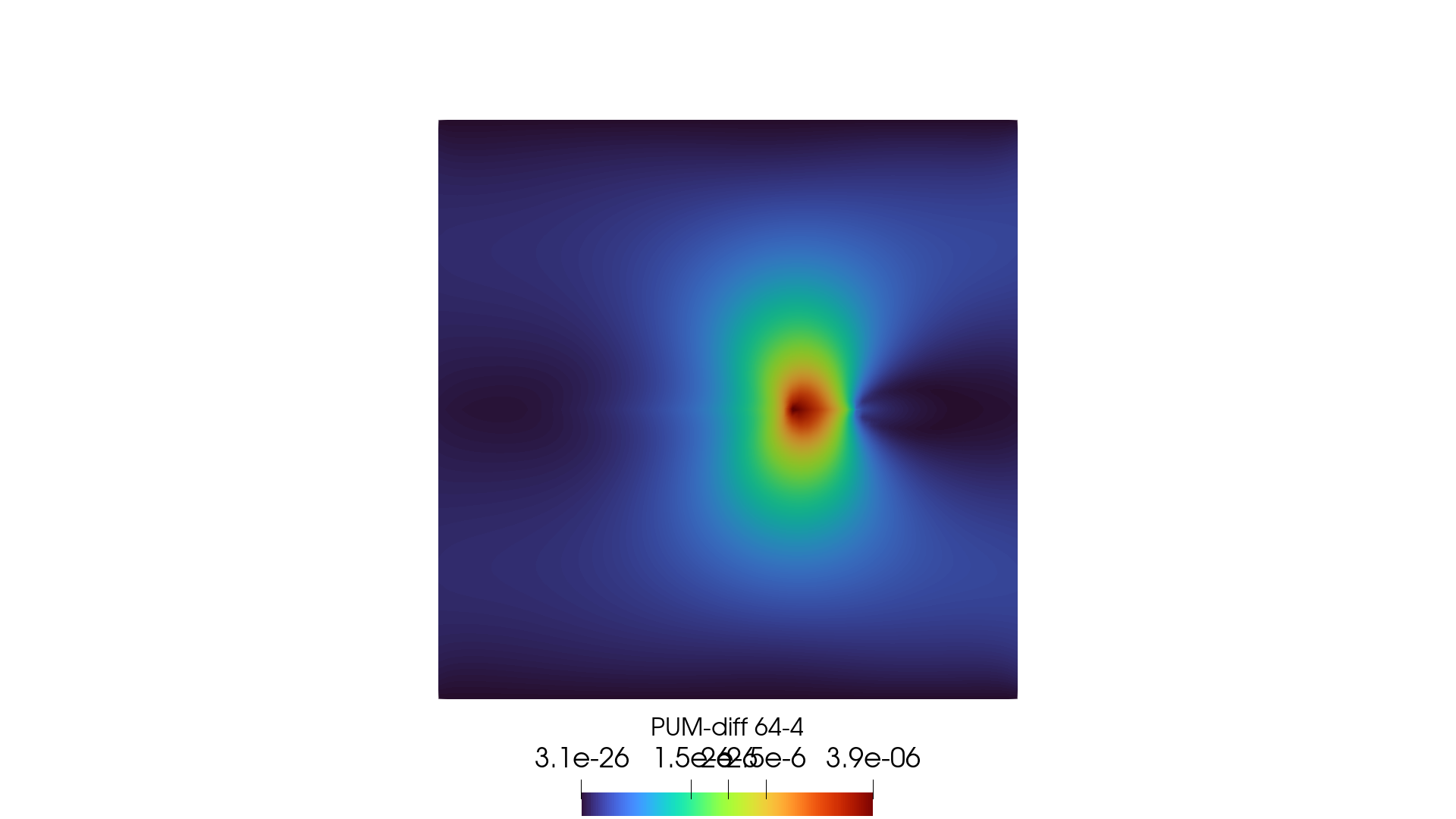}
    \includegraphics[trim=350 0 350 0, width=0.495\linewidth]{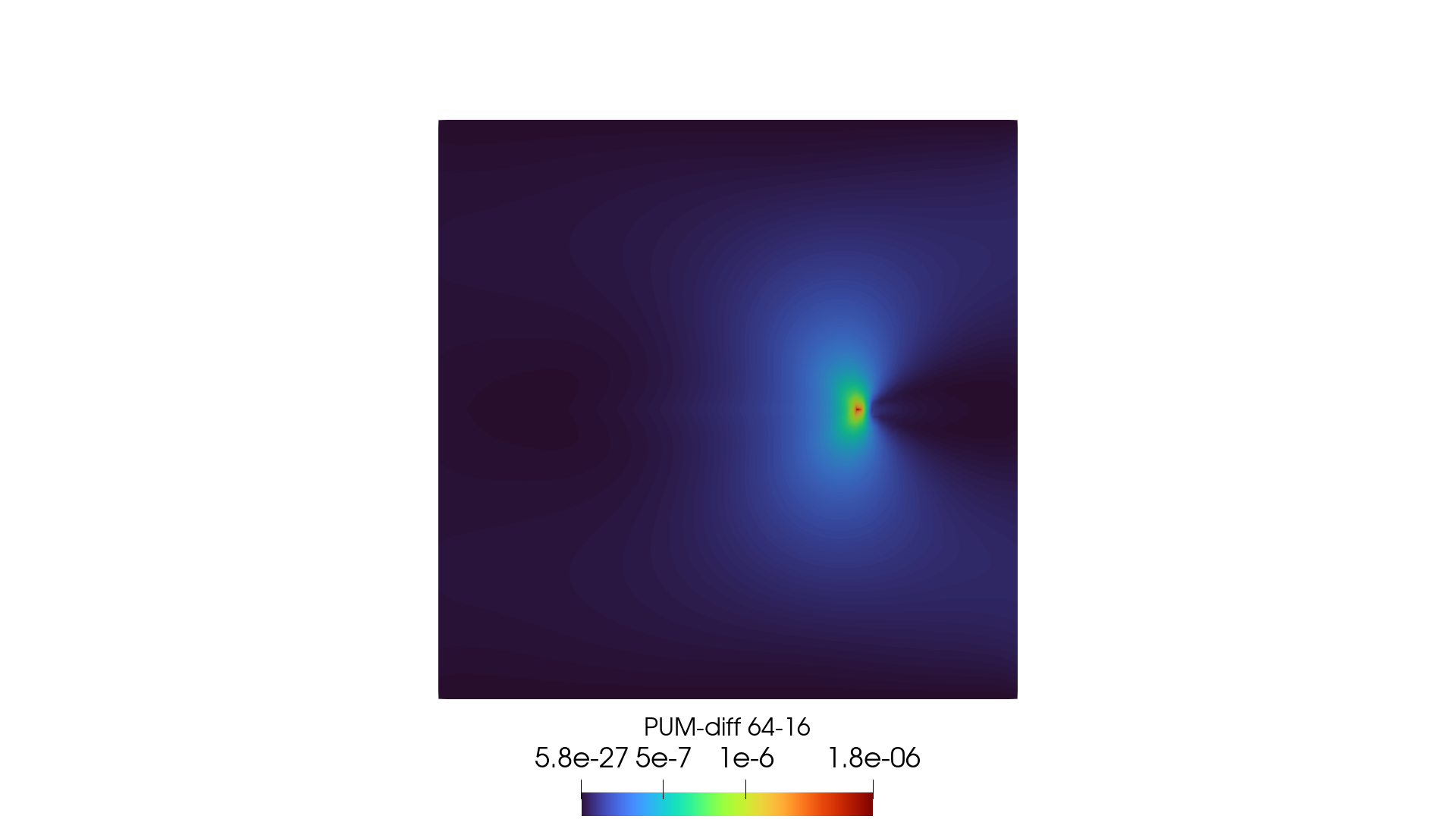}
    \includegraphics[trim=350 0 350 0, width=0.495\linewidth]{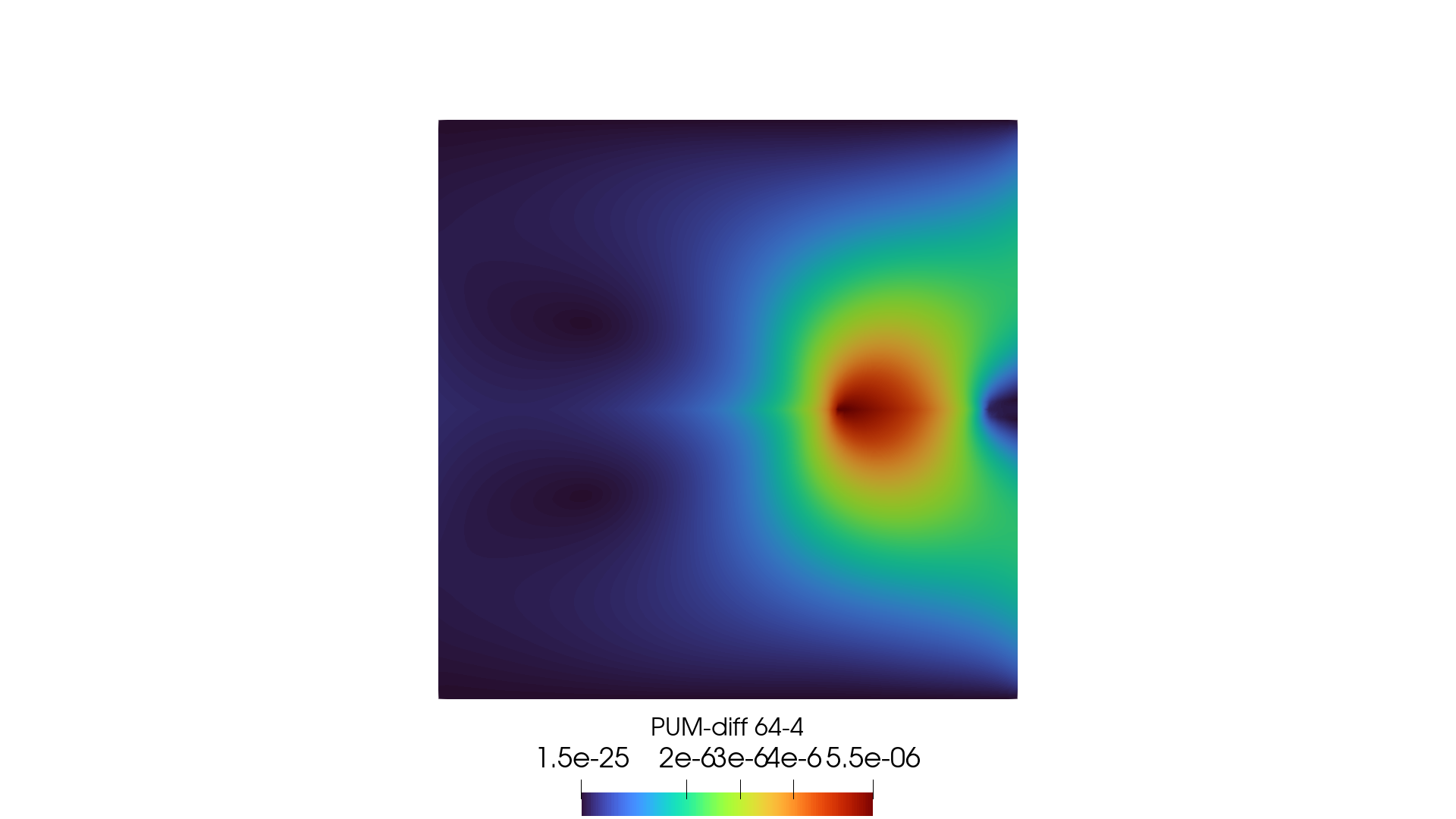}
    \includegraphics[trim=350 0 350 0, width=0.495\linewidth]{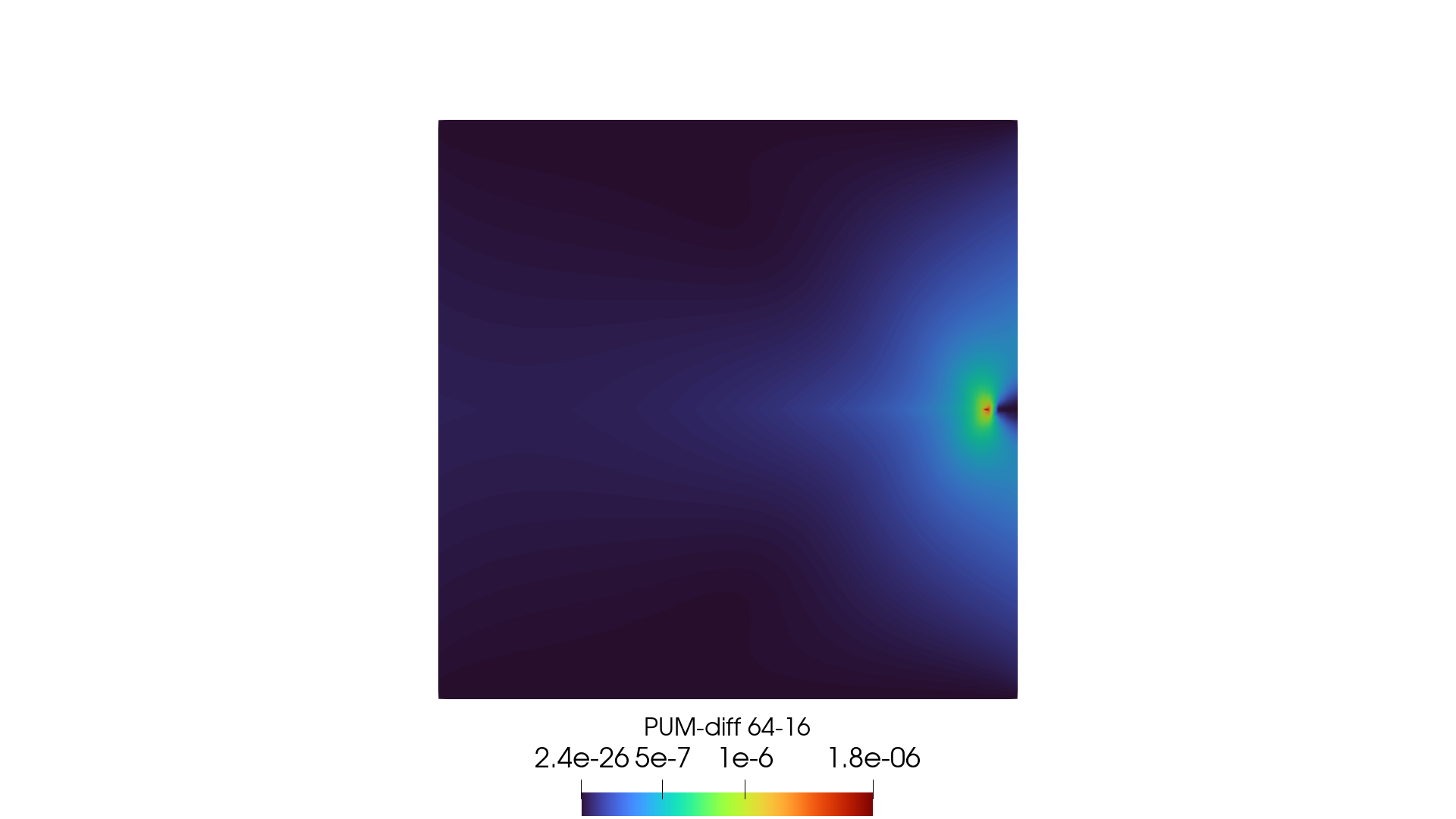}
    \caption{Differences between PUM solutions with different crack paths for third (top) and fifth (bottom) crack growth step in simple coupling problem. 
        Solution with crack path from PD solution with \(\hPd^3\) is taken as reference and difference to solution with path from \(\hPd^1\) (left) and \(\hPd^2\) (right) is shown.}
    \label{fig:global:solution}
\end{figure*}

\begin{table*}[tb]
    \centering
    \begin{tabular}{lll}\toprule
     & \multicolumn{2}{c}{Absolute Difference of the crack tip} \\
     Time step  & $\delta_1=0.5$ and $h_\text{PD}^1=\sfrac{\delta_1}{2}$  & $\delta_2=0.25$ and $h_\text{PD}^2=\sfrac{\delta_2}{4}$  \\\midrule
     43000  & 0.074 & 0.008    \\
     44000 & 0.15 & 0.024\\ 
     45000 & 0.20 & 0.016\\
     46000 & 0.4 & 0.015 \\\bottomrule
    \end{tabular}
    \caption{The absolute difference of the crack tip position, see Equation~\ref{eq:crack:tip:pd}, with respect to the finest PD discretization ($\delta_3 = 0.125$ and $h^2_\text{PD}=\sfrac{\delta_3}{8}$) and the constant PUM mesh.}
    \label{tab:crack:tip:difference}
\end{table*}
\textcolor{black}{
Table~\ref{tab:crack:tip:difference} shows the absolute difference for the extracted crack tip position
\begin{equation}
    e_\text{tip}(\delta_i,t) =    \left|\frac{crack\_pos(\delta_i,t) - crack\_pos(\delta_3,t)}{crack\_pos(\delta_3,t) }  \right|  \qquad i \in \{1, 2\}
    \label{eq:crack:tip:pd}
\end{equation}
at various load steps with respect to the finest PD discretization ($\delta_3 = 0.125$ and $h^2_\text{PD}=\sfrac{\delta_3}{8}$) while keeping the PUM mesh constant.
In addition, we computed the difference between global PUM solutions with different PD crack paths, see Figure~\ref{fig:global:solution}.
\begin{table}[tb]
    \centering
    \begin{tabular}{lll}\toprule
     & \multicolumn{2}{c}{Relative Difference} \\
     Time step  & $e_{\hPd^1}$  & $e_{\hPd^2}$  \\\midrule
     44000  & \num{1.2e-1} & \num{2.1e-2}    \\
     46000  & \num{2.5e-1} & \num{2.2e-2} \\\bottomrule
    \end{tabular}
    \caption{Relative difference~\eqref{eq:def:pum:path:error} between PUM solutions with different PD crack paths.}
    \label{tab:crack:u:difference}
\end{table}
The PUM solution \(u_3\) with crack path from PD discretization with \(\hPd^3\) is taken as reference, and we further compute the relative difference
\begin{equation}\label{eq:def:pum:path:error}
    e_{\hPd^i} := \sqrt{\frac{\sum_{j} (u_{3}(p_j) - u_i(p_j))^2 }{  (u_{3}(p_j))^2 }} \qquad i \in \{1, 2\}
\end{equation}
on points \(p_j\) of the visualization mesh, with respect to PUM solutions with crack paths from PD discretizations with mesh size \(\hPd^i\), results are shown in Table~\ref{tab:crack:u:difference}.
}
Note that the finest PD discretization had \num{1572350} nodes. 
This indicates that if the PD discretization has a good enough resolution, the crack position and displacement field does not change much. 
As a next step, is to compare against experimental data for more sophisticated comparisons.

\begin{figure*}[tb]
    \centering
    \subfloat[$\delta_1=0.5$ and $h_\text{PD}^1=\sfrac{\delta_1}{2}$\label{fig:square:crack:path:coarse} ]{
      \begin{tikzpicture}[scale=0.8]
    \begin{axis}[xmin=0,xmax=10,ymin=0,ymax=10]
    \draw [thick, draw=black]
        (axis cs: 0.,5) -- (axis cs: 5,5);
    \addplot[azure,mark=*] table [x=x1, y=y1, col sep=comma] {crack_pos_straigth.csv};
    \end{axis}
    \end{tikzpicture}
    }
    \subfloat[$\delta_2=0.25$ and $h_\text{PD}^2=\sfrac{\delta_2}{4}$\label{fig:square:crack:path:r1}]{
    \begin{tikzpicture}[scale=0.8]
    \begin{axis}[xmin=0,xmax=10,ymin=0,ymax=10]
    \draw [thick, draw=black]
        (axis cs: 0.,5) -- (axis cs: 5,5);
    \addplot[azure,mark=*] table [x=x1, y=y1, col sep=comma] {crack_pos_straigth_1.csv};
    \end{axis}
    \end{tikzpicture}
    }
    \\
    \subfloat[$\delta_3=0.125$ and $h_\text{PD}^3=\sfrac{\delta_3}{8}$\label{fig:square:crack:path:r2}]{
    \begin{tikzpicture}[scale=0.8]
    \begin{axis}[xmin=0,xmax=10,ymin=0,ymax=10]
    \draw [thick, draw=black]
        (axis cs: 0.,5) -- (axis cs: 5,5);
    \addplot[azure,mark=*] table [x=x1, y=y1, col sep=comma] {crack_pos_straigth_2.csv};
    \end{axis}
    \end{tikzpicture}
    }
    \caption{Extracted crack geometry from the global-local simulation for the initial PD mesh \protect\subref{fig:square:crack:path:coarse}, the refined mesh \protect\subref{fig:square:crack:path:r1},  and  two times refined meshes \protect\subref{fig:square:crack:path:r2}.}
    \label{fig:square:crack:path}
\end{figure*}

%%%%%%%%%%%%%%%%%%%%%%%%%%%%%%%%%%
\FloatBarrier
\section{Conclusion}
\label{sec:conclusion}
%%%%%%%%%%%%%%%%%%%%%%%%%%%%%%%%%%
In fracture mechanics problems, peridynamics allows for naturally developing and growing cracks, with the downside of high computational costs.
Partition of unity methods however require only a few degrees of freedom to model a known crack path.
The primary focus of this work has been to introduce a constructive multiscale enrichment approach that allows to combine the benefits of both methods and provide a proof of concept implementation of the proposed approach.
In numerical experiments, we have tested all steps necessary for the proposed scheme in isolation.
The presented results clearly show the validity and potential of the presented approach.
It is also clear, however, that in order to provide a proof of concept, not too complex geometries were considered.

\subsubsection*{Future work}
In the next part of the series, we want to further investigate the effect of applying a displacement or a force from the local problem as boundary conditions on the local PD problem, see Section~\ref{sec:numerical:results:coupling:pum} for some preliminary results.
Another aspect is to construct a suitable PD discretization in the local problem's domain employing the prescribed displacement, especially around the crack or discontinuity.
On the PUM side, we want to apply stable explicit time stepping schemes for enriched dynamic crack growth from the literature to avoid the small stability issues observed in the inclined crack example.

A further topic would be the automatic identification of the domain where to switch to the non-local model in a running PUM simulation, \emph{e.g.}\ by some stress or strain based damage model.
Ultimately, we want to combine all parts introduced in the series into an automated algorithm, such that information passes automatically between the PD and PUM codes.
Specifically, this requires automatic extraction of the crack path on the PD side and passing the boundary values from the PUM to the local PD problem.
Once the scheme is running, there are several more topics to investigate:
Measuring the impact of global and local time step choices.
Measuring the impact of different kinds of boundary conditions, with which we transfer information from the global PUM simulation to the local PD model.

%%%%%%%%%%%%%%%%%%%%%%%%%%%%%%%%%%
\section*{Supplementary materials}
%%%%%%%%%%%%%%%%%%%%%%%%%%%%%%%%%%
The NLMech/PeriHPX code~\cite{Diehl2020hpx,Jha2021} utilizing the C++ standard library for concurrency and parallelism (HPX)~\cite{Kaiser2020} is available on GitHub\footnote{\url{https://github.com/perihpx/}}.
NLMech/PeriHPX has the following dependencies: HPX 1.5.9, Blaze~\cite{iglberger2012high} 3.5, Blaze\_Iterative, YAML-CPP 0.6.3, hwloc 2.2.0, boost 1.73.0, jemalloc 5.1.0, gcc 9.3.0, gmsh~\cite{geuzaine2009gmsh} 4.7.1, PCL~\cite{Rusu_ICRA2011_PCL} 1.11.1, FLANN 1.9.1, and VTK 9.0.1.
The simulation input files are available on Zenodo~\cite{patrick_diehl_2021_5142369} or GitHub\footnote{\url{https://github.com/diehlpk/paperPUMPD}}.
The PUMA\footnote{\url{https://www.scai.fraunhofer.de/en/business-research-areas/meshfree-multiscale-methods/products/puma.html.}} software framework~\cite{schweitzer2017rapid} used for all PUM simulations is developed at Fraunhofer SCAI\@.

\section*{Acknowledgements}

PD thanks the Center of Computation \& Technology at Louisiana State University for supporting this work. This material is partially based upon work supported by the U. S. Army Research Laboratory and the U. S. Army Research Office under Contract/Grant Number W911NF-19-1-0245.

%\section*{Conflict of interest}

%The authors declare that they have no conflict of interest.

%\begin{acknowledgements}
%PD thanks the Center of
%\end{acknowledgements}

\bibliography{references}   % name your BibTeX data base

\begin{thebibliography}{10}

\bibitem{babuvska1997partition}
Ivo Babu{\v{s}}ka and Jens~M Melenk.
\newblock The partition of unity method.
\newblock {\em International journal for numerical methods in engineering},
  40(4):727--758, 1997.

\bibitem{birner2017global}
Matthias Birner and Marc~Alexander Schweitzer.
\newblock Global-local enrichments in puma.
\newblock In {\em International Workshop on Meshfree Methods for Partial
  Differential Equations}, pages 167--183. Springer, 2017.

\bibitem{bobaru2011adaptive}
Florin Bobaru and Youn~Doh Ha.
\newblock Adaptive refinement and multiscale modeling in 2d peridynamics.
\newblock {\em International Journal for Multiscale Computational Engineering},
  9(6), 2011.

\bibitem{bobaru2009convergence}
Florin Bobaru, Mijia Yang, Leonardo~Frota Alves, Stewart~A Silling, Ebrahim
  Askari, and Jifeng Xu.
\newblock Convergence, adaptive refinement, and scaling in 1d peridynamics.
\newblock {\em International Journal for Numerical Methods in Engineering},
  77(6):852--877, 2009.

\bibitem{bourdin2000numerical}
Blaise Bourdin, Gilles~A Francfort, and Jean-Jacques Marigo.
\newblock Numerical experiments in revisited brittle fracture.
\newblock {\em Journal of the Mechanics and Physics of Solids}, 48(4):797--826,
  2000.

\bibitem{bussler2017visualization}
Michael Bussler, Patrick Diehl, Dirk Pflüger, Steffen Frey, Filip Sadlo,
  Thomas Ertl, and Marc~Alexander Schweitzer.
\newblock Visualization of fracture progression in peridynamics.
\newblock {\em Computers \& Graphics}, 67:45--57, 2017.

\bibitem{chang2002explicit}
Shuenn-Yih Chang.
\newblock Explicit pseudodynamic algorithm with unconditional stability.
\newblock {\em Journal of Engineering Mechanics}, 128(9):935--947, 2002.

\bibitem{chang2007enhanced}
Shuenn-Yih Chang.
\newblock Enhanced, unconditionally stable, explicit pseudodynamic algorithm.
\newblock {\em Journal of Engineering Mechanics}, 133(5):541--554, 2007.

\bibitem{chen2011continuous}
Xi~Chen and Max Gunzburger.
\newblock Continuous and discontinuous finite element methods for a
  peridynamics model of mechanics.
\newblock {\em Computer Methods in Applied Mechanics and Engineering},
  200(9-12):1237--1250, 2011.

\bibitem{patrick_diehl_2021_5142369}
Patrick Diehl.
\newblock A fracture multiscale model for peridynamic enrichment within the
  partition of unity method: Part i, December 2017.
\newblock If you use this software, please cite it as below.

\bibitem{diehl_2020_emu}
Patrick Diehl.
\newblock Emu-nodal discretization, May 2020.

\bibitem{diehl2017extraction}
Patrick Diehl, Michael Bußler, Dirk Pflüger, Steffen Frey, Thomas Ertl, Filip
  Sadlo, and Marc~Alexander Schweitzer.
\newblock Extraction of fragments and waves after impact damage in
  particle-based simulations.
\newblock In {\em Meshfree Methods for Partial Differential Equations VIII},
  pages 17--34. Springer, 2017.

\bibitem{diehl2016bond}
Patrick Diehl, Fabian Franzelin, Dirk Pflüger, and Georg~C Ganzenmüller.
\newblock Bond-based peridynamics: a quantitative study of mode i crack
  opening.
\newblock {\em International Journal of Fracture}, 201(2):157--170, 2016.

\bibitem{Diehl2020hpx}
Patrick Diehl, Prashant~K. Jha, Hartmut Kaiser, Robert Lipton, and Martin
  Lévesque.
\newblock An asynchronous and task-based implementation of peridynamics
  utilizing hpx---the c++ standard library for parallelism and concurrency.
\newblock {\em SN Applied Sciences}, 2(12):2144, Dec 2020.

\bibitem{https://doi.org/10.1002/nme.7005}
Patrick Diehl and Robert Lipton.
\newblock Quasistatic fracture using nonlinear-nonlocal elastostatics with
  explicit tangent stiffness matrix.
\newblock {\em International Journal for Numerical Methods in Engineering}.

\bibitem{diehl_lipton_wick_tyagi_2021}
Patrick Diehl, Robert Lipton, Thomas Wick, and Mayank Tyagi.
\newblock A comparative review of peridynamics and phase-field models for
  engineering fracture mechanics.
\newblock {\em Computational Mechanics}, pages 1--35, 2022.

\bibitem{diehl2019review}
Patrick Diehl, Serge Prudhomme, and Martin Lévesque.
\newblock A review of benchmark experiments for the validation of peridynamics
  models.
\newblock {\em Journal of Peridynamics and Nonlocal Modeling}, 1(1):14--35,
  2019.

\bibitem{dipasquale2014crack}
Daniele Dipasquale, Mirco Zaccariotto, and Ugo Galvanetto.
\newblock Crack propagation with adaptive grid refinement in 2d peridynamics.
\newblock {\em International Journal of Fracture}, 190(1-2):1--22, 2014.

\bibitem{du2015robust}
Qiang Du and Xiaochuan Tian.
\newblock Robust discretization of nonlocal models related to peridynamics.
\newblock In {\em Meshfree methods for partial differential equations VII},
  pages 97--113. Springer, 2015.

\bibitem{duarte2000generalized}
C.~A. Duarte, I.~Babuška, and J.~T. Oden.
\newblock Generalized finite element methods for three-dimensional structural
  mechanics problems.
\newblock {\em Computers \& Structures}, 77(2):215--232, 2000.

\bibitem{duarte2007global}
C~Armando Duarte, Dae-Jin Kim, and Ivo Babuška.
\newblock A global-local approach for the construction of enrichment functions
  for the generalized fem and its application to three-dimensional cracks.
\newblock In {\em Advances in meshfree techniques}, pages 1--26. Springer,
  2007.

\bibitem{d2019review}
Marta D’Elia, Xingjie Li, Pablo Seleson, Xiaochuan Tian, and Yue Yu.
\newblock A review of local-to-nonlocal coupling methods in nonlocal diffusion
  and nonlocal mechanics.
\newblock {\em Journal of Peridynamics and Nonlocal Modeling}, pages 1--50,
  2021.

\bibitem{emmrich2007peridynamic}
Etienne Emmrich and Olaf Weckner.
\newblock The peridynamic equation and its spatial discretisation.
\newblock {\em Mathematical Modelling and Analysis}, 12(1):17--27, 2007.

\bibitem{FishBook}
Jacob Fish.
\newblock {\em Practical Multiscaling}.
\newblock Wiley, 2013.

\bibitem{GALVANETTO201641}
Ugo Galvanetto, Teo Mudric, Arman Shojaei, and Mirco Zaccariotto.
\newblock An effective way to couple fem meshes and peridynamics grids for the
  solution of static equilibrium problems.
\newblock {\em Mechanics Research Communications}, 76:41 -- 47, 2016.

\bibitem{geelen2020extended}
Rudy Geelen, Julia Plews, Michael Tupek, and John Dolbow.
\newblock An extended/generalized phase-field finite element method for crack
  growth with global-local enrichment.
\newblock {\em International Journal for Numerical Methods in Engineering},
  121(11):2534--2557, 2020.

\bibitem{geuzaine2009gmsh}
Christophe Geuzaine, Jean-François Remacle, et~al.
\newblock Gmsh: a three-dimensional finite element mesh generator with built-in
  pre-and post-processing facilities.
\newblock {\em International journal for numerical methods in engineering},
  79(11):1309--1331, 2009.

\bibitem{giannakeas2020coupling2}
Ilias~N Giannakeas, Theodosios~K Papathanasiou, Arash~S Fallah, and Hamid
  Bahai.
\newblock Coupling xfem and peridynamics for brittle fracture simulation: part
  ii—adaptive relocation strategy.
\newblock {\em Computational Mechanics}, 66(3):683--705, 2020.

\bibitem{giannakeas2020coupling1}
Ilias~N Giannakeas, Theodosios~K Papathanasiou, Arash~S Fallah, and Hamid
  Bahai.
\newblock Coupling xfem and peridynamics for brittle fracture simulation—part
  i: feasibility and effectiveness.
\newblock {\em Computational Mechanics}, 66(1):103--122, 2020.

\bibitem{gravouil2009explicit}
Anthony Gravouil, Thomas Elguedj, and Hubert Maigre.
\newblock An explicit dynamics extended finite element method. part 2:
  Element-by-element stable-explicit/explicit dynamic scheme.
\newblock {\em Computer Methods in Applied Mechanics and Engineering},
  198(30-32):2318--2328, 2009.

\bibitem{han2012coupling}
Fei Han and Gilles Lubineau.
\newblock Coupling of nonlocal and local continuum models by the arlequin
  approach.
\newblock {\em International Journal for Numerical Methods in Engineering},
  89(6):671--685, 2012.

\bibitem{iglberger2012high}
Klaus Iglberger, Georg Hager, Jan Treibig, and Ulrich Rüde.
\newblock High performance smart expression template math libraries.
\newblock In {\em 2012 International Conference on High Performance Computing
  \& Simulation (HPCS)}, pages 367--373. IEEE, 2012.

\bibitem{javili2019peridynamics}
Ali Javili, Rico Morasata, Erkan Oterkus, and Selda Oterkus.
\newblock Peridynamics review.
\newblock {\em Mathematics and Mechanics of Solids}, 24(11):3714--3739, 2019.

\bibitem{Jha2021}
Prashant~K. Jha and Patrick Diehl.
\newblock Nlmech: Implementation of finite difference/meshfree discretization
  of nonlocal fracture models.
\newblock {\em Journal of Open Source Software}, 6(65):3020, 2021.

\bibitem{CMPer-JhaLipton}
Prashant~K Jha and Robert Lipton.
\newblock Numerical analysis of nonlocal fracture models in holder space.
\newblock {\em SIAM Journal on Numerical Analysis}, 56(2):906--941, 2018.

\bibitem{CMPer-JhaLipton8}
Prashant~Kumar Jha and Robert Lipton.
\newblock Numerical convergence of finite difference approximations for state
  based peridynamic fracture models.
\newblock {\em Computer Methods in Applied Mechanics and Engineering},
  351:184--225, 2019.

\bibitem{CMPer-JhaLipton3}
Prashant~Kumar Jha and Robert Lipton.
\newblock Finite element convergence for state-based peridynamic fracture
  models.
\newblock {\em Communications on Applied Mathematics and Computation},
  2:93--128, 2020.

\bibitem{jin2021coupling}
Suyeong Jin, Young~Kwang Hwang, and Jung-Wuk Hong.
\newblock Coupling of non-ordinary state-based peridynamics and finite element
  method with reduced boundary effect.
\newblock {\em International Journal for Numerical Methods in Engineering},
  122(16):4033--4054, 2021.

\bibitem{Kaiser2020}
Hartmut Kaiser, Patrick Diehl, Adrian~S. Lemoine, Bryce~Adelstein Lelbach,
  Parsa Amini, Agustín Berge, John Biddiscombe, Steven~R. Brandt, Nikunj
  Gupta, Thomas Heller, Kevin Huck, Zahra Khatami, Alireza Kheirkhahan, Auriane
  Reverdell, Shahrzad Shirzad, Mikael Simberg, Bibek Wagle, Weile Wei, and
  Tianyi Zhang.
\newblock Hpx - the c++ standard library for parallelism and concurrency.
\newblock {\em Journal of Open Source Software}, 5(53):2352, 2020.

\bibitem{kilic2010coupling}
Bahattin Kilic and Erdogan Madenci.
\newblock Coupling of peridynamic theory and the finite element method.
\newblock {\em Journal of Mechanics of Materials and Structures},
  5(5):707--733, 2010.

\bibitem{kunin1975theory}
I.~A. Kunin.
\newblock {\em Elastic Media with Microstructure II: Three-Dimensional Models
  (Springer Series in Solid-State Sciences)}.
\newblock Springer, softcover reprint of the original 1st ed. 1983 edition, 1
  2012.

\bibitem{doi:10.1063/1.4971634}
Christopher~J. Lammi and Min Zhou.
\newblock Multi-scale peridynamic modeling of dynamic fracture in concrete.
\newblock {\em AIP Conference Proceedings}, 1793(1):100009, 2017.

\bibitem{lipton2014dynamic}
Robert Lipton.
\newblock Dynamic brittle fracture as a horizon limit of peridynamics.
\newblock {\em Journal of Elasticity}, 117(1):21--50, 2014.

\bibitem{lipton2016cohesive}
Robert Lipton.
\newblock Cohesive dynamics and brittle fracture.
\newblock {\em Journal of Elasticity}, 124(2):143--191, 2016.

\bibitem{LiptonJhaBdry}
Robert Lipton and Prashant~K. Jha.
\newblock Nonlocal elastodynamics and fracture.
\newblock {\em Nonlinear Differential Equations and Applications}, 28(Article
  number:23), 2021.

\bibitem{littlewood2016peridynamic}
David~John Littlewood, Stephen~D Bond, and Timothy Costa.
\newblock Peridynamic multiscale finite element methods.
\newblock Technical report, Sandia National Lab.(SNL-NM), Albuquerque, NM
  (United States), 2016.

\bibitem{liu2012coupling}
Wenyang Liu and Jung-Wuk Hong.
\newblock A coupling approach of discretized peridynamics with finite element
  method.
\newblock {\em Computer Methods in Applied Mechanics and Engineering},
  245:163--175, 2012.

\bibitem{macek2007peridynamics}
Richard~W Macek and Stewart~A Silling.
\newblock Peridynamics via finite element analysis.
\newblock {\em Finite elements in analysis and design}, 43(15):1169--1178,
  2007.

\bibitem{madenci2018coupling}
Erdogan Madenci, Atila Barut, Mehmet Dorduncu, and Nam~D Phan.
\newblock Coupling of peridynamics with finite elements without an overlap
  zone.
\newblock In {\em 2018 AIAA/ASCE/AHS/ASC Structures, Structural Dynamics, and
  Materials Conference}, page 1462, 2018.

\bibitem{melenk1996partition}
J.~M. Melenk and I.~Babuška.
\newblock The partition of unity finite element method: basic theory and
  applications.
\newblock {\em Computer methods in applied mechanics and engineering},
  139(1-4):289--314, 1996.

\bibitem{moes1999finite}
N.~Moës, J.~Dolbow, and T.~Belytschko.
\newblock A finite element method for crack growth without remeshing.
\newblock {\em International journal for numerical methods in engineering},
  46(1):131--150, 1999.

\bibitem{NIKPAYAM2019308}
Jaber Nikpayam and Mohammad~Ali Kouchakzadeh.
\newblock A variable horizon method for coupling meshfree peridynamics to fem.
\newblock {\em Computer Methods in Applied Mechanics and Engineering},
  355:308--322, 2019.

\bibitem{ONGARO2021113515}
Greta Ongaro, Pablo Seleson, Ugo Galvanetto, Tao Ni, and Mirco Zaccariotto.
\newblock Overall equilibrium in the coupling of peridynamics and classical
  continuum mechanics.
\newblock {\em Computer Methods in Applied Mechanics and Engineering},
  381:113515, 2021.

\bibitem{parks2008implementing}
Michael~L Parks, Richard~B Lehoucq, Steven~J Plimpton, and Stewart~A Silling.
\newblock Implementing peridynamics within a molecular dynamics code.
\newblock {\em Computer Physics Communications}, 179(11):777--783, 2008.

\bibitem{prudhomme2020treatment}
Serge Prudhomme and Patrick Diehl.
\newblock On the treatment of boundary conditions for bond-based peridynamic
  models.
\newblock {\em Computer Methods in Applied Mechanics and Engineering},
  372:113391, 2020.

\bibitem{Rahman_2014}
Rezwanur Rahman, John~T. Foster, and A.~Haque.
\newblock A multiscale modeling scheme based on peridynamic theory.
\newblock {\em International Journal for Multiscale Computational Engineering},
  12(3):223--248, 2014.

\bibitem{rahman2014multiscale}
Rezwanur Rahman, John~T Foster, and Anwarul Haque.
\newblock A multiscale modeling scheme based on peridynamic theory.
\newblock {\em International Journal for Multiscale Computational Engineering},
  12(3), 2014.

\bibitem{Rusu_ICRA2011_PCL}
Radu~Bogdan Rusu and Steve Cousins.
\newblock 3d is here: Point cloud library (pcl).
\newblock In {\em IEEE International Conference on Robotics and Automation
  (ICRA)}, pages 1--4, Shanghai, China, May 9-13 2011.

\bibitem{schweitzer2003phd}
M.~A. Schweitzer.
\newblock {\em A Parallel Multilevel Partition of Unity Method for Elliptic
  Partial Differential Equations}, volume~29 of {\em Lecture Notes in
  Computational Science and Engineering}.
\newblock Springer, 2003.

\bibitem{schweitzer2009algebraic}
M.~A. Schweitzer.
\newblock An algebraic treatment of essential boundary conditions in the
  particle--partition of unity method.
\newblock {\em SIAM Journal on Scientific Computing}, 31(2):1581--1602, 2009.

\bibitem{schweitzer2017rapid}
M.~A. Schweitzer and A.~Ziegenhagel.
\newblock Rapid enriched simulation application development with puma.
\newblock In {\em Scientific Computing and Algorithms in Industrial
  Simulations}, pages 207--226. Springer, 2017.

\bibitem{schweitzer2009adaptive}
M.A. Schweitzer.
\newblock An adaptive hp-version of the multilevel particle--partition of unity
  method.
\newblock {\em Comput. Methods Appl. Mech. Engrg.}, 198:1260--1272, 2009.

\bibitem{schweitzer2013variational}
Marc~Alexander Schweitzer.
\newblock Variational mass lumping in the partition of unity method.
\newblock {\em SIAM Journal on Scientific Computing}, 35(2):A1073--A1097, 2013.

\bibitem{schweitzer2014moving}
Marc~Alexander Schweitzer and Sa~Wu.
\newblock A moving least squares approach to the construction of discontinuous
  enrichment functions.
\newblock In {\em Singular Phenomena and Scaling in Mathematical Models}, pages
  347--360. Springer, 2014.

\bibitem{seleson2013force}
Pablo Seleson, Samir Beneddine, and Serge Prudhomme.
\newblock A force-based coupling scheme for peridynamics and classical
  elasticity.
\newblock {\em Computational Materials Science}, 66:34--49, 2013.

\bibitem{seleson2015concurrent}
Pablo Seleson, Youn~Doh Ha, and Samir Beneddine.
\newblock Concurrent coupling of bond-based peridynamics and the navier
  equation of classical elasticity by blending.
\newblock {\em International Journal for Multiscale Computational Engineering},
  13(2), 2015.

\bibitem{shojaei2016coupled}
A~Shojaei, T~Mudric, M~Zaccariotto, and U~Galvanetto.
\newblock A coupled meshless finite point/peridynamic method for 2d dynamic
  fracture analysis.
\newblock {\em International Journal of Mechanical Sciences}, 119:419--431,
  2016.

\bibitem{silling2020Couplingstresses}
SA~Silling.
\newblock Local-nonlocal coupling in emu/pdms.
\newblock {\em Sandia Report (SAND2020-11382)}, 2020.

\bibitem{silling2015variable}
Stewart Silling, David Littlewood, and Pablo Seleson.
\newblock Variable horizon in a peridynamic medium.
\newblock {\em Journal of Mechanics of Materials and Structures},
  10(5):591--612, 2015.

\bibitem{silling2000reformulation}
Stewart~A Silling.
\newblock Reformulation of elasticity theory for discontinuities and long-range
  forces.
\newblock {\em Journal of the Mechanics and Physics of Solids}, 48(1):175--209,
  2000.

\bibitem{silling2005meshfree}
Stewart~A Silling and Ebrahim Askari.
\newblock A meshfree method based on the peridynamic model of solid mechanics.
\newblock {\em Computers \& structures}, 83(17-18):1526--1535, 2005.

\bibitem{talebi2014computational}
Hossein Talebi, Mohammad Silani, Stephane Bordas, Pierre Kerfriden, and Timon
  Rabczuk.
\newblock A computational library for multiscale modeling of material failure.
\newblock {\em Computational Mechanics}, 53(5):1047--1071, 2014.

\bibitem{tian2014asymptotically}
Xiaochuan Tian and Qiang Du.
\newblock Asymptotically compatible schemes and applications to robust
  discretization of nonlocal models.
\newblock {\em SIAM Journal on Numerical Analysis}, 52(4):1641--1665, 2014.

\bibitem{parksyueyoutrask}
Nathaniel Trask, Huaiqian You, Yue Yu, and Michael~L. Parks.
\newblock An asymptotically compatible meshfree quadrature rule for nonlocal
  problems with applications to peridynamics.
\newblock {\em Computer Methods in Applied Mechanics and Engineering},
  343:151--165, 2019.

\bibitem{weckner2005numerical}
Olaf Weckner and Etienne Emmrich.
\newblock Numerical simulation of the dynamics of a nonlocal, inhomogeneous,
  infinite bar.
\newblock {\em J. Comput. Appl. Mech}, 6(2):311--319, 2005.

\bibitem{wu2020phase}
Jian-Ying Wu, Vinh~Phu Nguyen, Chi~Thanh Nguyen, Danas Sutula, Sina Sinaie, and
  St{\'e}phane~PA Bordas.
\newblock Phase-field modeling of fracture.
\newblock {\em Advances in applied mechanics}, 53:1--183, 2020.

\bibitem{xu2016multiscale}
Feifei Xu, Max Gunzburger, John Burkardt, and Qiang Du.
\newblock A multiscale implementation based on adaptive mesh refinement for the
  nonlocal peridynamics model in one dimension.
\newblock {\em Multiscale Modeling \& Simulation}, 14(1):398--429, 2016.

\bibitem{zaccariotto2018coupling}
Mirco Zaccariotto, Teo Mudric, Davide Tomasi, Arman Shojaei, and Ugo
  Galvanetto.
\newblock Coupling of fem meshes with peridynamic grids.
\newblock {\em Computer Methods in Applied Mechanics and Engineering},
  330:471--497, 2018.

\bibitem{zaccariotto2017enhanced}
Mirco Zaccariotto, Davide Tomasi, and Ugo Galvanetto.
\newblock An enhanced coupling of pd grids to fe meshes.
\newblock {\em Mechanics Research Communications}, 84:125--135, 2017.

\end{thebibliography}

\end{document}